\documentclass[manuscript]{acmart}
\usepackage{microtype}%if unwanted, comment out or use option "draft"

\usepackage{multicol,amsthm}
\usepackage{graphicx,url}
\usepackage{color,setspace,enumitem}
\usepackage{epstopdf}
\usepackage{caption}
\usepackage{times}
\usepackage{array}
\usepackage{color}

\usepackage[final]{changes}
\usepackage{algpseudocode}
\usepackage[ruled]{algorithm}
\usepackage{ioa_code}
\usepackage{scalerel,stackengine}
\usepackage[symbol]{footmisc}
\newtheorem{theorem}{Theorem}
\newtheorem{lemma}[theorem]{Lemma}

\newtheorem{remark}[theorem]{Remark}
\newtheorem{definition}[theorem]{Definition}

\newcommand{\prf}[1]{{}}

\newtheorem{property}{Property}

\renewcommand{\time}[1]{\ms{time}(#1)}

\newtheorem{Def}{Definition}[section]

\newcommand{\sacode}[5]
{ \vspace{.06in} \hrule \vspace{.06in} %///AS reduce vspace
 \noindent {\bf #1}: \\
 \footnotesize \noindent {\bf Signature:}\B \nobreak
 \normalsize \begin{quote} \nobreak #2 \end{quote}
 \footnotesize \noindent {\bf States:}\B \nobreak
 \begin{quote} \nobreak #3 \end{quote}
 \noindent {\bf Transitions:} \nobreak
 \vspace{-.2in} \nobreak
 \normalsize #4
 \vspace{-.06in} \hrule \vspace{.06in} %///AS reduce vspace
}

\newcommand{\act}[1]{%
    \relax\ifmmode
        \mathord{\mathcode`\-="702D\sf #1\mathcode`\-="2200}%
    \else
        $\mathord{\mathcode`\-="702D\sf #1\mathcode`\-="2200}$%
    \fi
}

\newcommand{\tup}[1]{%
    \relax\ifmmode
      \langle #1 \rangle%
    \else
        $\langle$#1$\rangle$%
    \fi
}

\newcommand{\seq}[1]{%
    \relax\ifmmode
      \langle \! \langle #1 \rangle \! \rangle%
    \else
        $\langle \! \langle$ #1 $\rangle \! \rangle$%
    \fi
}

\newcommand{\B}{\vspace*{-\smallskipamount}}

\newcommand{\FF}{\vspace*{\medskipamount}}

\newcommand{\TT}{\hspace*{2em}}

\newcommand{\N}{\mathbb N}

\newcommand{\ms}[1]{%
    \relax\ifmmode
        \mathord{\mathcode`\-="702D\it #1\mathcode`\-="2200}%
    \else
        {\it #1}%
    \fi
}

\newcommand{\lit}[1]{%
    \relax\ifmmode
        \mathord{\mathcode`\-="702D\sf #1\mathcode`\-="2200}%
    \else
        {\it #1}%
    \fi
}

\newcommand{\XDK}[1]{}% To keep stuff for DISC final
\newcommand{\remove}[1]{} % To remove stuff
\newcommand{\proofremove}[1]{} % To remove stuff
\newcommand{\uselater}[1]{} % To remove stuff

\makeatletter
\def\mainlistofsymbols{
  \normalsize
  \vspace*{1.5 em}
  \@starttoc{los}
}

\def\partonelistofsymbols{
  \normalsize
  \vspace*{1.5 em}
  \@starttoc{p1los}
}

\def\parttwolistofsymbols{
  \normalsize
  \vspace*{1.5 em}
  \@starttoc{p2los}
}

\def\l@symbol#1#2{\addpenalty{-\@highpenalty} \vskip 4pt plus 2pt
{\@dottedtocline{0}{0em}{8em}{#1}{#2}}}
\makeatother

\newcommand{\newhiddensym}[2]{%
}

\newcommand{\algIOA}[2]{\ifmmode{\text{#1}_{#2}}\else{$\text{#1}_{#2}$}\fi}
\newcommand{\EX}{\ifmmode{\xi}\else{$\xi$}\fi}
\newcommand{\EXF}{\ifmmode{\phi}\else{$\phi$}\fi}
\newcommand{\acts}{\alpha}
\renewcommand{\state}{\sigma}
\newcommand{\st}{\sigma}
\newcommand{\quo}[1]{Q_{#1}}
\newcommand{\inter}[1]{
	\ifmmode{\left(\bigcap_{\mathcal{Q}\in#1}\mathcal{Q}\right)}
	\else{$\left(\bigcap_{\mathcal{Q}\in#1}\mathcal{Q}\right)$}
	\fi
}
\newcommand{\idSet}{\mathcal{I}}
\newcommand{\wSet}{\mathcal{W}}
\newcommand{\rdSet}{\mathcal{R}}
\newcommand{\recSet}{\mathcal{G}}
\newcommand{\srvSet}{\mathcal{S}}
\newcommand{\cSet}{\mathcal{I}}

\newcommand{\confSet}{\mathcal{C}}
\newcommand{\servers}[1]{ #1.Servers}
\newcommand{\quorums}[1]{ #1.Quorums}
\newcommand{\consensus}[1]{ #1.Con}

\newcommand{\op}{\pi}
\newcommand{\wrt}{\omega}

\mathchardef\mhyphen="2D
\newcommand{\pr}{p}
\newcommand{\rdr}{r}
\newcommand{\rec}{rc}

\newcommand{\bef}{\rightarrow}

\newcommand{\vid}[1]{\ifmmode{\nu_{#1}}\else{$\nu_{#1}$}\fi}
\newcommand{\seen}{\ifmmode{seen}\else{$seen$}\fi}
\newcommand{\ABD}{{\sc ABD}}
\newcommand{\mwABD}{{\sc mw}\ABD{}}
\newcommand{\fast}{{\sc Fast}}
\newcommand{\ares}{{\sc Ares}}
\newcommand{\valSet}{{\mathcal V}}
\newcommand{\tsSet}{{\mathcal T}}
\newcommand{\tg}[1]{\tau_{#1}}

\newcommand{\maxts}[1]{\ifmmode{maxTS_{#1}}\else{$maxTS_{#1}$}\fi}
\newcommand{\maxtag}[1]{\ifmmode{maxTag_{#1}}\else{$maxTag_{#1}$}\fi}
\newcommand{\maxpair}[1]{\ifmmode{maxMPair_{#1}}\else{$maxMPair_{#1}$}\fi}
\newcommand{\mintag}[1]{\ifmmode{minTag_{#1}}\else{$minTag_{#1}$}\fi}
\newcommand{\maxps}{\ifmmode{maxPS}\else{$maxPS$}\fi}
\newcommand{\conftg}[1]{\ifmmode{confirmed_{#1}}\else{$confirmed_{#1}$}\fi}
\newcommand{\maxconftag}{\ifmmode{\ms{maxCT}}\else{$maxCT$}\fi}

\usepackage[binary-units=true]{siunitx}

\definechangesauthor[name=Nicolas, color=red]{NN}
\definechangesauthor[name=Viveck, color=green]{VC}
\definechangesauthor[name=Kishori, color=blue]{KK}
\newcommand{\nnfix}[1]{{\color{black} #1}}
\newcommand{\myemph}[1]{{\it #1}}
\newcommand{\nn}[1]{\added[id=NN]{#1}}
\newcommand{\nnrev}[2]{\replaced[id=NN]{#2}{#1}}

\newcommand{\kmk}[1]{\added[id=KK]{#1}}

\newcommand{\vc}[1]{\added[id=VC]{#1}}
\newtheorem*{theorem*}{{\bf Theorem}}
\newtheorem*{lemma*}{{\bf Lemma}}

\newcommand{\myparagraph}[1]{\smallskip\noindent{\textbf{#1}}}
\newcommand{\cvec}[2]{\mathbf{c}^{#1}_{#2}}
\newcommand{\atT}[2]{#1|_{#2}}
\newcommand{\status}[1]{#1.status}
\newcommand{\config}[1]{#1.cfg}
\newcommand{\GetTag}{{ \it{get-tag}}}

\newcommand{\PutData}{{ \it{put-data}}}

\newcommand{\GetData}{{ \it{get-data}}}

\newcommand{\Coded}{code\act{-}elems}

\newcommand{\gseq}{{\mathcal{G}_L}}

\newcommand{\smdelay}{d}
\newcommand{\lgdelay}{D}
\newcommand{\opdelay}[1]{T(#1)}
\newcommand{\opdelaymin}[1]{T_{min}(#1)}

\newcommand{\dap}[1]{{DAP(#1)}}

\algblockdefx[Operation]{Operation}{EndOperation}%
[2]{{\bf operation} $\act{#1}$(#2)}%
{{\bf end operation}}
\algblockdefx[Procedure]{Procedure}{EndProcedure}%
[2]{{\bf procedure} $\act{#1}$(#2)}%
{{\bf end procedure}}
\algblockdefx[Receive]{Receive}{EndReceive}%
[2]{{\bf Upon receive} (#1)$_{\text{ #2 }}${\bf from} $q$}%
{{\bf end receive}}

\newcommand\wwidehat[1]{%
\savestack{\tmpbox}{\stretchto{%
  \scaleto{%
    \scalerel*[\widthof{\ensuremath{#1}}]{\kern-.6pt\bigwedge\kern-.6pt}%
    {\rule[-\textheight/2]{1ex}{\textheight}}%WIDTH-LIMITED BIG WEDGE
  }{\textheight}% 
}{0.5ex}}%
\stackon[1pt]{#1}{\tmpbox}%
}

\newcommand{\daputdata}[2]{ {#1}.{\act{put-data}(#2)}}
\newcommand{\dagetdata}[1]{ {#1}.{\act{get-data}()}}
\newcommand{\dagettag}[1]{ {#1}.{\act{get-tag}()}}
\newcommand{\treas}{{\sc Treas}}

\newcommand{\cseq}[1]{\widehat{#1}}
\newcolumntype{P}[1]{>{\centering\arraybackslash}p{#1}}

\title{
	\ares{}:  Adaptive,  Reconfigurable,  Erasure coded, Atomic Storage 	
}

\author{Nicolas Nicolaou}
\email{nicolas@algolysis.com}
\affiliation{Algolysis Ltd, Limassol, Cyprus}
\author{Viveck Cadambe}
\email{vxc12@engr.psu.edu}
\affiliation{Pennsylvania State University, US}
\author{N. Prakash} 
\email{prakashn@mit.edu}
\affiliation{Intel Corp.}
\author{Andria Trigeorgi}
\email{aatrige01@cs.ucy.ac.cy}
\affiliation{University of Cyprus, Nicosia, Cyprus}
\author{Kishori M. Konwar}
\email{kishori@csail.mit.edu}
\author{Muriel Medard}
\email{medard@mit.edu}
\author{Nancy Lynch}
\email{lynch@csail.mit.edu}
\affiliation{Massachusetts Institute of Technology, USA}

	\thanks{This work was partially funded by the Center for Science of Information NSF Award CCF-0939370, 
	NSF Award CCF-1461559, AFOSR Contract Number: FA9550-14-1-0403, NSF CCF-1553248 and RPF/POST-DOC/0916/0090.}

\begin{document}

\begin{abstract}
Emulating a shared \myemph{atomic}, read/write storage system 
%using a set of distinct, often geographically dispersed processes, 
is a fundamental problem in distributed computing. Replicating atomic 
objects among a set of data hosts was the norm for traditional implementations (e.g., \cite{ABD96})
in order to guarantee the availability and accessibility of the data despite host failures.
As replication is highly storage demanding, recent approaches suggested the use 
of erasure-codes to offer the same fault-tolerance while optimizing storage usage at the hosts. 
Initial works focused on a fix set of data hosts. To guarantee longevity and scalability, 
a storage service should  be able to dynamically mask hosts failures by allowing new hosts to join, 
and failed host to be removed without service interruptions. This work presents the first erasure-code 
based atomic algorithm, called \ares{}, which allows the set of hosts to be modified in the course of 
an execution. \ares{} is composed of three main components: (i) a \emph{reconfiguration protocol},
(ii) a \emph{read/write protocol}, and (iii) a set of \emph{data access primitives}. The design of 
\ares{} is modular and is such to accommodate the usage of various erasure-code parameters on a 
per-configuration basis. We provide bounds on the latency of read/write operations, and analyze the 
storage and communication costs of the \ares{} algorithm.  

\end{abstract}

\maketitle

%\begin{IEEEkeywords}
%	atomic storage; erasure codes; fault tolerance;
%\end{IEEEkeywords}
%
\section{Introduction}
\label{sec:intro}

\nnrev{With  the rapid increase of computing power on portable devices, such as, smartphone, laptops, tablets, etc.,
and the near ubiquitous availability of Internet connectivity, our day to day lives are becoming increasingly 
dependent  on Internet-based applications. Most of these applications, 
%\nnrev{if not all,}{}
 rely on large volumes 
of data from a wide range of sources. %, and their performance improves with the easy accessibility of the data. 
%\nn{[NN: What do we mean here?]}  
%Today, data is gathered at an even faster pace from numerous sources of interconnected devices
%around the world. In order to keep abreast with this veritable tsunami of data, researchers, in both 
%industry and academia, are hurtling to invent new ways to increase  {the} capacity of durable, large-scale distributed
%storage systems, and the efficient ingestion and retrieval of data.
 %Currently, most of the data is stored in 
%\nnrev{and accessed from}{} 
%cloud-based storages,  
%\nnrev{in datacenters enabling data hungry applications consume these data
%through the Internet.}
%{offered by major providers like} Amazon, Dropbox, Google, etc. 
%\nnrev{are just a few of the providers of cloud-based durable 
%data storage for application developers to build data-driven applications with access 
%data programatically via APIs, without burdening themselves with the task of managing data.}{}
%
}{}
 
  Distributed Storage Systems (DSS)  store large
amounts of data in an affordable manner. Cloud vendors deploy hundreds to thousands of commodity machines, networked together to act as {a single} 
giant storage system. 
Component failures of  commodity devices,  and network delays are the norm, therefore, \nnrev{designing}{ensuring} consistent data-access and availability  at the same time is challenging. Vendors often solve availability by replicating data across multiple servers. \vc{These services use carefully constructed algorithms that ensure that these copies \nn{are} consistent, especially when they can be accessed concurrently by different operations.} \nn{The problem of keeping copies consistent becomes even more challenging
	when failed servers need to be replaced or new servers are added, without interrupting the service.  Any type of service interruption in a heavily used DSS usually translates to immense revenue loss. }

%\nn{The goal of this work is to provide a \myemph{reconfigurable}, \textit{erasure-coded},  \myemph{atomic} or \myemph{strongly consistent} ~\cite{HW90, Lynch1996} 
%	read/write data storage implementation, which will: (i)
%provide the illusion that data objects are accessed sequentially, even when each object is accessed by multiple operations concurrently, 
%(ii) allow the set of data hosts (servers) to change without affecting the operation of the service, (iii) tolerate crash failures, 
%and (iv) require low storage and communication resources.}

\nn{The goal of this work is to provide an algorithm for implementing strongly consistent \vc{(i.e., atomic/linearizable)}, fault-tolerant distributed read/write storage, with  
	low storage and communication footprint, and the ability to reconfigure the set of data hosts without service interruptions.} 
		
%	
%	
%	\myemph{reconfigurable}, \textit{erasure-coded},  \myemph{atomic} or \myemph{strongly consistent} ~\cite{HW90, Lynch1996} 
%	read/write data storage implementation, which will: (i)
%	provide the illusion that data objects are accessed sequentially, even when each object is accessed by multiple operations concurrently, 
%	(ii) allow the set of data hosts (servers) to change without affecting the operation of the service, (iii) tolerate crash failures, 
%	and (iv) require low storage and communication resources. These properties will make our algorithm attractive in practice.}
%

\myparagraph{Replication-based Atomic Storage.} 
A long stream of work used replication of data across multiple servers to implement atomic (linearizable) read/write objects in message-passing, asynchronous environments where servers (data hosts) may crash fail~\cite{FNP15, ABD96, CDGL04,  FL03,  FHN16,   GNS08,  GNS09, LS97}. \nn{\vc{A notable} replication-based algorithm appears in the work by Attiya, Bar-Noy and Dolev~\cite{ABD96} (we refer to as the ABD algorithm) 
	\vc{which implemented non-blocking atomic read/write data storage} via logical timestamps paired with values to order read/write operations.} 
%and in the work by Fan and Lynch~\cite{FL03} (which is referred to as the LDR algorithm). 
Replication based strategies, however, incur high storage and communication costs; for example, to store 1,000,000 objects each of size 1MB (a total size of $1$TB) across a $3$ server system, the ABD algorithm replicates the objects in all the $3$ servers,  which blows up the worst-case \myemph{storage cost} to $3$TB. Additionally, every write or read operation may need to transmit up to $3$MB of data (while retrieving an object value of size $1$MB), incurring high \myemph{communication cost}. 
%\nnrev{of $3$ MB. The communication cost, or simply the cost, associated with a read or write operation is the amount of total data in bytes that gets transmitted in the various messages sent as part of the operation.}{}
% Since the focus in this paper is on  large data objects, the storage and communication costs include only the total sizes of stable storage and messages dedicated to the data itself. \nn{[NN:Do we need the last sentence?]}

\myparagraph{Erasure Code-based Atomic Storage.} 
%\nnrev{Replication-based atomic memory emulations  suffer from high storage cost and 
%bandwidth with larger replication factor. On the other hand, higher replication factor increases data-durability in the presence of failures. 
%%Several commercial vendors, use erasure codes in their system for fault-tolerance and storage cost reduction,  for their systems which store immutable data.
%}{To avoid the high storage and communication costs stemmed from the use of replication, }
%erasure codes provide an alternative way to emulate fault-tolerant
%shared 
%atomic storage.
\nn{Erasure Coded-based DSS are extremely beneficial to save storage and  communication costs while maintaining similar fault-tolerance levels as in replication based DSS~\cite{GIZA2017}.}
% In comparison to replication, algorithms based on erasure codes significantly reduce both the storage and communication costs of the implementation. 
Mechanisms using an $[n, k]$ erasure code splits a value $v$ of size, say  $1$ unit, into $k$ elements, each of size $\frac{1}{k}$ units, \nnrev{creates $n$ \myemph{coded elements},
 and stores one coded element per server. The size of each coded element is also $\frac{1}{k}$ units, and thus the total storage cost across the $n$ servers is $\frac{n}{k}$ units.}{creates $n$ \myemph{coded elements} of the same size, and stores one coded element per server, for a total storage cost of $\frac{n}{k}$ units.} 
%For example, if we use 
\nnrev{So in our previous example, an $[n = 3, k = 2]$ code, will incur a storage cost of  $1.5$ TB, which is 2 times  lower than the storage needed by replication-based methods.
A similar reduction in  bandwidth used per operation, \nn{and thus in operation latency}, is also possible.}{So the $[n = 3, k = 2]$ code in the previous example will reduce the storage cost to 1.5TB and the communication cost 
to 1.5MB (improving also operation latency).} 
%in many erasure code-based algorithms for implementing atomicity.
 %A class of erasure codes known as
 Maximum Distance Separable (MDS) codes have the property that value $v$ can be reconstructed from any $k$ out of these $n$ coded elements\vc{; \nnrev{it is also worth noting}{note} that replication is a special case of MDS codes with $k=1$.} 
 \vc{\nnrev{The potential cost-savings in light of rapidly growing data volumes, combined}{In addition to the potential cost-savings, the suitability of erasure-codes for DSS is amplified} with the emergence of highly optimized erasure coding libraries, %optimized to specific hardware 
 \nn{that} reduce encoding/decoding overheads~\cite{burihabwa2016performance, intel-isal, EC-Cache}. 
 %has %particularly 
 %made erasure coding increasingly attractive in recent times. 
 In fact, an exciting recent body of systems and optimization works \cite{PARS, EC-Store, EC-Cache, WPS, xiang2016joint, joshi2017efficient, xiang2015multi,yu2018sp} have demonstrated that for several data stores, 
 \nn{the use of} erasure coding \nnrev{can have much}{results in} lower latencies than replication based \nn{approaches.}
 \nn{This is achieved} by allowing the system \nnrev{to more flexibility} to carefully tune erasure coding parameters, data placement strategies, and other system parameters
 \nn{that} improve %minimize \nn{operation} latency based on 
 workload characteristics -- such as load and spatial distribution. A complementary body of work has \nnrev{developed}{proposed} novel non-blocking algorithms that use erasure coding to provide an \nnrev{consistent}{} atomic storage over asynchronous message passing models \nnrev{have been proposed in}{}\cite{CT06, CadambeLMM17,  DGL08, SODA2016, radon,GIZA2017, Zhang2016}.}
 %, and used in  practice~\cite{GIZA2017, Zhang2016}.} 
%
Since erasure code-based algorithms, unlike \nn{their} replication-based counter parts, incur the additional burden of synchronizing the access of multiple pieces of coded-elements from the \textit{same version} of the data object, these algorithms \nnrev{are}{are quite} complex.

\myparagraph{Reconfigurable Atomic Storage.} %for Erasure-coded Algorithms.} %\red{We need to shorten this paragraph}
%\nnrev{Apart from storage cost and bandwidth efficient atomic storage, any  such  distributed storage systems require removal or addition  of the set of servers. }
%\nn{Although replication and erasure-codes may help the system survive server failures,
%	%the failure of a subset of servers, 
%	they do not suffice to ensure the liveness of the 
%service in a longer period where a larger number of servers may fail.}
%The gains on storage and operation latency, is a key-motivation to consider erasure-coded based algorithms for \myemph{reconfigurable} systems as well,
%where the set of servers may change 
%\myemph{Reconfiguration} \nnrev{operations}{is the process that}  allows addition or removal of servers %from a live system, \nnrev{or changing the underlying storage mechanisms or algorithms.}
%without affecting the normal operation 
%during the execution of the service.
\nn{{\it Configuration} refers to the set of storage servers that are collectively used to host the data and implement the DSS. %,  is called a .
{\it Reconfiguration} is the process of adding or removing \nnrev{configurations}{servers}  in a DSS. }
%reonfigurations that implements the same set of objects.}
%
In practice, reconfigurations are often desirable by system administrators~\cite{aguileratutorial}, for a wide range of 
purposes,  
	especially during system maintenance. As the set of storage servers becomes older and unreliable they are replaced with new ones to ensure data-durability. \vc{Furthermore, to scale the storage service to increased or decreased load,  larger (or smaller) configurations \nnrev{might be}{may} be needed to be 
	 deployed.}
	%\blue{In such atomic memory system, we consider three operations: \it{read}, \it{write} and \it{recon}}.
	 Therefore, in order to carry out such reconfiguration steps, in addition to the usual  \act{read} and \act{write} operations, an operation called \act{reconfig}  is  invoked  by  reconfiguration clients.
%Reconfiguration also allows the system administrator to enhance data survivability,  or even  scale up or down the  level of performance.
 %However, p
 Performing reconfiguration of a system, without service
interruption, is a very challenging task and an active area of research. RAMBO~\cite{LS02} and DynaStore~\cite{ AKMS09}  are two of the handful 
of algorithms~\cite{CGGMS09, GM15, G03, LVM15, SMMK2010, spiegelman:DISC:2017} that allows reconfiguration on live systems; \vc{all these algorithms are replication-based}. 
%Recently, the authors in~\cite{spiegelman:DISC:2017} presented a general framework for consensus-free reconfiguration algorithms. 

\nnrev{So far, none of the existing reconfiguration approaches
demonstrated the use of erasure-codes for fault-tolerance, 
	or provided any analysis of bandwidth and storage cost of such algorithm.}{}
%Thus, such algorithms do not benefit from the low storage overheads and low communication 
%cost offered when using erasure-codes.} %, even though some (e.g., \cite{spiegelman:DISC:2017}) may be able use them.}
% implicitly or explicitly, assume  a replication-based system in \nnrev{their}{each} configuration.
Despite the attractive prospects of creating strongly consistent DSS with low storage and communication costs\nnrev{ by employing erasure-codes}{}, so far,  \nnrev{there is}{} no algorithmic framework \nn{for reconfigurable atomic DSS }\nnrev{to reconfigure the underlying configurations without service
interruption}{employed erasure coding for fault-tolerance, 
or provided any analysis of bandwidth and storage costs}. Our paper fills this vital gap in algorithms literature, 
through the development of novel reconfigurable approach for atomic storage that use \emph{erasure codes} for fault-tolerance. 
From a practical viewpoint, our work may be interpreted as a bridge between the systems optimization works \cite{PARS, EC-Store, EC-Cache, WPS, xiang2016joint, joshi2017efficient, xiang2015multi,yu2018sp} and non-blocking erasure coded based consistent storage \cite{CT06, CadambeLMM17,  DGL08, SODA2016, radon, GIZA2017, Zhang2016}. Specifically, the use of our \emph{reconfigurable} algorithms would potentially enable a data storage service to dynamically shift between different erasure coding based parameters and placement strategies, as the demand characteristics (such as load and spatial distribution) change, without service interruption.

\myparagraph{Our Contributions.}
%In this work 
We develop a \myemph{reconfigurable}, \textit{erasure-coded},  \myemph{atomic} or \myemph{strongly consistent} ~\cite{HW90, Lynch1996} 
	read/write storage algorithm, called \ares{}. Motivated by many practical systems, \ares{} assumes clients and servers are separate processes \footnote{In practice, these processes can be on the same node or different nodes.} that communicate via logical point-to-point channels.

\vc{In contrast to the, replication-based reconfigurable algorithms \cite{LS02, AKMS09, CGGMS09, GM15, G03, LVM15, SMMK2010, spiegelman:DISC:2017}, where a configuration essentially corresponds to the set of servers that stores the data, the same concept
	% The concept of a \emph{configuration}, however, 
	for erasure coding need to be much more \nnrev{more rich}{involved}. In particular, in erasure coding, even if the same set of $n$ servers are used, a change in the value of $k$ defines a new configuration. Furthermore, several erasure coding based algorithms \cite{CadambeLMM17,  DGL08} have additional parameters that tune how many older versions each server store, which in turn influences the concurrency level allowed. Tuning of such parameters can also fall under the purview of reconfiguration. 
	%for erasure coding based algorithms. %Another challenge that we overcome is that 
}

\vc{
To accommodate these various reconfiguration requirements, \ares{} takes a modular approach\nnrev{, where the static atomic memory implementation is
	oblivious of the mechanics of the reconfiguration}. In \nn{particular}, \ares{} \nn{uses a} set of primitives, called \myemph{data-access primitives} (DAPs).
%, \nnrev{. These primitives, if implemented according to our specifications}{which if implemented to satisfy specific properties} can be combined to provide an abstraction of a static atomic memory storage implementation.
 \nn{A different implementation of the DAP primitives may be specified in each configuration.} 
	\nnrev{Each configuration is specified by its own implementation of the DAP primitives.}{} \nnrev{The reconfiguration service then uses}{\ares{}} uses DAPs as a ``black box'' to: % implement: 
		 (i) %a consensus-based reconfiguration protocol (similar to~\cite{LS02}) that uses DAPs to 
		 transfer the object state from one configuration to the next during \act{reconfig} operations, and
		 (ii) invoke \act{read}/\act{write} operations 
		 %implemented using DAPs 
		 on a single configuration. \vc{Given the DAP implementation for each configuration
		 	%, its DAP implementation satisfies specific properties,} 
		 	we show that \ares{} correctly implements a \textit{reconfigurable}, \textit{atomic} read/write storage.} 
}

		\begin{table*}[!h]
			{\scriptsize
			\begin{center}
				\begin{tabular}{|lP{1cm}P{1cm}cP{1cm}ccc|}
					\hline
					\textbf{Algorithm}  & \textbf{\#rounds /write}  & \textbf{\#rounds /read} & \textbf{Reconfig.}  & \textbf{Repl. or EC} & \textbf{Storage cost}  & \textbf{read bandwidth} & \textbf{write bandwidth}  \\ 
					\hline

					{\sc CASGC} \cite{CLMM14} & 3 & 2 & No  & EC & $(\delta +1)\frac{n}{k}$  & $\frac{n}{k}$   & $\frac{n}{k}$ \\
					{\sc  SODA} \cite{SODA2016} & 2 & 2 & No & EC & $\frac{n}{k}$ & $(\delta +1)\frac{n}{k}$  & $\frac{n^2}{k}$ \\  
					{\sc ORCAS-A} \cite{DGL08} & 3 & $\geq 2$ & No  & EC & $n$ & $n$ & $n$ \\ 
					{\sc ORCAS-B} \cite{DGL08} & 3 & 3 & No  & EC & $\infty$ & $\infty$ & $\infty$ \\ 
					{\sc ABD} \cite{ABD96} & 2 & 2 & No  & Repl. & $n$ & $2n$  & $n$ \\ 
					{\sc RAMBO} \cite{LS02} & 2 & 2 & Yes  & Repl. & $\geq n$ & $\geq n$  &$\geq n$ \\
					{\sc Dynastore} \cite{AKMS09} & $\geq 4$ & $\geq 4$ & Yes & Repl. & $\geq n$ & $\geq n$ & $\geq n$ \\
					{\sc SmartMerge} \cite{LVM15} & 2 & 2  & Yes  & Repl. & $\geq n$ & $\geq n$ & $\geq n$ \\
					%{\sc SpSn} \cite{} & &  & Dynamic & MWMR & & & \\
					\hline\hline
					\ares{} (this paper) & 2 & 2 & Yes & EC & $(\delta +1)\frac{n}{k}$ & $(\delta +1)\frac{n}{k}$  & $\frac{n}{k}$ \\ \hline
					%\treasmod & 2 & 2 & & No &  & $\frac{1}{ \lceil \frac{k}{\delta +1} \rceil}$ & \\ \hline
				\end{tabular}
			\end{center}
		}
			\caption{Comparison of \ares{} with previous algorithms emulating atomic Read/Write Memory for replication (Repl.) 
			and erasure-code based (EC) algorithms.  $\delta$ is the maximum number of concurrent writes with any read during the course of an execution of the algorithm. In practice, $\delta < 4$~\cite{GIZA2017}.
		%	\red{Should we not add communiation and storage cost? What is the difference between recon and adaptive?}
			}\label{tab:compare}
		\end{table*}	

	\remove{	\nn{For the ease of exposition and modularity of \ares{} we define a set of primitives, called data-access primitives (DAPs).}}
		%		,  we  then present the complete \ares{} algorithm, for emulating reconfigurable, atomic read/write storage.}
		%, a protocol that allows reconfiguration of the servers. 
%that emulates an atomic memory, 
%and is specifically suitable for implementing atomic memory service that uses erasure codes without interrupting the service. 
%		
		 %In particular, 
		  
		  %		 
%		 
%		 
%		 %is implemented in terms of  data access primitives (
%		 uses DAPs as a ``black box'' to implement read and write operations.
%		 
		
		\vc{The DAP primitives \nnrev{allow}{provide} \ares{} a much broader view of the notion of a configuration as compared to replication-based algorithms. Specifically, the DAP primitives may be parameterized, \nnrev{ by the choice}{following the parameters } of protocols used \nnrev{to implement the primitive and its parameters}{for their implementation} (e.g., erasure coding parameters, set of servers, quorum design, concurrency level, etc.). 
			While transitioning from one configuration to another, our modular construction, allows \ares{} to reconfigure between different sets of servers, quorum configurations, and erasure coding parameters. In principle, \ares{} even allows to reconfigure between completely different protocols as long as they can be interpreted/expressed in terms of the primitives; though in this paper, we only present one implementation of the DAP primitives to keep the scope of the paper reasonable. From a technical point of view, our modular structure \nn{makes} the atomicity proof of a complex algorithm (like \ares{}) easier}.
%		and is also \nn{able} to adaptively change DAP mechanisms in a per configuration basis.
		%the replication based transformations of  \mwABD{} and {\sc ldr}.
%        Such description simplifies the %show that our algorithm 
%	 atomicity proof,  as it is sufficient to show that the implemented primitives satisfy certain  properties.
%

%		\nn{To achieve reconfiguration, \ares{} implements a \textit{recon} operation,
%			that in its heart uses a consensus algorithm similar to ~\cite{LS02}. 
%			In particular, each configuration in \ares{} implements  a distributed consensus service (like Paxos~\cite{L98} or Raft~\cite{Raft}) 
%			 on its set of servers to install the next configuration. 
%			 %Note that the use of consensus may prevent 
%%			 
%%			 Note  that the use of distributed consensus we cannot 
%%			guarantee the termination of the 
%			a reconfiguration operation from terminating~\cite{FLP85}, however this does not affect the liveness and safety of 
%			 %but still guarantee  liveness of 
%			 read and write operations. Recall that in our setting, the 
%			clients are required to decide on one of the configurations proposed by any client, not a configuration consisting of servers proposed in various {\it recon} operations. 
%			It would be useful to find out  whether it is possible to achieve  liveness of reconfiguration operations of erasure-coded algorithm without relying on a ``consensus-like'' protocol.
%		}
%
	\vc{An important consideration in the design choice of  \ares{}, is to ensure that we gain/retain the advantages that 
	come with erasure codes -- cost of data storage and communication is low -- while having the flexibility to reconfigure the system. Towards this end,} 
	we present an erasure-coded implementation
	of DAPs which satisfy the necessary properties, and are used by 
	%	
	%	 we present an  implementation of the DAPs for an erasure-coded
	%		atomic algorithm, and we show that the DAPs satisfy the necessary properties to make them suitable to be adapted by \ares{}.
	%	Usage of those DAPs by 
	%As a consequence of our implementation of DAPs, 
	\ares{}  to yield
	%leads us to discover  \treas,  
	the first reconfigurable, \textit{erasure-coded},  read/write
	%with  cost-effective storage and communication,  
	%for emulating shared 
	atomic storage implementation, where \act{read} and \act{write} 
	operations complete in \textit{two-rounds}.
%
%
%
%we present an  implementation of the DAPs that utilize erasure-codes for an erasure-coded
%		atomic algorithm, and we show that the DAPs satisfy the necessary properties to make them suitable to be adapted by \ares{}.
%		Usage of those DAPs by 
%	As a consequence of our implementation of DAPs, \ares{}  becomes
%		leads us to discover  \treas,  
%		the first reconfigurable, \textit{erasure-coded},  read/write
% 		with  cost-effective storage and communication,  
% 		for emulating shared 
% 		atomic storage implementation, where each read and write 
% 		operation complete in \textit{two-rounds}.
% 		in a message-passing environment and in the presence of crash-failures. 
% 		%We call the resulted algorithm {\treas}. 
% 		%in terms of the  DAPs. Thus, 
% 		We show that the implemented DAPs satisfy the necessary properties, and thus are suitable to 
% 		be adapted by \ares{} 
% 		%can adopt the implemented DAPs to 
% 		yielding the first reconfigurable, \textit{erasure-coded} atomic storage implementation. 
%		
		%\blue{
			%\nn{[Still to examine: Latency analysis]}
			We provide the atomicity property %of ~\ares{} 
			and latency analysis for any 
			%reconfiguration 
			operation in \ares{}, along with the storage and communication costs resulting from the erasure-coded DAP implementation.
		%We provide the atomicity property of ~\ares{} and  latency analysis for any reconfiguration operation in \ares.
		In particular, we specify lower and upper bounds on the communication latency between the service participants,
		and we provide the necessary conditions 
		%relation between those bounds 
		to guarantee the termination of each \act{read/write} operation 
		while concurrent with \act{reconfig} operations.  
Table~\ref{tab:compare} compares \ares{} with a
%performance metrics of \treas{} with  a 
few well-known erasure-coded  and replication-based (static and reconfigurable) atomic memory algorithms. 
%and also the static and reconfigurable replication-based algorithms.
From the table we observe that \ares{} is the only algorithm to combine a dynamic behavior with the use of erasure codes,
%and also the only to allow adaptive change of the atomic algorithm per configuration, 
\kmk{while reducing the storage and communcation costs associated with
the read or write operations.}
%\kmkremove{On the downside, \ares{} still relies 
%on consensus to achieve reconfiguration, something however that does not affect its correctness and enables infinite 
%reconfigurations.}  
\kmk{Moreover, in ~\ares{} the  number of rounds per write and read is at least as good as in any of the remaining algorithms.}

\myparagraph{Document Structure.}
%The remainder of the manuscript consists of the following sections. 
Section~\ref{model} presents the model assumptions 
%for our setting 
and Section~\ref{ssec:dap}, the DAP  primitives.  
 In Section~\ref{sec:ares}, we present the implementation of the reconfiguration and read/write 
 protocols in \ares{} using the DAPs.
% our \ares{} framework and we describe the implementation of the reconfiguration and read/write 
%protocols using the DAPs. 
%for emulating shared atomic memory with erasure-codes where the system can undergo  reconfiguration, while it is live. 
 In Section~\ref{ssec:dap:impl}, we present an erasure-coded implementation of a set of  DAPs, which 
 can be used in every configuration of 
 %completes the description of 
 the \ares{} algorithm.
 Section~\ref{sec:performance} provides operation latency and cost analysis, and Section 
 \ref{sec:dap:flexible} the DAP flexibility. Section \ref{sec:evaluation} presents 
 an experimental evaluation of the proposed algorithms. 
%of read, write and reconfiguration operations.
% In Section~\ref{sec:transfer}, we describe an optimization or the reconfiguration mechanism of \ares{}.
%algorithm, we call  \aresopt{}. %algorithm.
 %Finally, in , 
 We conclude our work in Section~\ref{sec:conclusions}.
  Due to lack of space omitted proofs can be found in 
  %the extended version of our paper~
  \cite{ARES:Arxiv:2018}.

\section{Model and Definitions}\label{model}
A shared atomic storage, consisting of any number of individual objects,  can be emulated
by composing individual atomic memory objects. Therefore, herein we aim
in implementing a single atomic \textit{read/write} memory object. %on a set of servers.
{A read/write} object takes a value from a set $\valSet$. 
We assume a system consisting of four distinct sets of processes: 
a set $\wSet$ of writers, a set $\rdSet$ of readers, a set $\recSet$ of 
reconfiguration clients, and a set $\srvSet$ of servers. Let $\cSet = \wSet \cup\rdSet\cup\recSet$ 
be the set of clients. Servers host data elements (replicas or encoded data fragments).
Each writer is allowed to modify the value of a shared object, and each reader is allowed to obtain 
the value of that object. Reconfiguration clients attempt  to introduce new 
 configuration of servers to the 
system in order to mask transient errors and to ensure the longevity of the service. 
Processes communicate via \myemph{messages} through 
\myemph{asynchronous}, and \myemph{reliable} channels. 
%%Let $\idSet = \cSet\cup\reconSet\cup\srvSet$. 
%In a read/write object implementation, we assume that the object may take a value from a set $\valSet$. 
%Each writer is allowed to modify the value of the object, and each reader is allowed to obtain 
%the value of the object. Servers host data elements (replicas or encoded data fragments).
%%maintain encoded elements of the redundant object.
%
%We assume an \myemph{asynchronous} environment, where processes communicate
%by exchanging messages. The writer, any subset of readers, and up to 
%$f$ servers may \myemph{crash} without any notice.
%\ares{} allows the set of server host to be modified during the course of an execution for 
%masking transient or permanent failures of servers and preserve the longevity of the service.  
%\kmk{In the paper, we are interested only  in the fair executions of any algorithm.}

\myparagraph{Configurations.} 
\kmk{A \textit{configuration},  with a unique identifier from a set $\confSet$, is a data type that 
	describes the finite set of servers that are used to implement the atomic storage service. In our setting, 
	each configuration is also used to describe the way the servers are grouped into intersecting sets, called 
	\textit{quorums}, the consensus instance that is used as an external service to determine the next configuration, 
	and a set of data access primitives that specify 
	the interaction of the clients and servers in the configuration %for implementing the read/write operations 
	(see Section \ref{ssec:dap}). 
%	A configuration refers In our setting, a \textit{configuration} consists of a finite set of servers, describes 
%	the organization of the servers in intersecting sets, and specifies a consensus algorithm
%	implemented by the servers of the configuration.
}
	%Informally, in our setting, a configuration consists of a finite set of servers, each with an unique identifier, that
%$(i)$ collectively implements an atomic object; $(ii)$ implements an underlying quorum-based algorithm for implmenting the
%atomic object; and $(iii)$ implements some distributed consensus algorithm}
More formally, a configuration,  
%identified by a unique identifier 
 $c\in\confSet$, consists of: 
 %is a data type that describes explicitly: 
$(i)$ $\servers{c}\subseteq\srvSet$: a set of server identifiers; %that {belong} in $c$; 
$(ii)$ $\quorums{c}$: the set of quorums on $\servers{c}$, s.t. $\forall Q_1,Q_2\in\quorums{c}, Q_1,Q_2\subseteq\servers{c}$ and $Q_1\cap Q_2\neq \emptyset$; 
%$(iii)$ \nn{an underlying algorithm, $\algo{c}$, that implements an atomic memory (including related parameters);}
$(iii)$ $\dap{c}$: the set of primitives (operations at level lower than reads or writes) that clients in $\idSet$ may invoke on $\servers{c}$; 
% an underlying algorithm that implements atomic memory in $\servers{c}$, including related parameters; 
and $(iv)$ $\consensus{c}$: a consensus instance with the values from $\confSet$, %the set of all configuration identifiers, 
implemented and running on top of the servers in $\servers{c}$.
%, the set of servers in 
%some $c \in \confSet$ is denoted by $\servers{c}$.
We refer to a server $s \in \servers{c}$ as a \myemph{member} of  configuration $c$.
% and (iii) the data access primitives that are used to access the data in those servers.  
%we can formally define a quorum system $\quorums{c}$, with $c\in\confSet$, as follows:
%Consider a configuration identifier $c$.
%We define as $\servers{c}=\bigcup_{Q\in\quorums{c}} Q$ the set of servers that belong in the quorums
%f a quorum system $\quorums{c}$. 
%We refer to a server $s$ as a \myemph{member} of a configuration $c$ if $s\in \servers{c}$.
%
%\paragraph{Notations and Definitions}  We denote by $\mathcal{W}$, $\mathcal{R}$ and $\mathcal{S}$,  the set of writers, readers and servers, respectively.
%We denote by $\confSet$ the set of possible configurations of the systems. A configuration $c \in \confSet$ has a unique the configuration identifier   $c.conf$; a set of servers $c.servers$, such that, $c.Servers \subseteq \mathcal{S}$,
%$cseq[i].status\in\{F,P\}$, the pending or finalized status 
%
%The writer process $w$ may invoke write operations on the atomic object 
%by calling the $\act{write}(v)$ function whereas each reader may invoke 
%a read operation by calling the $\act{read}$ function. Each write event 
%returns an acknowledgement when successfully carried-out, whereas each read 
%operation returns the value of the atomic object. We assume that an 
%
%\nn{
%\blue{We assume that there is an instance of consensus protocol running on servers in $\srvSet{s}$.}
\nn{ The consensus instance $\consensus{c}$ in each configuration $c$ is used as a service that 
	returns the identifier of the configuration that follows $c$. }   
%	
%	
%	Note that due to the asynchrony of the channels the consensus instance $\consensus{c}$ may not terminate~\cite{FLP85}.
%	As we will see below this may affect the liveness of an ongoing reconfiguration and the liveness of read/write operations when 
%	not enough servers remain alive in the current configuration. The safety of read/write operations, however, is not affected.} 

\myparagraph{Executions.} An algorithm $A$ is a collection of processes, where process $A_p$
is assigned to \nnfix{process} $p\in\idSet\cup\srvSet$. The \textit{state}, of a process $A_\pr$ is determined over a
set of state variables, and the state $\state$ of $A$ is a vector that contains the state of
each process. Each process $A_\pr$ implements a set of actions. When an action $\acts{}$ occurs 
it causes the state of $A_\pr$ to change, say from 
some state $\state_p$ to some different state $\state_p'$. We call the triple $\tup{\state_p, \acts{}, \state_p'}$
a \textit{step} of $A_\pr$. Algorithm $A$ performs a step, when some process $A_\pr$ performs a step.
%: (i) receives a
%message, (ii) performs local computation, (iii) sends a message. 
%Each such action
%causes the state at $p$ to change. 
An action $\acts{}$ is \textit{enabled} in a state $\state$ if $\exists$ a step $\tup{\state, \acts{}, \state'}$
	to some state $\state'$.
An \textit{execution} is an alternating sequence of states
and actions of $A$ starting with the initial state and ending in a state. 
An execution $\EX$ 
%is \textit{well-formed} if any process invokes one operation at a time and it is 
\textit{fair} if enabled actions perform a step infinitely often. In the rest of the paper 
we consider executions that are fair and well-formed. A process
$\pr$ \textit{crashes} in an execution if it stops taking steps; otherwise $p$ is \textit{correct} or \textit{non-faulty}.
We assume a function $c.\mathcal{F}$ to describe the failure model of a configuration $c$.

\myparagraph{\kmk{Reconfigurable Atomic Read/Write Objects.}} \kmk{
	A reconfigurable atomic object supports three operations:  $\act{read}()$, $\act{write}(v)$ and $\act{reconfig}(c)$.
	A \act{read}() operation returns the value of the atomic object, $\act{write}(v)$ attempts to modify the value of 
	the object to $v\in\valSet$, and the $\act{reconfig}(c)$ that attempts to install a new configuration $c\in\confSet$.
 We assume \textit{well-formed} executions where each client may invoke \nnfix{one} operation ($\act{read}()$, $\act{write}(v)$ or $\act{reconfig}(c)$) 
at a time. 
%To allow reconfiguration of our atomic memory service, in addition to the usual reads or writes,  we assume an additional operation $reconfig(c)$, where $c$ is a new configuration,  that reconfig clients can invoke to install new configuration of servers or storage nodes.
%
% Reconfig operation invoked at a client,  upon completion,  returns  an {\it ok} response. 
% Note that a reconfig client does not invoke
% a new reconfig operation unless the previously invoked reconfig operations at it is complete.
% Note that any time serveral reconfig operations can be invoked at various clients, or while other read/write operations are underway.
%We need these properties in order to cope with the asynchronous nature of the network.
}

An implementation of a \act{read}/\act{write} or a \act{reconfig} operation contains an \textit{invocation} action 
	(such as a call to a procedure) and a \textit{response} action (such as a
	return from the procedure). An operation $\op$ is \textit{complete} in an execution, if it
	contains both the invocation and the \textit{matching} response actions for $\op$; otherwise $\op$
	is \textit{incomplete}. 
	We say that an operation $\op$ \textit{precedes} an operation $\op'$ in an execution $\EX$,
	denoted by $\op\bef\op'$, if the response step of $\op$ appears before the invocation
	step of $\op'$ in $\EX$. Two operations are \textit{concurrent} if neither precedes the other.
	An implementation $A$ of a read/write object satisfies the atomicity \nnfix{(linearizability \cite{HW90})} property
	if the following holds \cite{Lynch1996}. Let the set $\Pi$ contain all complete \nnfix{read/write} operations in
	any well-formed execution of $A$. 
	Then %for operations in $\Pi$ 
	there exists an irreflexive partial ordering $\prec$ satisfying the
	following:	
	\begin{itemize}
		%\item [\em P1.] No operation has infinitely many other 
		%			operations ordered before it.
		\item [\bf A1.] 
		%					The partial order is consistent with the 
		%					external order of invocation and responses, that is, there do 
		%					not exist operations $\op_1$ and $\op_2$, 
		%					such that $\op_1$ completes before $\op_2$ starts, 
		%					yet $\op_2 \prec \op_1$.
		For any operations $\pi_1$ and $\pi_2$ in $\Pi$,  if $\pi_1\bef\pi_2$, then it
		cannot be the case that $\pi_2\prec \pi_1$.
		\item[\bf A2.] 
		If $\pi\in\Pi$ is a write operation and $\pi'\in\Pi$ is any \nnfix{read/write} operation,  
		then either $\pi\prec \pi'$ or $\pi'\prec \pi$.
		%					All write operations are totally 
		%					ordered and every read operation is ordered with respect 
		%					to all the writes.
		\item[\bf A3.] 
		The value returned by a read operation is the value 
		written by the last preceding write operation according to
		$\prec$ (or the initial value if there is no such write).
		%		Every read operation ordered after any writes returns
		%the value of the last write preceding it in the partial order, and any
		%read operation ordered before all writes returns the initial value
		%of the object.
\end{itemize}

\myparagraph{Storage and Communication Costs.} We are interested in the \myemph{complexity} of each
read and write operation. The complexity of each operation $\op$ invoked by a process 
$\pr$, is measured with respect to three metrics, during the interval between the invocation 
and the response of $\op$: $(i)$ \myemph{number of communication round}, accounting the number of messages 
exchanged during $\op$, $(ii)$ \myemph{storage efficiency} (storage cost), accounting the maximum storage requirements for 
any single object at the servers during $\op$, and  $(iii)$ \myemph{message bit complexity} (communication cost)
which measures the size of the messages used during $\op$. 

We define the total storage cost as the size of the
data stored across all servers, at any point during the execution of the algorithm. The
communication cost associated with a read or write operation is the size of the total data that
gets transmitted in the messages sent as part of the operation. We assume that metadata,
such as version number, process ID, etc. used by various operations is of negligible size, and
is hence ignored in the calculation of storage and communication cost. Further, we normalize
both costs with respect to the size of the value $v$; in other words, we compute the costs
under the assumption that $v$ has size $1$ unit.

 \myparagraph{{\bf Erasure Codes}.} %In algorithm \treas{}, 
 We use an $[n, k]$  linear MDS code ~\cite{verapless_book} over a finite field $\mathbb{F}_q$ to encode and store the value $v$ among the $n$ servers. An $[n, k]$ MDS code has the property that any $k$ out of the $n$ coded elements can be used to recover (decode) the value $v$. For encoding, $v$ is divided
 %\footnote{In practice $v$ is a file, which is divided into many stripes based on the choice of the code, various stripes are individually encoded and stacked against each other. We omit details of represent-ability of $v$ by a sequence of symbols of $\mathbb{F}_q$, and the mechanism of data striping, since these are fairly standard in the coding theory literature.} 
 into $k$ elements $v_1, v_2, \ldots v_k$ with each element having  size $\frac{1}{k}$ (assuming size of $v$ is $1$). The encoder takes the $k$ elements as input and produces $n$ coded elements $e_1, e_2, \ldots, e_n$ as output, i.e., $[e_1, \ldots, e_n] = \Phi([v_1, \ldots, v_k])$, where $\Phi$ denotes the encoder. For ease of notation, we simply write $\Phi(v)$ to mean  $[e_1, \ldots, e_n]$. The vector $[e_1, \ldots, e_n]$ is  referred to as the codeword corresponding to the value $v$. Each coded element $c_i$ also has  size $\frac{1}{k}$. In our scheme we store one coded element per server. We use $\Phi_i$ to denote the projection of $\Phi$ on to the $i^{\text{th}}$ output component, i.e., $e_i = \Phi_i(v)$. Without loss of generality, we associate the coded element $e_i$ with server $i$, $1 \leq i \leq n$.

\myparagraph{{\bf Tags.}}
We use logical tags to order operations. A tag $\tg{}$ is defined as a pair $(z, w)$, where $z \in \mathbb{N}$ and $w \in \mathcal{W}$, an ID of a writer.
Let $\mathcal{T}$ be the set of all tags.
Notice that tags could be defined in any totally ordered domain and given that this domain is countably infinite, then 
there can be a direct mapping to the domain we assume. 
% and we denote by  $\mathcal{T}$  the set of all possible tags. 
For any  $\tg{1}, \tg{2} \in \mathcal{T}$ we define  $\tg{2} > \tg{1}$ if $(i)$ $\tg{2}.z > \tg{1}.z$ or $(ii)$ $\tg{2}.z = \tg{1}.z$ and $\tg{2}.w > \tg{1}.w$.

\section{Data Access Primitives}

\nn{In this section we introduce a set of  primitives, 	we refer to as  \textit{data access primitives (DAP)},
	which are invoked by the clients during \act{read/write/reconfig} operations and are defined 
 %implemented by the  set of servers  of 
 for any configuration $c$ in \ares{}. 
% We refer to such 
% primitives as the  \textit{data access primitives (DAP)} of the  configuration.
%
%describe an algorithm for implememting  atomic memory, say  	$A$, in a modular way, given  
%which can be associated with a configuration $c$.
 %To allow \ares{} to be modular we  introduce a set of  primitives, implemented on a set of servers 
%(configuration),  which we refer to as  \textit{data access
%primitives} (
The DAPs allow us:  $(i)$ to describe 
%an atomic memory algorithm, specifically~
\ares{} in  a \myemph{modular} manner, 
%$(ii)$ to adaptively change DAP implementation per configuration (thus changing the operation protocol), 
and $(ii)$ a cleaner reasoning about the correctness of \ares{}.
%implementation of an atomic memory service, specifically~\ares{}.
%Additionally, we also require some properties  for the  DAPs to prove the atomicity property of ~\ares{}.
}

%\nn{In this section we present simple abstractions, we call 
	%the next section, we present the \treas{} algorithm  \nnrev{based on  of}{using} three 
%	\textit{data access primitives} (DAP), which can be used to describe an algorithm for implememting  atomic memory, say  
%	$A$, in a modular way, given  
%which can be associated with 
%a configuration $c$.} 

\nn{We define three data access primitives for each $c\in\confSet$:}
%in the context of $c$ are: 
$(i)$ $\daputdata{c}{\tup{\tg{},v}}$, via which 
a client can ingest the tag value pair $\tup{\tg{},v}$ in to the configuration $c$;
 $(ii)$ $\dagetdata{c}$, used to retrieve the most up to date 
 tag and vlaue pair stored in the configuration $c$; 
 and $(iii)$ $\dagettag{c}$, used to retrieve the most up to date tag for an object stored in a configuration $c$.
  \nn{More formally,} assuming  a tag $\tg{}$ from 
  %$\tg{}$ be a logical timestamp from 
  a set of totally ordered tags $\tsSet$, 
  %over a comparison function $<_{\tg{}}$, $v$ be a value within the 
   a value $v$ from a domain 
   %of the distributed atomic object 
   $\valSet$, and a configuration $c$ from 
  %$c$ be the identifier of 
  a set of identifiers $\confSet$, the  
  %We define below the 
  three primitives are defined 
  %over a configuration $c\in\confSet$, tag $\tg{}\in\tsSet$, and a value $v\in\valSet$ 
  as follows:
  %which we collectively refer to as the \myemph{data-primitives}:
  %	\kmkrev{\nn{The following theorem captures the main result of this subsection.}} {Below we provide the specification and properties  of the DAPs.}
  
    \begin{definition}[Data Access Primitives] %Let $c$, $c \in C$,  be any configuration identifier  and  any read, write or client that invokes any of the following interfaces:
  	Given a configuration identifier $c\in\confSet$, any non-faulty client process $\pr$ may invoke the following data access primitives during an execution $\EX$, where $c$ 
  	is added to specify the configuration specific implementation of the primitives:
  	\begin{enumerate}
  		\item [$D1$:]  $\dagettag{c}$ that returns  a tag $\tg{}\in\tsSet$;  
  		%completes by responding with some tag $t \in \mathcal{T}$
  		\item[$D2$:] $\dagetdata{c}$ that  
  		%the response is though returning 
  		returns a tag-value pair $(\tg{}, v) \in \tsSet \times \valSet$, 
  		\item[$D3$:] $\daputdata{c}{\tup{\tg{}, v}}$ which accepts the tag-value pair $(\tg{}, v) \in \tsSet \times \valSet$ as argument. 
  	\end{enumerate} 
  		%returns a tag-value pair $(t, v) \in \mathcal{T} \times \mathcal{V}$
  \end{definition}

  \kmk{In order for the DAPs to be useful in designing the \ares{} algorithm we further require the following consistency
   properties. As we see later in Section \ref{sec:correct}, the safety property of \ares{} 
   %hinges 
   holds, given that these properties hold for the DAPs in each configuration.} %as we see later in Section \ref{sec:safety:b}.}
 
 \begin{property}[DAP Consistency Properties]\label{property:dap}  In an execution $\EX$ 
 	we say that a DAP operation in an execution $\EX$ is complete if both the invocation and the 
 	matching response step  appear in $\EX$. 
 	If $\Pi$ is the set of complete DAP operations in execution $\EX$ then for any $\phi,\pi\in\Pi$: 
 	%be an execution of some algorithm that executes the data-primitives 
 \begin{enumerate}
 \item[ C1 ]  If $\phi$ is  $\daputdata{c}{\tup{\tg{\phi}, v_\phi}}$, for $c \in \confSet$, $\tup{\tg{\phi}, v_\phi} \in\tsSet\times\valSet$, % and $v_1 \in \valSet$,
 and $\pi$ is $\dagettag{c}$ (or  $\dagetdata{c}$) 
 %in $\EX$ such that 
 that returns $\tg{\pi} \in \tsSet$ (or $\tup{\tg{\pi}, v_{\pi}} \in \tsSet \times \valSet$) and $\phi$ completes \nn{before $\pi$ is invoked} in $\EX$, then $\tg{\pi} \geq \tg{\phi}$.
 \item[ C2 ] \sloppy If $\phi$ is a $\dagetdata{c}$ that returns $\tup{\tg{\pi}, v_\pi } \in \tsSet \times \valSet$, 
 then there exists $\pi$ such that $\pi$ is $\daputdata{c}{\tup{\tg{\pi}, v_{\pi}}}$ and $\phi$ did not complete before the invocation of $\pi$. 
 If no such $\pi$ exists in $\EX$, then $(\tg{\pi}, v_{\pi})$ is equal to $(t_0, v_0)$.
 \end{enumerate} \label{def:consistency}
 \end{property}\vspace{-0.5em}

\nn{In Section \ref{ssec:dap:impl} we show how to implement a set of DAPs, where erasure-codes are used to reduce storage and communication costs. 
	%We show that 
	Our DAP implementation satisfies Property \ref{property:dap}.}

   \nn{
 	As noted earlier, expressing \ares{}
 	%an atomic memory implementation  algorithm 
 	in terms of the DAPs 
 	%primitives 
 	allows one to achieve a modular 
	design. Modularity enables the usage of different DAP
	implementation per configuration, during any execution of \ares{},
	as long as the DAPs implemented in each configuration satisfy Property \ref{property:dap}.
	For example, the DAPs in a configuration $c$ may be implemented using replication,
	while the DAPs in the next configuration say $c'$, may be implemented using erasure-codes.}	
\nnfix{
%	This enables the use of different coding algorithms based on the application 
% 		or even per configuration in the same system
%		without affecting the correctness of \ares{}.
		Thus, a system may use a scheme that offers higher fault tolerance (e.g. replication)
	when storage is not an issue, while switching to a more storage efficient (less fault-tolerant)
scheme when storage gets limited. }

%	This in turn, allows the usage of different coding algorithms and 
%	code parameters in each individual configuration without affecting 
%	the correctness of \ares{}.

	\nn{
		In Section \ref{sec:dap:flexible}, we show that the presented DAPs are not only suitable
	for algorithm \ares{}, but can also be used to implement a large family of atomic read/write
	storage implementations. By describing an algorithm $A$ according to a simple algorithmic 
	template (see Algorithm~\ref{algo:atomicity:generic1}), we show that $A$ preserves safety (atomicity) if 
	the used DAPs satisfy Property \ref{property:dap}, and $A$ preserves liveness (termination),
	if every invocation of the used DAPs terminate,  under the failure model assumed.
}

\label{ssec:dap}

\section{The \ares{} Protocol}
\label{sec:ares}

\algblockdefx[Operation]{Operation}{EndOperation}%
[2]{{\bf operation} $\act{#1}$(#2)}%
{{\bf end operation}}
\algblockdefx[Procedure]{Procedure}{EndProcedure}%
[2]{{\bf procedure} $\act{#1}$(#2)}%
{{\bf end procedure}}
\algblockdefx[Receive]{Receive}{EndReceive}%
[2]{{\bf Upon receive} (#1)$_{\text{ #2 }}${\bf from} $q$}%
{{\bf end receive}}

In this section, we describe \ares{}.
%As opposed to its predecessors \cite{LS02, AKMS09}, \ares{} 
In the presentation of  \ares{} algorithm
%, compared to its predecessors \cite{LS02, AKMS09}, 
we decouple the reconfiguration service from the shared memory emulation, by utilizing
the DAPs presented in Section \ref{ssec:dap}. This allows \ares{},
to handle both the reorganization of the servers that host the data, as well as utilize 
a different atomic memory implementation per configuration. It is also important to 
note that \ares{} adopts a client-server architecture and separates the reader, writer 
and reconfiguration processes from the server processes that host the object data.
\kmk{More precisely, \ares{} algorithm comprises  of three major components: $(i)$ The reconfiguration protocol which consists
 of invoking, and subsequently installing new configuration via the \act{reconfig} operation by recon clients.
 $(ii)$ The read/write protocol for executing the \act{read} and \act{write} operations invoked by readers and writers.
 $(iii)$ The implementation of the DAPs for each installed configuration 
 that respect 
 %the consistency properties  (
 Property~\ref{property:dap} and 
 which are used by the 
 %on top of which the 
 \act{reconfig}, \act{read} and 
 \act{write} operations. } %are implemented.}

\subsection{Implementation of the Reconfiguration Service.}
%\vspace{-.5em}
\label{ssec:recbox} 
In this section, we describe the reconfiguration service in \ares{}.
The service relies on an underlying sequence of configurations (\kmk{already proposed or installed 
by \act{reconfig} operations}), 
%which can be  
%\textit{updated} by reconfiguration clients and \textit{read} by any client in $\idSet$. 
%Our service revolves around the idea that configurations are stored in the form of
%Configurations are stored 
%The set of configurations proposed by various reconfiguration clients, by invoking reconfiguration operations, are stored  in the form of 
in the from of a  ``distributed list'', which we refer to as the \myemph{global configuration sequence (or list)} $\gseq$. 
Conceptually, $\gseq$ represents
%The data type \myemph{configuration sequence}  
an ordered list of pairs $\langle c, status \rangle$, where $c$ is a configuration identifier ($ c \in \confSet$),  
 and a binary state variable $status \in \{F, P\}$
%. The variable $status$ associated with $c$,
that denotes whether $c$ is \myemph{finalized} ($F$) or is still \myemph{pending} ($P$). 
\nn{Initially,  $\gseq$ contains a single element, say $\tup{c_0, F}$, 
	%denote the first element of $\gseq$,
	 which is known to every participant in the service.}

\nn{ To facilitate the creation of $\gseq$, each}
server in $\servers{c}$ maintains a local variable $nextC  \in  \{\confSet \cup \{\bot\}\}\times\{P,F\}$, %(to point to the next configuration in $\gseq$), $nextC \in  \{\confSet \cup \{\bot\}\}\times\{P,F\}$,
which is used to point to the configuration that follows $c$ in $\gseq$. 
Initially, at any server  $nextC = \tup{\bot, F}$. Once $nextC$ it is set to a value %in $\confSet$ 
it is never altered.  As we show below, 
at any point in the execution of~\ares{} and in any configuration $c$, the 
%set of values, that are not equal to $\bot$, stored in 
$nextC$ variables of the non-faulty servers  in $c$ that are not equal to $\bot$ agree, i.e., 
 $\{s.nextC : s \in \servers{c} \wedge s.nextC\neq \bot\}$ is either empty of has only one element.
%We use the notation $|\cseq{seq}|$ to denote the length of a sequence.

 Clients discover the configuration that follows a $\tup{c,*}$
in the sequence by contacting a subset of servers in $\servers{c}$ and collecting their $nextC$ variables. 
%Each server in $\servers{c}$ has a variable $nextC$ (one for each configuration), $nextC \in  \{\confSet \cup \{\bot\}\}\times\{P,F\}$,
%which is used to point to the configuration that follows $c$ in $\gseq$. 
Every client in $\idSet$ maintains a local variable $cseq$ that is expected to  be some subsequence of 
$\gseq$.  Initially, at every client the value of  $cseq$ is $\tup{c_0,F}$.
We use the notation $\cseq{x}$ (a caret over some name) to denote state variables
that assume values from the domain $\{\confSet \cup \{\bot\}\}\times\{P,F\}$. %\nn{[NN: By element you mean the pair here?]} 
 %or $\cseq{config}$, or $\cseq{c}$, etc.

Reconfiguration clients may introduce new configurations,
each associated with a unique configuration identifier from $\confSet$.
 Multiple clients may concurrently attempt to introduce 
different configurations for same next link  in  $\gseq$.
\ares{} uses consensus to resolve such conflicts: 
\nn{a subset of servers in $\servers{c}$, in each configuration $c$,  
%collectively 
implements a distributed consensus service (such as 
  Paxos~\cite{L98}, RAFT~\cite{Raft}) , denoted by $\consensus{c}$. }
The  reconfiguration service consists of two major components: 
$(i)$ \myemph{sequence traversal}, responsible of discovering a  recent configuration in $\gseq$, and 
$(ii)$  \myemph{reconfiguration operation} that installs new configurations in $\gseq$.  

\begin{algorithm*}[!ht]
	%\hrule \F
	\begin{algorithmic}[2]
		\begin{multicols}{2}	{\small
				\Procedure{read-config}{$seq$}
				\State $\mu = \max(\{j: seq[j].status = F\})$	\label{line:readconfig:final}
				%\State $\nu = |cseq|$
				\State $\cseq{c} \gets seq[\mu]$ %.cfg$
				%\State {\bf send} $(\text{{\sc read-config}}, recon_i)$ to each   $s\in \bigcup_{\mu \leq i \leq \nu} \servers{currCfg}$
				% \While{$\cseq{c} \neq \bot$}
				%\State {\bf send} $(\text{\act{read-config}}, recon_i)$ to each $s\in \servers{c}$
				%\State {\bf until}  $\forall j,  \mu \leq j\leq \nu$  $\wedge$  
				%$\exists\quo{},  \quo{} \in \quorums{cseq[j].cfg}$ s.t. $ \forall s\in\quo{},  recon_i$  receives $cseq_s$ from $s$ 
				%\State {\bf until} $\exists\quo{},  \quo{}\in\quorums{c}$ s.t. $\forall s\in\quo{}, recon_i$  receives $nextC_s$ from $s$
				%\State $ell \gets \max_{cseq'\text{ received }}(|cseq'|)$
				
				%Andria: an alternative solution:
				% \While{$\cseq{c} \neq \bot$}
				% \State $ \at{\cseq{c}} \gets$\act{get-next-config}$(\cseq{c}.cfg)$ 
				% \If{$ \at{\cseq{c}} \neq\bot$} 
				% \State $\mu\gets \mu+1$				\label{line:readconfig:increment}
				% \State $seq[\mu] \gets \at{\cseq{c}}$	\label{line:readconfig:assign}
				% \State \act{put-config}$(seq[\mu-1].cfg, seq[\mu])$ 	\label{line:readconfig:put}
			 %   \label{line:newconfig:assign}
				% \EndIf
				% \EndWhile
				% \State {\bf return} $seq$
				% \EndProcedure

                \While{$\cseq{c} \neq \bot$}
				% \While{$\cseq{c} \neq \bot$}
				\State $ \cseq{c} \gets$\act{get-next-config}$(\cseq{c}.cfg)$ 
				\If{$ \cseq{c} \neq\bot$} 
				\State $\mu\gets \mu+1$				\label{line:readconfig:increment}
				\State $seq[\mu] \gets \cseq{c}$	\label{line:readconfig:assign}
				\State \act{put-config}$(seq[\mu-1].cfg, seq[\mu])$ 	\label{line:readconfig:put}
				% \State \at{\sout{$\cseq{c} \gets seq[\mu]$}} \label{line:newconfig:assign}
				\EndIf
				\EndWhile
				\State {\bf return} $seq$
				\EndProcedure

				\Statex
				
				\Procedure{get-next-config}{$c$}
				\State {\bf send} $(\text{{\sc read-config}})$ to each $s\in \servers{c}$
				\State {\bf until} $\exists\quo{},  \quo{}\in\quorums{c}$ s.t. $rec_i$ receives $nextC_s$ from $\forall s\in\quo{}$
				\If{$\exists s\in \quo{}\text{ s.t. } \status{nextC_s} = F$} 
					\State {\bf return} $nextC_s$
				\ElsIf{$\exists s\in \quo{}\text{ s.t. } \status{nextC_s} = P$} 
						\State {\bf return} $nextC_s$
					\Else
						\State {\bf return} $\bot$
				\EndIf 
				\EndProcedure
				
				\Statex
				
				\Procedure{put-config}{$c, nextC$}
				\State {\bf send} $(\text{{\sc write-config}}, nextC)$ to each $s\in \servers{c}$
				\State {\bf until} $\exists\quo{},  \quo{}\in\quorums{c}$ s.t. $rec_i$ receives {\sc ack} from $\forall s\in\quo{}$
				\EndProcedure	
		}\end{multicols}	
	\end{algorithmic}
	%\hrule \B
	\caption{Sequence traversal at each process $\pr\in\idSet$ of algorithm \ares.}
	\label{algo:parser}
	\vspace{-1em}
\end{algorithm*}

\myparagraph{Sequence Traversal.} 
Any \act{read/write/reconfig} operation $\op$ utilizes the sequence traversal mechanism  to discover the 
latest state of the global configuration sequence, as well as to ensure that such a state is discoverable
by any subsequent operation $\op'$. 
\kmk{See Fig.~\ref{fig:reconfig} for an example execution in the 
case of a reconfig operation.} \nn{In a high level, a client starts by 
collecting the $nextC$ variables from a quorum of servers in a configuration $c$,
such that $\tup{c,F}$ is the last  
finalized configuration in that client's local $cseq$ variable (or $c_0$ 
if no other finalized configuration exists). If any server $s$
returns a $nextC$ variable such that $nextC.cfg\neq\bot$,
then the client $(i)$ adds $nextC$ in its local $cseq$, $(ii)$ propagates $nextC$ 
in a quorum of servers in  $\servers{c}$, and $(iii)$ 
repeats this process in the configuration $nextC.cfg$. 
The client terminates when all servers reply with $nextC.cfg=\bot$.} 
More precisely, the sequence parsing consists of three actions (see Alg.~\ref{algo:parser}):  
%$(i)$ \act{get-next-config}(), to discover the next configuration, 
%$(ii)$ \act{put-config}(), which \nn{writes back $nextC$ to a quorum of servers},
%%makes sure the at least a majority of the servers in a configuration has $nextC$ to the same configuration 
%and  $(iii)$ \act{read-config}(), which finally returns \nn{the updated configuration sequence}. %a recent configuration in $\gseq$.
%%All three actions are part of the \textit{read/write/reconfig} operations. 
%We do present their specification and implementations 
%% here and we just 
%%refer to them during the description of the \textit{read/write/reconfig} operations. In high level, a $\act{read-config}$ 
%%action starts with a given configuration sequence and tries to append it with the latest installed configurations, a 
%%$\act{put-config}$ action propagates a configuration id to the servers of a given configuration, and lastly a
%%$\act{get-next-config}$ returns the configuration that follows a given configuration. More precisely the three 
%%actions are implemented 
%as follows (Alg.~\ref{algo:parser}):
%%\begin{itemize}
%%	\item 

\act{get-next-config}$(c)$:
	The action $\act{get-next-config}$ returns the configuration that follows $c$ in $\gseq$.
	During  \act{get-next-config}$(c)$, a client sends {\sc read-config}
	messages to all the servers in $\servers{c}$, and waits for replies containing $nextC$
	%. Once a server receives such a message responds with the value  of its $nextC$ variable. 
	%Once it receives replies 
	from a quorum in $\quorums{c}$. If there exists a reply with 
	%that contains a 
	$nextC.cfg\neq\bot$ 
	the action returns $nextC$; otherwise it returns $\bot$.  
	
%	\item 
\act{put-config}$(c, c')$:  
The $\act{put-config}(c, c')$ action propagates $c'$ to a quorum of servers in $\servers{c}$.
     During the action, the client  sends $( \mbox{{\sc write-config}}, c')$ messages,  
	to  the servers in $\servers{c}$ and waits for each server $s$ in some quorum $Q\in\quorums{c}$ to respond. 
	
%	\item 
\act{read-config}$(seq)$: 
	A $\act{read-config}(seq)$  sequentially traverses the installed configurations 
	%in $\gseq$ \nnrev{and  attempts  to set the status of the last configuration in $\gseq$ to $F$}{
	in order to discover the latest state of the sequence $\gseq$. 
	%This action accepts a configuration sequence $seq$ as an input and 
	%traverses the ``links'' between configurations to establish the latest form of the global configuration sequence. 
	At invocation, the client starts with the 
	last finalized configuration $\tup{c, F}$ in the given $seq$ (Line A\ref{algo:parser}:\ref{line:readconfig:final}), 
	%say $c=\config{\cseq{c_{\mu}}}$, 
	and  enters a loop to  traverse  $\gseq$ by  invoking  %$\act{get-next-config}(c)$
	$\act{get-next-config}()$, which returns the next configuration, assigned to $\cseq{c}$.
	While  $\cseq{c} \neq \bot$, then: (a) $\cseq{c}$ is appended at the end of the sequence $seq$;
	(b) a %$\act{put-config}(c, \cseq{c})$
	$\act{put-config}$ action is invoked to inform a quorum of servers in $\servers{c}$  to update the value of their
 $nextC$ variable to $\cseq{c}$.
%  \sout{;
% 	and (c) variable $c$ is set to $\config{\cseq{c}'}$}. %and $c’ = \act{get-next-config}(c_r)$. 
%The	$\act{put-config}(c, c_r)$ action is done in order to ensure that subsequent operations will retrieve a link to configuration $c_r$ from $\servers{c}$. 
	%the see the links as the current reconfiguration operations.
	If %$\cseq{c}' = \bot$
	$\cseq{c} = \bot$ the loop terminates and the action  \act{read-config} returns $seq$. 
%\end{itemize}

\begin{algorithm*}[!h]
	%\hrule \F
	\begin{algorithmic}[2]
		\begin{multicols}{2}{\small
				\State at each reconfigurer $rec_i$ 
				\State {\bf State Variables:}
				%\State  $\tg{}\in\N^+\times\wSet,~v\in V$
				\State  $cseq[] s.t. cseq[j]\in\confSet\times\{F,P\}$ with members:
				%\State $cseq[j].cfg\in\confSet$, the configuration identifier
				%\State $cseq[j].status\in\{F,P\}$, the configuration status 
				\State {\bf Initialization:} 
				\State $cseq[0] = \tup{c_0,F}$
				%\State $tg{}\gets \tup{0,\bot}, v \gets \bot, cseq[0] = \tup{c_0,F}$
				
				\Statex		
				
				\Operation{reconfig}{c} 
				%\State $wCounter\gets wCounter+1$
				\If {$c \neq \bot$} 		\label{line:install:valid}
				\State $cseq\gets$\act{read-config}$(cseq)$
				\State $cseq \gets \text{\act{add-config}}(cseq, c)$ %\Comment{Read the latest configuration sequence}
				\State $\text{\act{update-config}}(cseq)$
				\State $cseq\gets\text{\act{finalize-config}}(cseq)$
				\EndIf
				\EndOperation
				
				%				\Statex
				%				
				%				
				%				\Procedure{read-config}{$seq$}
				%				\State $\mu = \max(\{j: seq[j].status = F\})$	\label{line:readconfig:final}
				%				%\State $\nu = |cseq|$
				%				\State $c \gets seq[\mu].cfg$
				%				%\State {\bf send} $(\text{{\sc read-config}}, recon_i)$ to each   $s\in \bigcup_{\mu \leq i \leq \nu} \servers{currCfg}$
				%				\While{$c \neq \bot$}
				%				%\State {\bf send} $(\text{\act{read-config}}, recon_i)$ to each $s\in \servers{c}$
				%				%\State {\bf until}  $\forall j,  \mu \leq j\leq \nu$  $\wedge$  
				%				%$\exists\quo{},  \quo{} \in \quorums{cseq[j].cfg}$ s.t. $ \forall s\in\quo{},  recon_i$  receives $cseq_s$ from $s$ 
				%				%\State {\bf until} $\exists\quo{},  \quo{}\in\quorums{c}$ s.t. $\forall s\in\quo{}, recon_i$  receives $nextC_s$ from $s$
				%				%\State $ell \gets \max_{cseq'\text{ received }}(|cseq'|)$
				%				\State $nextC \gets$\act{read-next-config}$(c)$ 
				%				\If{$nextC.cfg\neq\bot$} 
				%				\State $\mu\gets \mu+1$				\label{line:readconfig:increment}
				%				\State $seq[\mu] \gets nextC$	\label{line:readconfig:assign}
				%				\State \act{put-config}$(seq[\mu-1].cfg, seq[\mu])$
				%				\State $c \gets seq[\mu].cfg$
				%				\Else
				%				\State $c \gets \bot$
				%				\EndIf
				%				\EndWhile
				%				\State {\bf return} $seq$
				%				\EndProcedure
				
				\Statex	
				\Procedure{add-config}{$seq$, $c$}
				%\State $\mu\gets\max(\{j: cseq[j].status = F\})$
				%\State $\nu \gets |cseq|$
				%\State $\mu'\gets\max(\{j: cseq'[j].status = F\})$
				\State $\nu \gets |seq|$
				\State $c' \gets seq[\nu].cfg$
				\State $d\gets \consensus{c'}.propose(c)$
				\State $seq[\nu+1]\gets \tup{d,P}$ 				\label{line:addconfig:assign}
				\State $\act{put-config}(c', \tup{d,P})$		\label{line:addconfig:put}
				\State  {\bf return} $seq$
				\EndProcedure
				
				\Statex
				
				\Procedure{update-config}{$seq$}
				\State $\mu\gets\max(\{j: seq[j].status = F\})$
				\State $\nu\gets |seq|$ 
				
				\State $M \gets \emptyset$
				\For{$i=\mu:\nu$}
				\State $\tup{t, v}  \gets \dagetdata{\config{seq[i]}}$
				\State $M  \gets M \cup  \{ \tup{\tg{}, v} \}$ \label{line:reconfig:max}
				\EndFor
				\State $\tup{\tg{},v} \gets \max_{t} \{ \tup{t, v}: \tup{t, v} \in M\}$
				%\State $\tup{\tg{},v} \gets \text{\act{get-data}}(cseq, \mu, \nu)$
				\State            	$\config{seq[\nu]}.\act{put-data}(\tup{\tg{},v})$
				% $seq[\nu].\act{put-data}(\tup{\tg{},v})$
				\EndProcedure
				
				\Statex

				\Procedure{finalize-config}{$seq$}
				\State $\nu = |seq|$
				\State $seq[\nu].status \gets F$	\label{line:status:finalize}
				\State $\act{put-config}(seq[\nu-1].cfg, seq[\nu])$
				\State \textbf{return} $seq$ 
				\EndProcedure

		}\end{multicols}	
	\end{algorithmic}
	%\hrule \B
	\caption{Reconfiguration protocol of algorithm \ares.}
	\label{algo:reconfigurer}
	\vspace{-1em}
\end{algorithm*}

\begin{algorithm*}[!ht]
	%\hrule \F
	\begin{algorithmic}[2]
		\begin{multicols}{2}{\small
			\State at each server $s_i$ in configuration $c_k$
			\State{\bf State Variables:}
			\State  $\tg{}\in\N \times\wSet$, initially, $\tup{0,\bot}$
			\State $v\in V$, intially, $\bot$
			\State $nextC\in \confSet\times \{P,F\}$, initially $\tup{\bot,P}$
			%\State  $msgType\in\{~seen\subseteq\mathcal{V}\cup\{w\}$	
			%\State{\bf Initialization:}
			%\State $\tg{}\gets \tup{0,\bot}, v \gets \bot$
			
			\Statex
			
			\Receive{{\sc read-config}}{$s_i,c_k$}
			\State send $nextC$ to $q$
			\EndReceive
			
			\Statex
			
			\Receive{{\sc write-config}, $cfgT_{in}$}{$s_i,c_k$}
			\If{$nextC.cfg=\bot~\vee~nextC.status=P$} \label{line:server:finalize}
			\State $nextC\gets cfgT_{in}$
			\EndIf
			\State send {\sc ack} to $q$
			\EndReceive
			
%			\Statex	
%			
%			\Receive{{\sc query-tag}}{$s_i,c_k$} %\Comment{Called upon reception of a message}
%				\State $\act{handle-get-tag(c_k)}$
%				%\State send $\tg{}$ to $q$
%			\EndReceive
%			
%			\Statex
%			
%			\Receive{{\sc query}}{$s_i,c_k$}
%				\State $\act{handle-get-data(c_k)}$
%				%\State send $\tup{\tg{}, v}$ to $q$
%			\EndReceive
%			
%			\Statex
%			
%			\Receive{{\sc write}, $\tup{\tg{in}, v_{in}}$}{$s_i,c_k$}
%				\State $\act{handle-put-data(c_k)}$
%%				\If {$\tg{in}> \tg{}$} 	\label{line:server:tg-comparison}
%%					\State  $\tup{\tg{},v}\gets \tup{\tg{in},v_{in}}$ \label{line:server:update}
%%				\EndIf
%%				\State  send  {\sc ack} to $q$ 	\label{line:server:reply}
%			\EndReceive
%	
			
		}\end{multicols}	
	\end{algorithmic}
	%\hrule \B
	\caption{Server protocol of algorithm \ares.}
	\label{algo:server}
	\vspace{-1em}
\end{algorithm*}

\myparagraph{Reconfiguration operation.}
%Consensus is used at the heart of reconfiguration, in order to establish the order on the proposed configurations. 
%	In each configuration $c$, a consensus protocol runs on the servers in  $\servers{c}$, and any reconfiguration client can  
%	propose a configuration $c'\neq c$ on the consensus instance of configuration $c$.
%In particular, 
A reconfiguration operation $\act{reconfig}(c)$, $c \in \mathcal{C}$,  invoked by any reconfiguration client 
$rec_i$, attempts to append $c$ to $\gseq$. \kmk{The set of server processes in $c$ are not a part of any other configuration different 
from $c$.} \nn{In a high-level, $rec_i$ first executes a sequence traversal to discover 
the latest state of $\gseq$. Then it attempts to add the new configuration $c$, at the end of the 
discovered sequence by proposing $c$ in the consensus instance of the last configuration in the sequence. 
The client accepts and appends the decision of the consensus instance (that might be different than $c$).
Then it attempts to transfer the latest value of the read/write object to the latest installed configuration. 
Once the information is transferred, $rec_i$ finalizes the last configuration in its local sequence and 
propagates the finalized tuple to a quorum of servers in that configuration.}
The operation consists of four phases, executed consecutively by $rec_i$ (see Alg.~\ref{algo:reconfigurer}): 
%$(i)$  $\act{read-config}$, reads the recent global configuration sequence;  
%$(ii)$ $\act{add-config}$, attempts to append a new configuration $c$;
% $(iii)$ $\act{update-config}$, scans for the most recent object value (w.r.t. their tags)
%  in the set of configurations in its local state variable $cseq$, and writes this value to the most recent 
%  configuration in $cseq$, $(iv)$ $\act{finalize-config}$, sets the $status$ component 
%  of the last tuple  in the local $cseq$ to $F$. More precisely:

$\act{read-config}(seq)$: The phase $\act{read-config}(seq)$ at $rec_i$, reads the recent global configuration 
sequence as described in the sequence traversal. 

$\act{add-config}(seq, c)$: The $\act{add-config}(seq, c)$ attempts to append a new configuration $c$ to the end of  \nnrev{$\gseq$}{$seq$ (client's view of $\gseq$)}. Suppose the last configuration in $seq$ is $c'$ (with status either $F$ or $P$), then in order to decide the 
most recent configuration, $rec_i$ invokes $\consensus{c'}.propose(c)$, on the consensus object associated with configuration $c'$. 
Let  $d\in\confSet$ be the configuration identifier decided by the consensus service.
If $d \neq c$,  this  implies that another (possibly concurrent) reconfiguration operation, invoked  by a reconfigurer $rec_j\neq rec_i$, proposed and succeeded $d$  as the configuration to follow $c'$.
%imply that $d$ is a another configuration already added into the global configuration sequence, possibly due to another reconfiguration operation by some reconfiguration client.
 In this case, $rec_i$  adopts $d$ as it own propose configuration, by  adding $\tup{d, P}$ to the end of its local $cseq$ (entirely ignoring $c$),
 using the operation $\act{put-config}(c', \tup{d, P})$, and returns the extended configuration $seq$.
 %and continues executing the remaining phases of the reconfiguration operation. 
 %Next, via the 
\remove{
\nn{[NN: We have talked about the action below so we can remove this description.]}

$\act{put-config}(c', \tup{d, P})$:
Once the $cseq$ is appended, $rec_i$ invokes the 
action $c'.\act{put-config}(\tup{d, P})$, to send $\tup{d,P}$ to a quorum of servers in $\servers{c'}$. 
%and awaits
% responses from a quorum; and upon receiving 
%such a message sets the value $nextC$ to $\tup{d, P}$ and responds with an acknowledgement to $rec_i$. 
Finally, $cseq$ is updated to the extended configuration sequence $seq'$ returned by the $\act{add-config}$ action. 
}

$\act{update-config}(seq)$:
Let us denote by $\mu$ the index of the last configuration in the local sequence $cseq$, at $rec_i$, 
such that its corresponding status is 
$F$; and $\nu$ denote the last index of $cseq$.  Next $rec_i$ invokes $\act{update-config}(cseq)$, which 
gathers the tag-value pair corresponding to 
the maximum tag in each of the configurations in $\cseq{cseq[i]}$ for $\mu \leq i \leq \nu$,
and transfers that pair to the configuration that was added by the $\act{add-config}$ action. 
The $\act{get-data}$ and $\act{put-data}$ DAPs are used to transfer the value of the object to the new configuration, 
and they are implemented \nnrev{respectively}{with respect} to the 
%atomic algorithm that is used in each of the 
configuration that is accessed. 
%During the  execution of the procedure  $\act{get-data}(cseq[i])$, $rec_i$ sends requests the tag-value  to all servers  in  $cseq[i]$ and awaits responses from a quorum.  
Suppose $\tup{t_{max}, v_{max}}$  is the tag value pair corresponding to the highest tag among the responses from all the $\nu - \mu + 1$ configurations. Then, 
 $\tup{t_{max}, v_{max}}$  is written to the configuration $d$ via the invocation of  
 $\config{\cseq{cseq[\nu]}}.\act{put-data}(\tup{\tg{max},v_{max}})$.
 
 {\footnotesize
	\begin{figure}[!t]
		\begin{center}
			\includegraphics[width=0.5\textwidth]{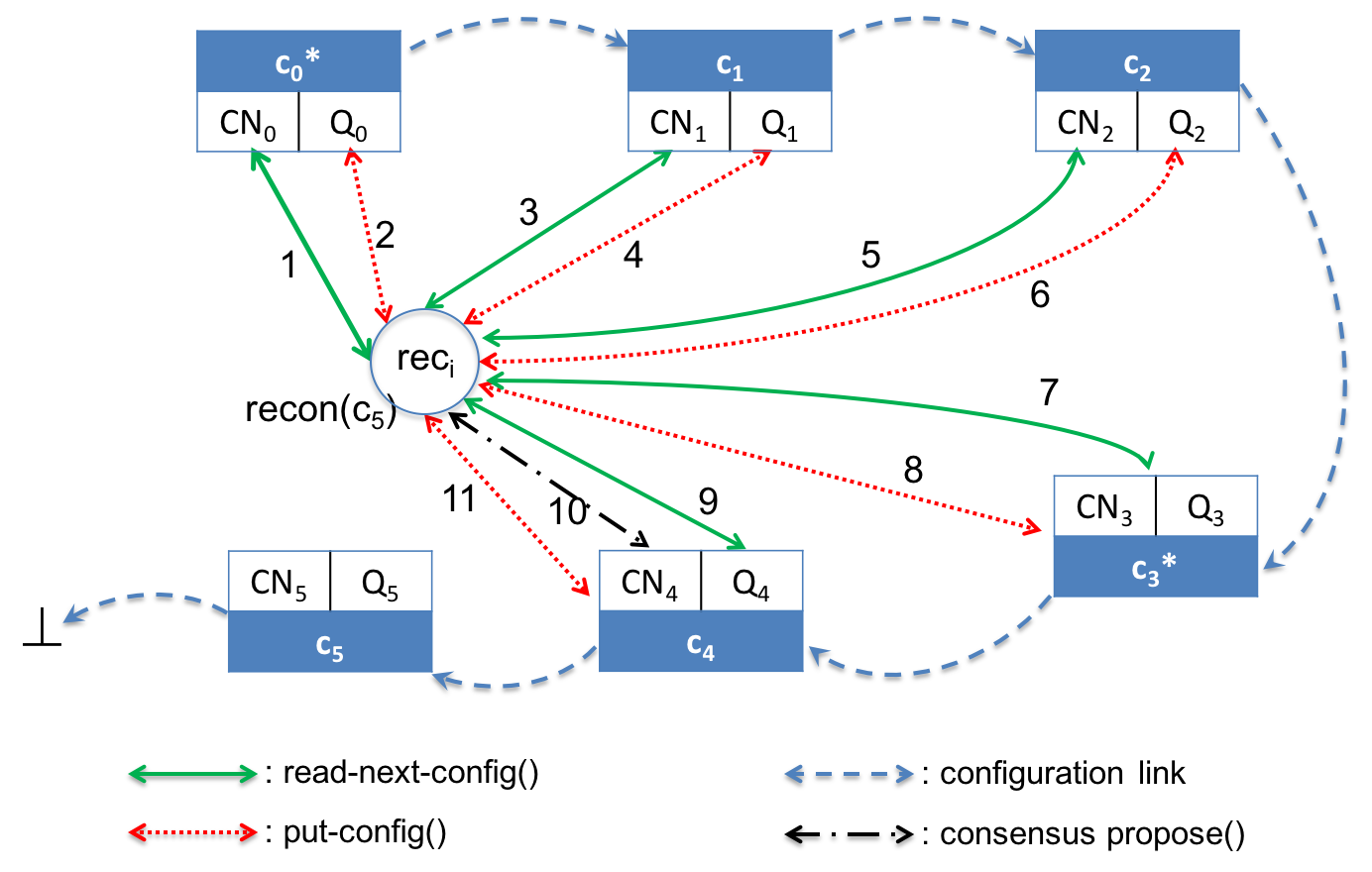}
			\caption{Illustration of an execution of the reconfiguration steps.}
			\label{fig:reconfig}
		\end{center}
		\vspace{-2.3em}
	\end{figure}
}

\sloppy$\act{finalize-config}(cseq)$:
Once the tag-value pair is transferred, in the last phase of the reconfiguration operation,  $rec_i$ executes %the procedure call 
$\act{finalize-config}(cseq)$, 
%during which, it 
to update the status of the last configuration in $cseq$, say $d = \config{\cseq{cseq[\nu]}}$, to $F$. 
%and writes to a quorum of the penultimate configuration of $cseq$, i.e., $cseq[\nu -1]$ and completes the reconfiguration operation.
The reconfigurer $rec_i$ informs a quorum of servers in the previous configuration $c=\config{\cseq{cseq[\nu-1]}}$, i.e.  in some $Q \in \quorums{c}$, 
about the change of status, by executing the $\act{put-config}(c, \tup{d,F})$ action. 

\myparagraph{Server Protocol.}
Each server  responds to requests from clients (Alg.~\ref{algo:server}). 
A server waits for \nnrev{five}{two} types of messages: {\sc read-config} and  {\sc write-config}.
% {\sc query}, {\sc query-tag}, and {\sc write} messages. 
When a {\sc read-config } message is received for a particular configuration $c_k$, then the server returns $nextC$ variables of the servers in $\servers{c_k}$. %\nn{[NN:again this notation]} 
A {\sc write-config} message attempts to update the $nextC$ variable of the server with a particular tuple $cfgT_{in}$.
A server changes the value of its local $nextC.cfg$ in two cases: (i) $nextC.cfg=\bot$, or (ii) $\status{nextC}= P$.

Fig.~\ref{fig:reconfig} 
%\nn{[NN: this is wrong numbering. probably has to do with the numbering of algorithms]} 
illustrates an example execution of a reconfiguration operation
$\act{recon}(c_5)$. %and the traversal of a concurrent write operation. 
In this example, the reconfigurer $rec_i$ goes through a number of configuration queries (\act{read-next-config})
before it reaches configuration $c_4$ in which a quorum of servers replies with $nextC.cfg=\bot$. 
There it proposes $c_5$ to the consensus object of $c_4$ ($\consensus{c_4}.propose(c_5)$ on arrow 10), and 
once $c_5$ is decided, $\act{recon}(c_5)$ completes after executing $\act{finalize-config}(c_5)$.

\begin{algorithm*}[!ht]
	%\hrule \F
	\begin{algorithmic}[2]
		\begin{multicols}{2}{\small
				\Part{Write Operation}
				\State at each writer $w_i$ 
				%\State {\bf State Variables:}
				%\State  $\tg{},\tg{max}\in\N^+\times\wSet,~v\in V, terminate\in\{true,false\}$
				%\State  $cseq[~]$, array with elements in $\confSet\times\{F,P\}$ with members:
				%\State\T $cseq[i].conf\in\confSet$, the configuration identifier
				%\State\T $cseq[i].status\in\{F,P\}$, the pending or finalized status 
				%\State {\bf Initialization:} 
				%\State $\tg{}\gets \tup{0,w_i}$
				%\State $v \gets \bot$
				%\State $cseq[0] = \tup{c_0,F}, terminate={\bf false}$
				\State {\bf State Variables:}
				%\State  $\tg{}\in\N^+\times\wSet,~v\in V$
				\State  $cseq[] s.t. cseq[j]\in\confSet\times\{F,P\}$ with members:
				%\State\T $cseq[j].cfg\in\confSet$, the configuration identifier
				%\State\T $cseq[j].status\in\{F,P\}$, the configuration status 
				\State {\bf Initialization:} 
				\State $cseq[0] = \tup{c_0,F}$
				
				\Statex		
				
				\Operation{write}{$val$}, $val \in V$ 
				%\State $wCounter\gets wCounter+1$
				\State $cseq\gets$\act{read-config}($cseq$)  \label{line:writer:readconfig} %\Comment{Read the latest configuration sequence}
				\State $\mu\gets\max(\{i: cseq[i].status = F\})$ \label{line:writer:lastfin}
				\State $\nu\gets |cseq|$ 
				\For{$i=\mu:\nu$}
				\State $\tg{max} \gets \max(\config{cseq[i]}.\act{get-tag}(), \tg{max})$  \label{line:writer:max}
				\EndFor
				\State $\tup{\tg{},v} \gets \tup{ \tup{\tg{max}.ts+1, \wrt_i}, val}$ \label{line:writer:increment}
				\State $done \gets false$
				\While{{\bf not} $done$} \label{line:writer:whilebegin}
				\State $\config{cseq[\nu]}.$\act{put-data}$(\tup{\tg{},v})$ \label{line:writer:prop}
				\State $cseq\gets$\act{read-config}($cseq$)
				\If{$|cseq| = \nu$}
				\State $done \gets  true$
				\Else
				\State $\nu\gets |cseq|$ \label{line:writer:whileend}
				\EndIf
				\EndWhile
				\EndOperation
				\EndPart
				
				\Part{Read Operation}
				\State at each reader $\rdr_i$ 
				\State {\bf State Variables:}
				\State  $cseq[] s.t. cseq[j]\in\confSet\times\{F,P\}$ with members:
				%			\State\T $cseq[j].cfg\in\confSet$, the configuration identifier
				%			\State\T $cseq[j].status\in\{F,P\}$, the configuration status 
				\State {\bf Initialization:} 
				\State $cseq[0] = \tup{c_0,F}$
				
				\Statex
				
				\Operation{read}{ } 
				%\State $wCounter\gets wCounter+1$
				\State $cseq\gets$\act{read-config}($cseq$)  \label{line:reader:readconfig}%\Comment{Read the latest configuration sequence}
				\State $\mu\gets\max(\{j: cseq[j].status = F\})$ \label{line:reader:lastfin}
				\State $\nu\gets |cseq|$ 
				\For{$i=\mu:\nu$} \label{line:rw:getdata:start}
				\State $\tup{\tg{},v} \gets \max(\config{cseq[i]}.\act{get-data}(), \tup{\tg{},v})$ \label{line:reader:max}
				\EndFor \label{line:rw:getdata:end}
				\State $done\gets {\bf false}$
				\While{{\bf not} $done$}  \label{line:reader:whilebegin}
				\State $\config{cseq[\nu]}.\act{put-data}(\tup{\tg{},v})$ \label{line:reader:prop}
				\State $cseq\gets$\act{read-config}($cseq$)
				\If{$|cseq| = \nu$}
				\State $done \gets  true$
				\Else
				\State $\nu\gets |cseq|$ \label{line:reader:whileend}
				\EndIf
				\EndWhile
				\State {\bf return} $v$
				\EndOperation
				\EndPart

		}\end{multicols}	
	\end{algorithmic}
	%\hrule \B
	\caption{Write and Read protocols at the clients for \ares.}
	\label{algo:writer}
	\vspace{-1em}
\end{algorithm*}
\subsection{Implementation of Read and Write operations.}
The read and write operations in \ares{} are expressed in terms of the DAP primitives  (see 
Section \ref{ssec:dap}). 
%A read  consists of  an execution of  $\act{get-data}$ primitive followed by a $\act{put-data}$ primitive,
%while a write consists of calls to $\act{get-tag}$ and  $\act{put-data}$ primitives. 
This provides the flexibility to \ares{} to use different implementation of DAP primitives in different configurations, without changing the basic structure of  \ares{}. 
At a high-level, a \act{write} (or \act{read})  operation is executed where the client: $(i)$ obtains the \textit{latest configuration sequence} by using the 
$\act{read-config}$ action of the reconfiguration service, $(ii)$ queries  the configurations, in $cseq$, starting from the last finalized configuration
to the end of the discovered configuration 
sequence, for the latest $\tup{tag,value}$ pair with a help of $\act{get-tag}$ (or $\act{get-data}$) operation \nn{as specified for each configuration}, and 
$(iii)$ repeatedly propagates  a new $\tup{tag', value'}$ pair (the largest $\tup{tag,value}$ pair)   with $\act{put-data}$ in the last configuration of its local sequence, until no additional configuration is observed. 
%Now
%{\sc get} and {\sc put} primitives 
%we  describe the execution 
In more detail, the algorithm of a  read or write operation $\pi$ is as follows (see Alg.~\ref{algo:writer}): 
%The server keeps a passive behavior and replies to any message it receives. 

%Here we present the read and write operations in more detail. 
%As we presented in Section \ref{sec:primitives}, for an algorithm to 
%preserve atomicity it needs to satisfy some properties on the 
%implementation of its {\sc get} and {\sc put} primitives. 

%Those primitives are implemented in our replication algorithm as follows:
%\begin{itemize}
%	\item {\sc get-tag}$(c)$: Send {\sc query-tag} messages to all the servers in $\servers{c}$ and wait 
%	for each server $s$ in a quorum $Q\in\quorums{q}$ to reply with its local $\tg{s}$. Find and return the 
%	maximum of those tags. 
%	\item {\sc get-data}$(c)$: Send {\sc query} messages to all the servers in $\servers{c}$ and wait 
%	for each server $s$ in a quorum $Q\in\quorums{q}$ to reply with its local $\tup{\tg{s}, v_s}$ pair. 
%	Find the maximum of those tags, say $\tg{s'}$, and return the pair received by $s'$, $\tup{\tg{s'},v_{s'}}$.
%	\item {\sc put-data}$(c, \tup{\tg{},v})$: Send {\sc write-tag} messages, along with the  $\tup{\tg{},v}$ pair, 
%	to all the servers in $\servers{c}$ and wait for each server $s$ in a quorum $Q\in\quorums{q}$ to reply with an {\sc ack}. 
%\end{itemize}

A write (or read) operation is invoked at a client  $\pr$ when line~Alg.~\ref{algo:writer}:\ref{line:writer:readconfig} %for the writer  
(resp. line~Alg.~\ref{algo:writer}:\ref{line:reader:readconfig}) is executed.  
%for reader),
At first, $\pr$ issues a $\act{read-config}$ action to obtain the latest 
introduced configuration in $\gseq$, in both operations. 
%Note that each process contains a state 
%variable $cseq$ which initially contains the initial configuration $\tup{c_0,F}$. 
%Starting with a guess, $cseq$, for the latest global configuration sequence,
%As shown in the reconfiguration service, 
%process $\pr$ queries a quorum in each configuration $\config{cseq[i]}$, for $\max(j:\status{cseq[j]}=F)\leq i \leq |cseq|$,
%to discover newly introduced and/or newly finalized configurations, during the $\act{read-config}$ action. 

%We assume that each configuration has enough active servers to allow 
%$\pr$ to make progress	even when some configuration is has been replaced 
%by a newer configuration. 
%In case a system wants to garbage collect or decommission inactive servers,
%then a service similar to DNS can be used to provide an estimate of 
%the latest finalized configuration. Notice that such a service would affect 
%the liveness and not the safety of our current implementation so we do 
%not get into details on how such a service can be implemented in this paper. 

\nn{If $\op$ is a \act{write} %(resp. Alg.~\ref{algo:writer}:\ref{line:reader:lastfin} if $\op$ is a  read), 
	$\pr$ detects the last finalized entry in $cseq$, 
	say $\mu$, and performs a $cseq[j].conf.\act{get-tag}()$ action,
	for $\mu\leq j\leq|cseq|$ (line Alg.~\ref{algo:writer}:\ref{line:writer:lastfin}). Then $\pr$ discovers the
	\textit{maximum tag} among all the returned tags ($\tg{max}$),
	and it increments the maximum tag discovered 
	(by incrementing the integer part of $\tg{max}$), generating a new tag, say $\tg{new}$. It assigns 
	$\tup{\tg{}, v}$ to $\tup{\tg{new}, val}$, where $val$ is the
	value he wants to write (Line Alg.~\ref{algo:writer}:\ref{line:writer:increment}).}

\nn{
if $\op$ is a \act{read}, 
$\pr$ detects the last finalized entry in $cseq$, 
say $\mu$, and performs a  $cseq[j].conf.\act{get-data}()$ action, 
for $\mu\leq j\leq|cseq|$ (line Alg.~\ref{algo:writer}:\ref{line:reader:lastfin}). Then $\pr$ discovers the
\textit{maximum tag-value} pair ($\tup{\tg{max},v_{max}}$) among the replies, 
and assigns $\tup{\tg{}, v}$ to $\tup{\tg{max},v_{max}}$.}

%In lines Alg.~\ref{algo:writer}:\ref{line:writer:lastfin} if $\op$ is a write (resp. Alg.~\ref{algo:writer}:\ref{line:reader:lastfin} if $\op$ is a  read), 
% $\pr$ detects the last finalized entry in $cseq$, 
%say $\mu$, and performs a $cseq[j].conf.\act{get-tag}()$ action if $\op$ is a \act{write}, or $cseq[j].conf.\act{get-data}()$ action if $\op$ is a \act{read}, 
%for $\mu\leq j\leq|cseq|$. Then $\pr$ discovers the
%maximum tag among all the returned tags ($\tg{max}$) or tag-value pairs ($\tup{\tg{max},v_{max}}$) respectively.
%%sends \textit{query} messages to all the members of each configuration $cseq[j].conf$, for $\mu\leq j\leq|cseq|$.
%%Notice that any configuration $cseq[j].cfg$,  for $\mu< j\leq|cseq|$, is in a pending state and thus $cseq[j].status = P$.
%%When the client receives replies from a quorum in each configuration $cseq[j].cfg$, it discovers the maximum 
%%$\tup{\tg{max},v_{max}}$ among those replies (Lines A\ref{algo:writer}:\ref{line:writer:max} and A\ref{algo:reader}:\ref{line:reader:max}). 
%If $\op$ is a \act{write}, $\pr$ increments the maximum tag discovered 
%(by incrementing the integer part of $\tg{max}$), generates a new tag, say $\tg{new}$, and assigns 
%$\tup{\tg{}, v}$ to $\tup{\tg{new}, val}$, where $val$ is the
%value he wants to write (Line Alg.~\ref{algo:writer}:\ref{line:writer:increment}). If $\op$ is a \act{read}, then $\pr$ assigns 
%$\tup{\tg{}, v}$ to $\tup{\tg{max},v_{max}}$, i.e.,  the maximum discovered tag-value pair.
%

Once specifying the $\tup{\tg{}, v}$ to be propagated, both reads and writes 
% if $\op$ is a read, $\pr$ 
 %does the following until no new configuration is discovered: (i) 
execute the $\config{cseq[\nu]}.\act{put-data}(\tup{\tg{}, v})$ action, where $\nu=|cseq|$, 
%to a quorum in the last configuration in $cseq$, 
followed by executing  $\act{read-config}$ action, to examine whether new configurations were 
introduced in $\gseq$. The repeat these steps until no new configuration is discovered (lines  Alg.~\ref{algo:writer}:\ref{line:writer:whilebegin}--\ref{line:writer:whileend},
or lines  Alg.~\ref{algo:writer}:\ref{line:reader:whilebegin}--\ref{line:reader:whileend}).
%retrieving a config
Let $cseq'$ be the sequence returned by the $\act{read-config}$ action. 
If $|cseq'| = |cseq|$ then no new configuration is introduced, and
the read/write operation terminates; otherwise, $\pr$ sets $cseq$ to $cseq'$ and repeats the two actions. 
Note,  in an execution of  \ares,  two consecutive $\act{read-config}$ 
operations that return $cseq'$ and $cseq''$ respectively must hold that $cseq'$ is a prefix of $cseq''$,
and hence $|cseq'|=|cseq''|$ only if $cseq' = cseq''$.  Finally, if $\pi$ is a read operation the value with the highest
tag discovered is returned to the client.

\nnfix{\myparagraph{Discussion} \ares{} shares similarities with previous algorithms like RAMBO \cite{GLS03} and 
	the framework  in \cite{spiegelman:DISC:2017}. The reconfiguration technique used in \ares{} ensures the prefix property 
	on the configuration sequence (resembling a blockchain data structure \cite{N08bitcoin}) while the array structure in RAMBO allowed 
	nodes to maintain an incomplete reconfiguration history. 
	%While in RAMBO recon operation needed to write in both the old and the new configurations due to the possible non-continuity of the 
	%known configurations in the config array, in ARES the prefix property of the configuration sequence 
	%allows each recon operation to write only to the latest configuration it discovered. 
	On the other hand, the DAP usage, 
	%enables \ares{} to be oblivious on the underlying atomic memory implementation in each configuration. This 
	exploits a different viewpoint compared to \cite{spiegelman:DISC:2017}, allowing implementations of 
	atomic read/write registers without relying on strong objects, like ranked registers \cite{GD05}.	
	%	where the authors focus to provide a general framework 
	%	for reconfigurations that led to the use of read-write-modify objects, like ranked registers \cite{GD05},
	%	for the implementation of linearizable read/write objects.  
}

\section{Implementation of the DAP{s}}\label{ssec:dap:impl}
\label{sec:dap:ec}

			In this section, we present an implementation of the DAPs,  that satisfies the properties in Property~\ref{property:dap},  for a configuration $c$,  with $n$ servers 
			 using a $[n, k]$ MDS coding scheme for storage. We implement an instance of the algorithm in a 
			%Atomicity is always guaranteed. 
 configuration of  $n$ server processes. 
 We store each coded element $c_i$, corresponding to an object  at server $s_i$, where $i=1, \cdots, n$.
 The implementations of DAP primitives used in \ares{} are shown  
%\nnrev{by implementing the}{the} 
%DAP primitives \nnrev{as}{are implemented} 
in Alg.~\ref{fig:casopt}, and the servers' responses in Alg.~\ref{fig:casopt:server}.
			\begin{algorithm*}[!ht]
				\begin{algorithmic}[2]
					{\small
					\begin{multicols}{2}
							\State{ at each process $\pr_i\in\idSet$}
							%\remove{
%										{\scriptsize
%				%\Part{Generic Algorithm $A_1$}
%				\Operation{read}{} 
%				%\State $wCounter\gets wCounter+1$
%				\State $\tup{t, v} \gets \dagetdata{c}$
%				\State $\daputdata{c}{ \tup{t,v}}$
%				\State return $ \tup{t,v}$
%				\EndOperation
%				\Statex
%				\Operation{write}{$v$} 
%				%\State $wCounter\gets wCounter+1$
%				\State $t \gets \dagettag{c}$
%				\State $t_w \gets inc(t)$
%				\State $\daputdata{c}{\tup{t_w,v}}$
%				\EndOperation
%				%\EndPart
%			}%}
	
							\Statex
							\Procedure{c.get-tag}{}
							%	\State {\bf send} $(\text{\act{query}},\rdr)$ to every server $s\in \bigcup_{cseq[i]}members(\qs_{cseq[i].conf})$
							\State {\bf send} $(\text{{\sc query-tag}})$ to each  $s\in \servers{c}$
							\State {\bf until}   $\pr_i$ receives $\tup{t_s}$ from $\left\lceil \frac{n + k}{2}\right\rceil$ servers in $\servers{c}$
				% 			\State $t_{max} \gets \max(\{t_s : \text{ received } \tup{t_s,v_s} \text{ from } s \})$
				            \State $t_{max} \gets \max(\{t_s : \text{ received } t_s \text{ from } s \})$
							\State {\bf return} $t_{max}$
							\EndProcedure
							
							\Statex
							
							\Procedure{c.get-data}{}
							%	\State {\bf send} $(\text{{\sc query}},\rdr)$ to every server $s\in \bigcup_{cseq[i]}members(\qs_{cseq[i].conf})$
								\State {\bf send} $(\text{{\sc query-list}})$ to each  $s\in \servers{c}$
								\State {\bf until}    $\pr_i$ receives $List_s$ from each server $s\in\srvSet_g$ s.t. $|\srvSet_g|=\left\lceil \frac{n + k}{2}\right\rceil$ and  $\srvSet_g\subset \servers{c}$ 
								\State  $Tags_{*}^{\geq k} = $ set of tags that appears in  $k$ lists	\label{line:getdata:max:begin}
								\State  $Tags_{dec}^{\geq k} =$ set of tags that appears in $k$ lists with values
								\State  $t_{max}^{*} \leftarrow \max Tags_{*}^{\geq k} $
                                \State  $t_{max}^{dec} \leftarrow \max Tags_{dec}^{\geq k} $ \label{line:getdata:max:end}
								\If{ $t_{max}^{dec} =  t_{max}^{*}$} 
								    \State  $v \leftarrow $ decode value for $t_{max}^{dec}$
								\EndIf
								%\State $List_M \triangleq \bigcup_{s \in \srvSet_g}  List_s$
								%$\State  $\forall t$, $List_M(t) \triangleq \{ (t, v): (t,v) \in List_M \}$  
								%\State $\forall t$, $T(t') \triangleq \{t: (t,v) \in List_M(t) \wedge t \geq t' \}$
								%\State $t_r \gets \max \{t : (t, *) \in List_M ~\wedge |List_M(t)| \geq k~\wedge |T(t)| \leq \delta \}$
								%\State $v_s\gets \text{decode from }  List_M(t_{r}))$
								\State {\bf return} $\tup{t^{dec}_{max},v}$
							\EndProcedure
							
							\Statex				
							
							\Procedure{c.put-data}{$\tup{\tg{},v})$}
								\State $\Coded = [(\tg{}, e_1), \ldots, (\tg{}, e_n)]$, $e_i = \Phi_i(v)$
								\State {\bf send} %$(\text{{\sc write}},
							$(\text{{\sc PUT-DATA}},
							\tup{\tg{},e_i})$ to each $s_i \in \servers{c}$
								\State {\bf until} $\pr_i$ receives {\sc ack} from $\left\lceil \frac{n + k}{2}\right\rceil$ servers in $\servers{c}$
							\EndProcedure
							%\EndPart

%							\Part{write($v$)}\EndPart
%							\Part{\underline{\GetTag}} {
%								\State  Send  $(\QueryTag)$ to all servers $\mathcal{S}$.
%								\State  Await responses from majority
%								\State  Select the max tag  $t^*$
%							}\EndPart
%							\Statex
%							\Part{\underline{\PutData}} {
%								\State $t_w = (t^{*}.z + 1, w)$.  
%								\State $\Coded = [(t_w, c_1), \ldots, (t_w, c_n)]$, $c_i = \Phi_i(v)$
%								\State Send  $(\CodedElementTag, \Coded)$ to all servers $\mathcal{S}$.
%								\State Terminate after $\left\lceil \frac{n + k}{2}\right\rceil$ acks
%							}	\EndPart
							
%							\Statex
%							\Part{read}\EndPart
%							\Part{\underline{\GetData}} {
%								\State  Send $(\QueryList)$ to all servers $\mathcal{S}$.
%								\State  Wait for $\left\lceil \frac{n+k}{2}\right\rceil$ $Lists$ 
%								\State  Select the max tag, $t_r$, the corresponding value, $v_r$, is decodable using the $Lists$; additionally   $t_r$ is among the highest distinct $\delta$ tags received in any $Lists$.
%							}\EndPart	
%							\Statex
%							\Part{\underline{\PutData}} {
%								\State $\Coded = [(t_r, c_1), \ldots, (t_r, c_n)]$, $c_i = \Phi_i(v_r)$
%								\State Send $(\CodedElementTag, \Coded)$ to all servers $\mathcal{S}$.
%								\State Wait for $\left\lceil \frac{n + k}{2}\right\rceil$ acks
%								\State Return $v_r$
%							}	\EndPart
%							
%							
					\end{multicols}
				}
				\end{algorithmic}	
				\caption{DAP implementation 
					%for  template $A_1$ to implement 
					for  \ares{}. }
				\label{fig:casopt}
				\vspace{-1em}
			\end{algorithm*}

	\begin{algorithm*}[!ht]
	\begin{algorithmic}[2]
		{\small
		\begin{multicols}{2}
				\State{at each server $s_i \in \mathcal{S}$ in configuration $c_k$}
				\Statex
				\State{\bf State Variables:}%{ 										
					%\Statex $(t_{loc}, v_{loc}) \in \mathcal{T} \times {\mathcal V}$, initially   $(t_0, v_0)$
					%\Statex $status \in \{active, repair\}$, initially $active$
					\Statex $List \subseteq  \mathcal{T} \times \mathcal{C}_s$, initially   $\{(t_0, \Phi_i(v_0))\}$
				%}\EndPart
			
			\Statex
			\Receive{{\sc query-tag}}{$s_i,c_k$}
				\State $\tg{max} = \max_{(t,c) \in List}t$
				\State Send $\tg{max}$ to $q$
			\EndReceive
			\Statex

			\Receive{{\sc query-list}}{$s_i,c_k$}
				\State Send $List$ to $q$
			\EndReceive
\State
			\Receive{{\sc put-data}, $\tup{\tg{},e_i}$}{$s_i,c_k$}
				\State $List \gets List \cup \{ \tup{\tg{}, e_i}  \}$ 
				\If{$|List| > \delta+1$}
					\State $\tg{min}\gets\min\{t: \tup{t,*}\in List\}$
				%	\Statex
                                              \Statex  ~~~~~~~~/* remove the coded value and retain the tag */
					%\State $List \gets List \backslash~\{\tup{\tg{},e}: \tg{}=\tg{min} ~\wedge~\tup{\tg{},e}\in List\} \cup \{  (  \tg{min}, \bot)  \}$\label{line:server:removemin}
					\State $List \gets List \backslash~\{\tup{\tg{},e}: \tg{}=\tg{min} ~\wedge \tup{\tg{},e}\in List\}$
					\State $List \gets List  \cup \{  (  \tg{min}, \bot)  \}$\label{line:server:removemin}
				\EndIf
				\State  Send {\sc ack} to $q$
			\EndReceive
			
%				\Statex
%				\Part {\underline{\GetTagResp,recv $\QueryTag$ from writer $w$}} {
%					%\If{ $status = active$ }
%					\State $t^* = \max_{(t,c) \in List}t$
%					\State Send $t^*$ to $w$
%					\Statex %\EndIf
%				}\EndPart
%				%										\Statex
%				\Part {\underline{\GetDataResp, recv $\QueryList$ from reader $r$}} {
%					%\If{ $status = active$ }
%					\State Send  $List$ to $r$
%					%\EndIf
%				}\EndPart
%				%	
%				\Statex
%				\Part{ \underline{\PutDataResp, recv $\CodedElementTag, (t, c_i)$ from $p$ }}{
%					%\If{$status = active$}
%					\State $List \leftarrow List \cup \{ (t, c_i)  \}$ 
%					\If{ $|List| > \delta + 1$ } 
%					\State  Retain the (tag, coded-element) pairs for the $\delta +1 $ highest tags in $List$, and delete the rest.
%					\EndIf 
%					\State  Send ack to $p$.
%					%\EndIf
%					
%				}\EndPart
				\end{multicols}
			}
	\end{algorithmic}	
	\caption{The response protocols at  any server $s_i \in {\mathcal S}$ in  
					\ares{} for client requests.}\label{fig:casopt:server}
					\vspace{-1em}
\end{algorithm*}		
 Each server $s_i$ stores one  state variable,  $List$,  which is a set of up to $(\delta + 1)$  (tag, coded-element) pairs. Initially the set at $s_i$ contains a single element, $List = \{ (t_0,  \Phi_i(v_0)\}$.   Below we describe the implementation of the DAPs.
%
%				At a high-level, the algorithm (see Fig.~\ref{fig:casopt}) is a natural generalization of the $ABD$ algorithm accounting for the fact that we use MDS codes.
	
$\dagettag{c}$: A  client,  during the execution of a  $\dagettag{c}$ primitive, queries all the servers in $\servers{c}$ for the highest tags in their  $Lists$, and awaits responses from $\left\lceil \frac{n+k}{2} \right\rceil$ servers.
% with $k \geq \frac{2n}{3}$. 
A server upon receiving the {\sc get-tag} request, 
responds to the client with the highest tag, as $\tg{max} \equiv \max_{(t,c) \in List}t$. 
Once the client receives the tags from $\left\lceil \frac{n+k}{2} \right\rceil$ servers,  it selects  the highest  tag $t$ and returns it . 
							
 $c.\act{put-data}(\tup{t_w, v})$: During the  execution of the primitive  $c.\act{put-data}(\tup{t_w, v})$,  a client 
 % computes the coded elements for each of the $n$ servers, and 
 sends the  pair  $(t_w, \Phi_i(v))$ to each server $s_i\in\servers{c}$.  
 When a server $s_i$ receives a message $(\text{\sc put-data}, t_w, c_i)$ , it adds the pair in its local $List$, 
 trims the pairs with the smallest tags exceeding the length $(\delta+1)$ of the $List$ , and replies 
 with an ack to the client.
 %
 %Every time a $(\text{\sc put-data}, t_w, c_i)$  message arrives at a server $s_i$, 
 %from a writer, 
 %the pair gets added to the $List$. As the size of the $List$ at each $s_i$ is bounded by $(\delta+1)$, then following an insertion in the $List$, $s_i$ trims the coded-elements associated with the smallest tags. 
 In particular, $s_i$ replaces the coded-elements of the older tags with $\bot$, and maintains only the coded-elements associated with the 
 	$(\delta+1)$ highest tags in the $List$ (see Line Alg.~\ref{fig:casopt:server}:\ref{line:server:removemin}).
 %which is then garbage collected to keep tag and coded-element pairs of the highest  $(\delta+1)$ tags, and by replacing the coded-elements of the older tags with $\bot$,  a symbol that signifies garbage-collected coded-elements. 
  The client completes the primitive operation after getting acks from $\left\lceil \frac{n+k}{2} \right\rceil$ servers.
			
	$\dagetdata{c}$:	A  client, during the execution of a  $\dagetdata{c}$ primitive, queries all the servers in $\servers{c}$ for their  local variable $List$, and awaits responses from $\left\lceil \frac{n+k}{2} \right\rceil$ servers. Once the client receives $Lists$ from $\left\lceil \frac{n+k}{2} \right\rceil$ servers,  it selects the highest  tag $t$, such that: $(i)$ its corresponding value $v$ is decodable from the coded elements in the lists; and $(ii)$ $t$ is the highest tag seen from the responses of at least $k$ $Lists$ 
			(see lines Alg.~\ref{fig:casopt}:\ref{line:getdata:max:begin}-\ref{line:getdata:max:end}) and returns the pair $(t, v)$. 
Note that in the case where anyone of the above conditions is not satisfied the corresponding read operation does not complete.

\subsection{Safety (Property~\ref{property:dap})  proof of the DAP{s}}
\label{sec:safety:daps}
%\vspace{-1.em}
\myparagraph{Correctness.} 
In this section we are concerned with only one configuration $c$, consisting of a set of servers 
%$\mathcal{S}$
$\servers{c}$.
%, and a set of reader and writer clients $\mathcal{R}$ and $\mathcal{W}$, respectively. In other words, 
%in such static system the sets $\mathcal{S}$, $\mathcal{R}$ and $\mathcal{W}$ are fixed, and 
We assume that at most $f \leq \frac{n-k}{2}$ servers from $\servers{c}$ may crash.  
Lemma~\ref{casflex:data-access:consistent} states that the DAP implementation 
 satisfies the  consistency properties Property~\ref{property:dap}  which will be used to 
%of \treas{}, \nn{and in turn by Theorem \ref{atomicity:A1}} these 
imply the atomicity of the \ares{} algorithm. 
%which implies the atomicity city properties and consequently the
%atomicity property 
%(Theorem~\ref{thm:atomicity_radonc}).			
%\myparagraph{Liveness and Safety Conditions.}\blue{
%The \treas{} algorithm we present satisfy \myemph{wait-free termination} (Liveness) and \myemph{atomicity} (Safety).
%}
	%Due to lack of space the proof of the following Theorem is produced in the Appendix.	
\label{sec:primitives}

 \begin{theorem}[Safety]\label{casflex:data-access:consistent}
Let $\Pi$ a set of complete DAP operations of Algorithm \ref{fig:casopt} in a configuration $c\in\confSet$,
$\act{c.get-tag}$, $\act{c.get-data}$ and $\act{c.put-data}$,
of an execution $\EX$. Then, every pair of operations $\phi,\op\in\Pi$ satisfy Property \ref{property:dap}.
% The data-access primitives, i.e., $\act{get-tag}$, $\act{get-data}$ and $\act{put-data}$ primitives implemented in any configuration  $c$
% in this section satisfy Property~\ref{property:dap}.
\end{theorem}

\begin{proof}
As mentioned above we are concerned with only configuration $c$, and therefore, in our proofs it suffices to examine only one
configuration. Let $\EX$ be some execution of \ares{}, then we consider two cases for $\pi$ for proving property $C1$:  $\pi$ is a  $\act{get-tag}$, or $\pi$ is a $\act{get-data}$ primitive. 

 %\item[ C1 ]  If $\phi$ is a  $\daputdata{c}{\tup{\tg{\phi}, v_\phi}}$, for $c \in \confSet$, $\tg{1} \in\tsSet$ and $v_1 \in \valSet$,
 %and $\pi$ is a $\dagettag{c}$ (or a $\dagetdata{c}$) 

 %that returns $\tg{\pi} \in \tsSet$ (or $\tup{\tg{\pi}, v_{\pi}} \in \tsSet \times \valSet$) and $\phi$ completes before $\pi$ in $\EX$, then $\tg{\pi} \geq \tg{\phi}$.
Case $(a)$: $\phi$ is   $\daputdata{c}{\tup{\tg{\phi}, v_\phi}}$ and  $\pi$ is a $\dagettag{c}$ returns $\tg{\pi} \in \tsSet$. Let $c_{\phi}$ and $c_{\pi}$ denote the clients that invokes $\phi$ and $\pi$ in $\EX$. Let $S_{\phi} \subset \mathcal{S}$ denote the set of $\left\lceil \frac{n+k}{2} \right \rceil$ servers that responds to $c_{\phi}$, during $\phi$. Denote by $S_{\pi}$ the set of $\left\lceil \frac{n+k}{2} \right \rceil$ servers that responds to $c_{\pi}$, during $\pi$.  Let $T_1$ be a point in execution $\EX$ 
after the completion of $\phi$ and before the invocation of $\pi$. Because $\pi$ is invoked after $T_1$, therefore, at $T_1$ each of the servers in $S_{\phi}$ contains $t_{\phi}$ in its $List$ variable. Note that, once a tag is added to $List$, it is never removed. Therefore, during $\pi$, any server in $S_{\phi}\cap S_{\pi}$ responds with $List$ containing $t_{\phi}$ to $c_{\pi}$. Note that since  $|S_{\phi}| = |S_{\pi}| =\left\lceil \frac{n+k}{2} \right \rceil $ implies
				 $| S_{\phi} \cap S_{\pi} | \geq k$, and hence $t^{dec}_{max}$ at $c_{\pi}$, during $\pi$ is at least as large as $t_{\phi}$, i.e., $t_{\pi} \geq t_{\phi}$. Therefore, it suffices to prove our claim with respect to the tags and the decodability of  its corresponding value.

Case $(b)$: $\phi$ is   $\daputdata{c}{\tup{\tg{\phi}, v_\phi}}$ and  $\pi$ is a $\dagetdata{c}$ returns $\tup{\tg{\pi}, v_{\pi}} \in \tsSet \times \valSet$. 
As above, let $c_{\phi}$ and $c_{\pi}$ be the clients that invokes $\phi$ and 
$\pi$. Let $S_{\phi}$ and $S_{\pi}$ be the set of servers that responds to $c_{\phi}$ and $c_{\pi}$, respectively. Arguing as above, 
 $| S_{\phi} \cap S_{\pi} | \geq k$ and every server in  $S_{\phi} \cap S_{\pi} $ sends $t_{\phi}$ in response to $c_{\phi}$, during 
 $\pi$, in their $List$'s and hence $t_{\phi} \in Tags_{*}^{\geq k}$. Now, because $\pi$ completes in $\EX$, hence we have 
 $t^*_{max} = t^{dec}_{max}$. Note that $\max Tags_{*}^{\geq k} \geq \max Tags_{dec}^{\geq k}$ so 
  $t_{\pi} \geq \max Tags_{dec}^{\geq k} = \max Tags_{*}^{\geq k} \geq t_{\phi}$. Note that each tag is always associated with 
  its corresponding value $v_{\pi}$, or the corresponding coded elements $\Phi_s(v_{\pi})$ for $s \in \mathcal{S}$.

Next, we prove the $C2$ property of DAP for the \ares{} algorithm. Note that the initial values of the $List$ variable in each servers $s$ in $\mathcal{S}$ is 
$\{ (t_0, \Phi_s(v_{\pi}) )\}$. Moreover, from an inspection of the steps of the algorithm, new tags in the $List$ variable of any servers of any servers is introduced via $\act{put-data}$ operation. Since $t_{\pi}$ is returned by a $\act{get-tag}$ or 
$\act{get-data}$ operation then it must be that either $t_{\pi}=t_0$ or $t_{\pi} > t_0$. In the case where $t_{\pi} = t_0$ then we have nothing to prove. If $t_{\pi} > t_0$ then there must be a $\act{put-data}(t_{\pi}, v_{\pi})$ operation $\phi$. To show that for every $\pi$ it cannot be that $\phi$ completes before $\pi$, we adopt by a contradiction. Suppose for every $\pi$, $\phi$ completes before $\pi$ begins, then clearly $t_{\pi}$ cannot be returned $\phi$, a contradiction.
\end{proof}
			
	\remove{
				\begin{theorem}[Atomicity]  \label{thm:atomicity_radonc}
					Any well-formed and fair execution of \treas{},  is atomic.
				\end{theorem}
		}
	\myparagraph{Liveness.} \label{sec:treas_liveness}
    To reason about the liveness of the proposed DAPs, we define a concurrency parameter $\delta$ which  captures all the  $\act{put-data}$ operations that overlap with the $\act{get-data}$, until the time the client has all data needed to attempt decoding a value. However, we ignore those $\act{put-data}$ operations that might have started in the past, and never completed yet, if their tags are less than that of any $\act{put-data}$ that completed before the  $\act{get-data}$  started. This allows us to ignore $\act{put-data}$ operations due to failed clients, while counting concurrency, as long as the failed $\act{put-data}$ operations are followed by a successful $\act{put-data}$ that completed before the $\act{get-data}$ started. 				
\kmk{In order to define the amount of concurrency  in  our specific implementation of the DAPs presented in this section the}  following definition captures the $\act{put-data}$ operations that overlap with the $\act{get-data}$, until  the client has all data required to  decode the value.
				
\begin{definition}[Valid $\act{get-data}$ operations]
A $\act{get-data}$  operation $\pi$ from a process $p$ is \myemph{valid}  if 
%the associated client 
$p$ does not crash until the reception of $\left\lceil \frac{n+k}{2} \right\rceil$ responses during the{\GetData} phase. 
\end{definition}

				\begin{definition}[$\act{put-data}$ concurrent with a valid $\act{get-data}$] \label{defn:concurrent}
					Consider a valid $\act{get-data}$ operation $\pi$ from a process $p$. 
					Let $T_1$ denote the point of initiation of $\pi$. For $\pi$, let $T_2$ denote the earliest point of time during the execution when $p$ 
					%the associated client 
					receives all the $\left\lceil \frac{n+k}{2} \right\rceil$ responses.
					% For a valid repair,  let $T_2$ denote the point of time during the execution when the repair completes, and takes the associated server back to the active state. 
					Consider the set $\Sigma = \{ \phi: \phi$ is any $\act{put-data}$ operation that completes before $\pi \text{ is initiated} \}$, and let $\phi^* = \arg\max_{\phi \in \Sigma}tag(\phi)$. Next, consider the set $\Lambda = \{\lambda:  \lambda$  is any $\act{put-data}$ operation that starts before $T_2 \text{ such that } tag(\lambda) > tag(\phi^*)\}$. We define the number of $\act{put-data}$ concurrent with the valid $\act{get-data}$  $\pi$ to be the cardinality of the set $\Lambda$.
				\end{definition}
							
Termination (and hence liveness)  of the DAPs is guaranteed in an execution $\EX$, provided that a process 
	no more than $f$ servers in $\servers{c}$ crash, and no more that $\delta$ $\act{put-data}$ may be concurrent at any point in $\EX$. 
	%in  property of an algorithm,  we mean that 
	If the failure model is satisfied, then any operation invoked by a non-faulty client will collect the necessary replies
	% process terminates  
	independently of the progress of any other client process in the system. Preserving $\delta$ on the other hand,
	ensures that any operation will be able to decode a written value. These are captured in the following theorem:

\begin{theorem}[Liveness]  \label{thm:liveness_radonc}
Let $\EX$ be well-formed and fair execution of DAPs, with an $[n, k]$ MDS code, where $n$ is the number of servers out of which no more than $\frac{n-k}{2}$ may crash, 
		%and $k  > n/3$,
and $\delta$ be the maximum number of $\act{put-data}$ operations concurrent with any valid $\act{get-data}$ operation. 
Then any $\act{get-data}$ and $\act{put-data}$ operation $\op$ invoked by a process $\pr$  terminates in $\EX$ if $\pr$
does not crash between the invocation and response steps of $\op$.\vspace{-.5em}
\end{theorem}
	
\begin{proof}
Note that in the read and write operation the  $\act{get-tag}$ and $\act{put-data}$ operations initiated by any non-faulty client  always complete. Therefore, the liveness property with respect to any write operation is clear because it uses only  $\act{get-tag}$ and $\act{put-data}$ operations of the DAP. So, we focus on proving the liveness property of any read operation $\pi$, 
specifically,   the  $\act{get-data}$ operation completes. Let $\EX$ be and execution of \ares{} and let 
$c_{\omega}$ and $c_{\pi}$ be the clients that invokes the write operation $\omega$ and read operation $\pi$, respectively.
				
Let $S_{\omega}$ be the set of 
$\left\lceil \frac{n+k}{2} \right \rceil$ servers that responds to $c_{\omega}$, in the $\act{put-data}$ operations, in $\omega$.
Let $S_{\pi}$ be the set of $\left\lceil \frac{n+k}{2} \right \rceil$ servers that responds to  $c_{\pi}$ during the  $\act{get-data}$ step of $\pi$. Note that in $\EX$ at the point execution $T_1$, just before the execution of  $\pi$, none of the write operations in 
$\Lambda$ is complete. Observe that,  by algorithm design, the coded-elements corresponding to  $t_{\omega}$ are garbage-collected from the $List$ variable of a server only if more than $\delta$ higher tags are introduced by subsequent writes into the server.  Since the number of concurrent writes  $|\Lambda|$, s.t.  $\delta > | \Lambda |$ the corresponding value of tag $t_{\omega}$ is not garbage collected in $\EX$, at least until execution point $T_2$  in  any of the servers in $S_{\omega}$.

Therefore, during the execution fragment between the execution points $T_1$ and $T_2$ of the execution $\EX$, the tag and coded-element pair is present in the $List$ variable of every in $S_{\omega}$ that is active. As a result, the tag and coded-element pairs, $(t_{\omega}, \Phi_s(v_{\omega}))$ exists in the $List$ received from any $s \in S_{\omega} \cap S_{\pi}$ during operation $\pi$. Note that since $|S_{\omega}| = |S_{\pi}| =\left\lceil \frac{n+k}{2} \right \rceil $ hence $|S_{\omega} \cap S_{\pi} | \geq k$ and hence 
$t_{\omega} \in Tags_{dec}^{\geq k} $, the set of decode-able tag, i.e., the value $v_{\omega}$ can be decoded by $c_{\pi}$ in $\pi$, which demonstrates that $Tags_{dec}^{\geq k}  \neq \emptyset$. Next we want to argue that 
$t_{max}^* = t_{max}^{dec}$ via a contradiction: we assume  $ \max Tags_{*}^{\geq k}  >  \max Tags_{dec}^{\geq k}  $. Now, consider any tag $t$, which  exists due to our assumption,  such that,  $t \in Tags_{*}^{\geq k} $,  $t \not\in Tags_{dec}^{\geq k} $ and $t > t_{max}^{dec}$.
Let $S^k_{\pi} \subset S$ be any subset of $k$ servers that responds with $t^*_{max}$ in their $List$ variables to $c_{\pi}$. Note that since $k >  n/3$ hence $|S_{\omega} \cap S^k_{\pi}|  \geq \left\lceil \frac{n+k}{2} \right \rceil +  \left\lceil \frac{n+1}{3} \right \rceil \geq 1$, i.e., $S_{\omega} \cap S^k_{\pi} \neq \emptyset$. Then $t$  must be in some servers in $S_{\omega}$ at $T_2$ and since $t > t_{max}^{dec} \geq t_{\omega}$. 
Now since $|\Lambda| < \delta$ hence $(t, \bot)$ cannot be in any server at $T_2$  because there are not enough concurrent write operations (i.e., writes in $\Lambda$) to garbage-collect the coded-elements corresponding to tag $t$, which also holds  for tag  $t^{*}_{max}$. In that case, $t$ must be in $Tag_{dec}^{\geq k}$, a contradiction.
\end{proof}

\section{Correctness of \ares{}}
\label{sec:correct}
In this section, we prove that \ares{} correctly implements an atomic, read/write, shared storage service. 
%We also provide an analysis of its storage and communication costs, and the latency of read and write operations. 
The correctness of \ares{} highly depends on the way the configuration 
sequence is constructed at each client process.
Also, atomicity is ensured if the DAP implementation in each configuration $c_i$
%relies on the satisfaction of 
%hinges on 
satisfies Property~\ref{property:dap}.
Thus, we begin by showing some critical properties preserved by the reconfiguration service proposed in \ares{} in subsection \ref{sec:safety:recon}, and then we proof
the correctness of \ares{} in subsection \ref{sec:safety:atomic} 
when those properties hold and the DAPs used in 
each configuration satisfy Property \ref{property:dap}.

% of the  used in an execution of~\ares{}. 
%Due to lack of space proofs are omitted and can be found in the 
%% of the following Theorem is produced in the 
%extended version of the paper~\cite{ARES:Arxiv:2018}.	

We proceed by first introducing some definitions and notation, that we use in the proofs that follow. 
%then by presenting some properties that are satisfied 
%	by the reconfiguration service in any execution, and then we show that given these properties our algorithm satisfies 
%	the safety (atomicity) conditions. 

	\myparagraph{Notations and definitions.}
	For a server $s$, we use the notation $\atT{s.var}{\state}$ to refer to the value of the state variable $var$, in $s$, at a state $\state$ of an  execution $\EX$. 
	If server  $s$ crashes at a state $\state_f$ in an execution $\EX$ then $\atT{s.var}{\state}\triangleq\atT{s.var}{\state_f}$ for any state variable $var$ and for 
	any state $\state$ that appears after $\state_f$ in $\EX$. 
	%refers to the value of $v$ at $s$ at the state just before crashes. In other words, $\atT{s.v}{T}  \triangleq \atT{s.v}{\hat{T}}$, where $\hat{T}$ is the latest point in the execution, such that, $(a)$ $\hat{T} \leq T$ and $(b)$ $s$ is non-faulty.

    We define as the tag of a configuration $c$ the smallest tag among the maximum tags found in each quorum of $c$. This is essentially the smallest tag that an operation may witness when receiving replies from a single quorum in $c$. More formally:

	\begin{definition}[Tag of a configuration]  Let  $c \in \mathcal{C}$ be a configuration, $\state$ be a state in some execution $\EX$ then 
		we define the tag of $c$ at state $\state$ as  
		$ \atT{tag(c)}{\state} \triangleq \min_{Q \in \quorums{c}} \max_{s \in Q}~\atT{(s.tag}{\state}).$
		We  drop the suffix $|_\state$, and simply denote as $tag(c)$,  when the state  is clear from the context.
	\end{definition}
	Next we provide the notation to express the configuration sequence witnessed by a process $p$ in a state $\state$ (as $ \atT{\pr.cseq}{\state}$), the last finalized configuration in that sequence
	(as $\mu(\cvec{\pr}{\state})$), and the length of that sequence (as $\nu(\cvec{\pr}{\state})$). More formally:
	
	\begin{definition}
		Let $\sigma$ be any point in an execution of \ares{} and suppose we use the notation $\cvec{\pr}{\state}$ for $ \atT{\pr.cseq}{\state}$,  i.e., the $cseq$ variable at process $p$ at the state $\state$. %be the value of a configuration sequence vector at a process $\pr$ at some state  $\st$ in an execution $\EX$. 
		Then we define as $ \mu(\cvec{\pr}{\state})  \triangleq  \max\{ i : \cvec{\pr}{\state}[i].status = F\}$ 
		and $ \nu(\cvec{\pr}{\state}) \triangleq |\cvec{\pr}{\state}|$, where $|\cvec{\pr}{\state}|$ is the length of the  configuration vector 
		$\cvec{\pr}{\state}$. % that are not equal to $\bot$.  
	\end{definition}
	
	Last, we define the prefix operation on two configuration sequences. 
	
	\begin{definition} [Prefix order]
		Let $\mathbf{x}$ and $\mathbf{y}$ be any two configuration sequences. We say that $\mathbf{x}$ is a prefix of $\mathbf{y}$, denoted by 
		$\mathbf{x} \preceq_p  \mathbf{y}$, if $\config{\mathbf{x}[j]}=\config{\mathbf{y}[j]}$, for all $j$ such that $\mathbf{x}[j]\neq\bot$.
	\end{definition}

% \subsection{Safety (Property~\ref{property:dap})  proof of the DAP{s}}
% \label{sec:safety:daps}
% \input{ssec_dap_correct_v1.tex}

\subsection{Reconfiguration Protocol Properties}
\label{sec:safety:recon}
In this section we analyze the properties that we can achieve through our reconfiguration algorithm. In high-level, we do show that the following properties are preserved: 
\begin{itemize} 
\item[i] {\bf configuration uniqueness:} the configuration sequences in any two processes have identical configuration at any place $i$,
\item[ii] {\bf sequence prefix:} the configuration sequence witnessed by an operation is a prefix of the sequence witnessed by any succeeding operation, and 
\item[iii] {\bf sequence progress:} if the configuration with index $i$ is finalized during an operation, then a configuration $j$, for $j\geq i$, will be finalized
for a succeeding operation.
\end{itemize}

The first lemma shows that any two configuration sequences have the same configuration identifiers
in the same indexes. 

\begin{lemma}
\label{lem:consconf}
	For any reconfigurer $r$ that invokes an $\act{reconfig}(c)$ action in an execution $\EX$ 
	of the algorithm, If $r$ chooses to install $c$ in index $k$ of its local $r.cseq$ vector, then $r$ invokes 
	the $Cons[k-1].propose(c)$ instance over configuration $r.cseq[k-1].cfg$.
\end{lemma}

\begin{proof}
	It follows directly from the algorithm. 
\end{proof}

\begin{lemma}
	\label{lem:server:monotonic}
	If a server $s$ sets $s.nextC$ to $\tup{c,F}$ at some state $\st$ in an execution $\EX$ 
	of the algorithm, then $s.nextC = \tup{c,F}$ for any state $\st'$ that appears after $\st$ in $\EX$.
\end{lemma}

\begin{proof}
	Notice that a server $s$ updates its $s.nextC$ variable for some specific configuration $c_k$ in a state $\state$ only when it receives a {\sc write-conf} message. This is either the first {\sc write-conf} message received at $s$ for $c_k$ (and thus $nextC=\bot$), or 
	$s.nextC = \tup{*,P}$ and the message received contains a tuple $\tup{c,F}$. 
	%received a tuple $\tup{c,P}$ and before $\st$ received the tuple . 
% 	A {\sc write-conf} is received at $s$ for a configuration $c_k$, whenever a reconfigurer invokes a $\act{put-config}(c_k,*)$ action.
% 	By Observation \ref{obs:consensus}
% 	$c=c'$ as $s$ updates the $s.nextC$ of the same configuration $c_k$. 
	Once the tuple becomes equal to 
	$\tup{c,F}$ then $s$ does not satisfy the update condition for $c_k$, and hence in any state $\st'$ after $\st$
	it does not change $\tup{c,F}$.
\end{proof}

\begin{lemma}[Configuration Uniqueness]
\label{lem:unique}

	%Let $\st_1$ and $\st_2$ be any two states of an execution $\EX$ of the algorithm,
	%and $\pr, q$ two participating processes. 
	%be the state after the response action of an operation $\op_1$ from process $p$,
	%and $\st_2$ be the state after the first $\act{read-config}$ call of an operation $\op_2$ from $q$.
	For any processes $\pr, q\in \idSet$ and any states $\st_1, \st_2$ in an execution $\EX$, it must hold that 
	$\config{\cvec{\pr}{\st_1}[i]}=\config{\cvec{q}{\st_2}[i]}$,  $\forall i$ s.t. 
	$\config{\cvec{\pr}{\st_1}[i]},\config{\cvec{q}{\st_2}[i]}\neq \bot$.
\end{lemma}

\begin{proof}
	The lemma holds trivially for $\config{\cvec{\pr}{\st_1}[0]}=\config{\cvec{q}{\st_2}[0]}=c_0$. 
	So in the rest of the proof we focus in the case where $i > 0$. Let us assume 
	w.l.o.g. that $\st_1$ appears before $\st_2$ in $\EX$.
	
	According to our algorithm a process $\pr$ sets $\pr.cseq[i].cfg$ to a configuration 
	identifier $c$ in two cases: (i) either it received $c$ as the result of the consensus 
	instance in configuration $\pr.cseq[i-1].cfg$, or (ii) $\pr$ receives $\config{s.nextC} = c$ from 
	a server $s\in\servers{\config{\pr.cseq[i-1]}}$. Note here that (i) is possible only 
	when $\pr$ is a reconfigurer and attempts to install a new configuration. On the 
	other hand (ii) may be executed by any process in any operation that invokes the 
	$\act{read-config}$ action. We are going 
	to proof this lemma by induction on the configuration index.

	\emph{Base case:} The base case of the lemma is when $i=1$. 
	Let us first assume that $p$ and $q$ receive $c_p$ and $c_q$, as the result of the consensus instance at $\pr.cseq[0].cfg$
	and $q.cseq[0].cfg$ respectively. By Lemma \ref{lem:consconf}, since both processes want to install a configuration 
	in $i=1$, then they have to run $Cons[0]$ instance over the configuration stored in their local $cseq[0].cfg$ variable. 
	Since $\pr.cseq[0].cfg=q.cseq[0].cfg=c_0$ then 
	both $Cons[0]$ instances run over the same configuration $c_0$ and 
	%according to Observation \ref{obs:consensus}  
	thus by the aggreement property the have to 
	decide the same value, say $c_1$. Hence $c_p=c_q=c_1$ and $\pr.cseq[1].cfg=q.cseq[1].cfg=c_1$.
	 
	 Let us examine the case now where $p$ or $q$ 
	assign a configuration $c$ they received from some server $s\in\servers{c_0}$. According to the
	algorithm only the configuration that has been decided by the consensus instance on 
	$c_0$ is propagated to the servers in $\servers{c_0}$. If $c_1$ is the decided configuration, then 
	$\forall s\in\servers{c_0}$ such that $s.nextC(c_0)\neq\bot$, it holds that $s.nextC(C_0) = \tup{c_1,*}$.
	So if $p$ or $q$ set $\pr.cseq[1].cfg$ or $q.cseq[1].cfg$ to some received configuration, then 
	$\pr.cseq[1].cfg = q.cseq[1].cfg = c_1$ in this case as well. 
	
     \emph{Hypothesis:} We assume  that 
	$\cvec{\pr}{\st_1}[k]=\cvec{q}{\st_2}[k]$  for some $k$, $k \geq 1$.
	
	%\noindent{\bf Induction Hypothesis:} 
	\emph{Induction Step:}  We need to show that the lemma holds for $i=k+1$.
	If both processes retrieve $\config{\pr.cseq[k+1]}$ and $\config{q.cseq[k+1]}$ through consensus, 
	then both $\pr$ and $q$ run consensus
	over the previous configuration. Since according to our hypothesis 
	$\cvec{\pr}{\st_1}[k]=\cvec{q}{\st_2}[k]$ then both process will receive the same
	decided value, say $c_{k+1}$, and hence $\pr.cseq[k+1].cfg=q.cseq[k+1].cfg=c_{k+1}$. Similar to the base case,
	a server in $\servers{c_k}$ only receives the configuration $c_{k+1}$ decided by the consensus instance run over $c_k$. 
	So processes 
	$\pr$ and $q$ can only receive $c_{k+1}$ from some server in $\servers{c_k}$ 
	%even if the processes update their $\pr.cseq[k+1].cfg$ or $q.cseq[k+1].cfg$ with a received 
	%configuration that will be equal to 
	so they can only assign $\pr.cseq[k+1].cfg=q.cseq[k+1].cfg=c_{k+1}$ at Line A\ref{algo:reconfigurer}:\ref{line:readconfig:assign}.
	That completes the proof. 
\end{proof}

Lemma \ref{lem:unique} showed that any two operations store the same
configuration in any cell $k$ of their $cseq$ variable. It is not known however 
if the two processes discover the same number of configuration ids. In the following
lemmas we will show that if a process learns about a configuration in a cell $k$ 
then it also learns about all configuration ids for every index $i$, such that $0\leq i\leq k-1$.

\begin{lemma}
\label{lem:confmonotonic}
	In any execution $\EX$ of the algorithm , If for any process $\pr\in\idSet$, $\cvec{\pr}{\st}[i]\neq\bot$ in some state $\st$ in $\EX$,
	then $\cvec{\pr}{\st'}[i]\neq\bot$ in any state $\st'$ that appears after $\st$ in $\EX$. 
\end{lemma}

\begin{proof}
	A value is assigned to $\cvec{\pr}{*}[i]$ either after the invocation of a consensus instance, or while executing
	the $\act{read-config}$ action. Since any configuration proposed for installation cannot be $\bot$ (A\ref{algo:reconfigurer}:\ref{line:install:valid}), 
	and since there is at least one configuration proposed in the consensus instance (the one from $\pr$), then by the validity of the consensus
	service the decision will be a configuration $c\neq\bot$. Thus, in this case $\cvec{\pr}{*}[i]$ cannot be $\bot$.
	Also in the $\act{read-config}$ procedure, $\cvec{\pr}{*}[i]$ is assigned to a value different than $\bot$ according
	to Line A\ref{algo:reconfigurer}:\ref{line:readconfig:assign}. Hence, if $\cvec{\pr}{\st}[i]\neq\bot$ at state $\st$ 
	then it cannot become $\bot$ in any state $\st'$ after $\st$ in execution $\EX$.
\end{proof}

\begin{lemma}
\label{lem:nogaps}
	Let $\st_1$ be some state in an execution $\EX$ of the algorithm. Then for any 
	process $\pr$, if $k = max\{i: \cvec{\pr}{\st_1}[i]\neq \bot\}$, then 
	$\cvec{\pr}{\st_1}[j]\neq \bot$, for $0\leq j < k$.
\end{lemma}
\begin{proof}
	Let us assume to derive contradiction that there exists $j < k$ such that 
	$\cvec{\pr}{\st_1}[j]=\bot$ and $\cvec{\pr}{\st_1}[j+1]\neq\bot$.
	Consider first that $j = k-1$ and that $\st_1$ is the state immediately 
	after the assignment of a value to $\cvec{\pr}{\st_1}[k]$, say $c_k$. 
	Since $\cvec{\pr}{\st_1}[k]\neq\bot$, then $\pr$ assigned $c_k$ to $\cvec{\pr}{\st_1}[k]$ 
	in one of the following cases: 
	(i) $c_k$ was the result of the consensus instance, or
	(ii) $\pr$ received $c_k$ from a server during a $\act{read-config}$ action.
	The first case is trivially impossible as according to Lemma \ref{lem:consconf} 
	$\pr$ decides for $k$ when it runs consensus over configuration $\config{\cvec{\pr}{\st_1}[k-1]}$. 
	Since this is equal to $\bot$, then we cannot run consensus over a non-existent set of 
	processes. 	In the second case, $\pr$ assigns $\cvec{\pr}{\st_1}[k] = c_k$  in Line A\ref{algo:parser}:\ref{line:readconfig:assign}.
	The value $c_k$ was however obtained when $\pr$ invoked $\act{get-next-config}$ on 
	configuration $\config{\cvec{\pr}{\st_1}[k-1]}$. In that action, $\pr$ sends {\sc read-config}
	messages to the servers in $\servers{\config{\cvec{\pr}{\st_1}[k-1]}}$ and waits until a quorum 
	of servers replies. Since we assigned $\cvec{\pr}{\st_1}[k] = c_k$ it means that $\act{get-next-config}$
	terminated at some state $\st'$ before $\st_1$ in $\EX$, and thus: 
	(a) a quorum of servers in $\servers{\config{\cvec{\pr}{\st'}[k-1]}}$
	replied, and (b) there exists a server $s$ among those that replied with $c_k$. 
	According to our assumption however, $\cvec{\pr}{\st_1}[k-1] = \bot$ at $\st_1$. 
	So if state $\st'$ is before $\st_1$ in $\EX$, %is the state after the response step of $\act{get-next-config}$, 
	then by Lemma \ref{lem:confmonotonic}, it follows that $\cvec{\pr}{\st'}[k-1] = \bot$. This however 
	implies that $\pr$ communicated with an empty configuration, and thus no server replied to $\pr$. 
	This however contradicts the assumption that a server replied with $c_k$ to $\pr$. 
	
	Since any process traverses the configuration sequence starting from the initial 
	configuration $c_0$, then with a simple induction and similar reasoning we can show that 
	$\cvec{\pr}{\st_1}[j]\neq \bot$, for $0\leq j\leq k-1$.
\end{proof}

We can now move to an important lemma that shows that any \act{read-config} action 
returns an extension of the configuration sequence returned by any previous \act{read-config} action. 
First, we show that the last finalized configuration observed by any \act{read-config} action is at least as 
recent as the finalized configuration observed by any subsequent \act{read-config} action. 
%Using this lemma we will then show that when 

\begin{lemma}
	\label{lem:config:propagation}
	If at a state $\st$ of an execution $\EX$ of the algorithm $\mu(\cvec{\pr}{\st}) = k$ % = \max\{i: \cvec{\pr}{\st}[i]\neq\bot\}$
	for some process $\pr$, then for any element $0\leq j < k$, $\exists Q\in \quorums{\config{\cvec{\pr}{\st}[j]}}$
	such that $\forall s\in Q, s.nextC(\config{\cvec{\pr}{\st}[j]})= \cvec{\pr}{\st}[j+1]$. 
\end{lemma}
\begin{proof}
	This lemma follows directly from the algorithm. Notice that whenever a process assigns a value to 
	an element of its local configuration (Lines  A\ref{algo:parser}:\ref{line:readconfig:assign} and 
	A\ref{algo:reconfigurer}:\ref{line:addconfig:assign}), it then propagates this value to a quorum of the 
	previous configuration (Lines  A\ref{algo:parser}:\ref{line:readconfig:put} and 
	A\ref{algo:reconfigurer}:\ref{line:addconfig:put}). So if a process $\pr$ assigned $c_j$ to an 
	element $\cvec{\pr}{\st'}[j]$ in some state $\st'$ in $\EX$, then $\pr$ may assign a value 
	to the $j+1$ element of $\cvec{\pr}{\st''}[j+1]$ only after $\act{put-config}(\config{\cvec{\pr}{\st'}[j-1]},\cvec{\pr}{\st'}[j])$
	occurs. During $\act{put-config}$ action, $\pr$ propagates $\cvec{\pr}{\st'}[j]$ in a quorum 
	$Q\in\quorums{\config{\cvec{\pr}{\st'}[j-1]}}$. Hence, if $\cvec{\pr}{\st}[k]\neq\bot$, then $\pr$ 
	propagated each $\cvec{\pr}{\st'}[j]$, for $0<j\leq k$ to a quorum of servers $Q\in\quorums{\config{\cvec{\pr}{\st'}[j-1]}}$.
	And this completes the proof. 
%	
%	Let examine a single element $\cvec{\pr}{\st'}[j]$ that is added in the configuration vector at some state $\st'$ 
%	that appears before $\st$ in $\EX$. Whenever a configuration is assigned a process $\pr$ invokes $\act{put-config}$
%	action before adding the next configuration. According to this action, the process sends the discovered configuration
%	and sends it to a quorum of servers. As this process is repeated for every element inserted then the lemma follows. 
\end{proof}

\begin{lemma}[Sequence Prefix]
	\label{lem:prefix}
Let $\op_1$ and $\op_2$ two 
%read/write/install operations 
completed \act{read-config} actions invoked by processes $\pr_1, \pr_2\in\idSet$ 
respectively, such that $\op_1\bef\op_2$ in an execution $\EX$. Let $\st_1$ be the state after the response 
step of $\op_1$ and $\st_2$ the state after the response step 
%termination of the first $\act{read-config}$ 
of $\op_2$. Then 
$\cvec{\pr_1}{\st_1}\preceq_p\cvec{\pr_2}{\st_2}$.
\end{lemma}

\begin{proof}
	Let $\nu_1 = \nu(\cvec{\pr_1}{\st_1})$ and $\nu_2 = \nu(\cvec{\pr_2}{\st_2})$.
	By Lemma \ref{lem:unique} for any $i$ such that $\cvec{\pr_1}{\st_1}[i]\neq\bot$ and 
	$\cvec{\pr_2}{\st_2}[i]\neq\bot$, then $\config{\cvec{\pr_1}{\st_1}[i]}=\config{\cvec{\pr_2}{\st_2}[i]}$.
	Also from Lemma \ref{lem:nogaps} we know that for $0\leq j\leq \nu_1, \cvec{\pr_1}{\st_1}[j] \neq \bot$, 
	and $0\leq j\leq \nu_2, \cvec{\pr_2}{\st_2}[j] \neq \bot$. So if we can show that $\nu_1\leq \nu_2$ then the lemma follows. 
	
	Let $\mu = \mu(\cvec{\pr_2}{\st'})$ %, or simply $\mu$, 
	be the last finalized element which $\pr_2$ established in the beginning of 
	the $\act{read-config}$ action $\op_2$ (Line A\ref{algo:reconfigurer}:\ref{line:readconfig:final}) at some state $\st'$ before $\st_2$. 
	It is easy to see that $\mu\leq \nu_2$. If $\nu_1 \leq \mu$ then $\nu_1\leq \nu_2$ and 
	the lemma follows. Thus, it remains
	to examine the case where $\mu < \nu_1$. Notice that since $\op_1\bef\op_2$ then $\st_1$ appears before 
	$\st'$ in execution $\EX$. By Lemma \ref{lem:config:propagation}, we know that by $\st_1$, 
	%in all the configurations $\cvec{\pr_1}{\st_1}[j]$, for $0\leq j < \nu_1$,  
	$\exists Q\in\quorums{\config{\cvec{\pr_1}{\st_1}[j]}}$, for $0\leq j < \nu_1$,   such that 
	$\forall s\in Q, s.nextC = \cvec{\pr_1}{\st_1}[j+1]$. Since $\mu < \nu_1$, then it must be the case 
	that $\exists Q\in \quorums{\config{\cvec{\pr_1}{\st_1}[\mu]}}$ such that $\forall s\in Q, s.nextC = \cvec{\pr_1}{\st_1}[\mu+1]$.
	But by Lemma \ref{lem:unique}, we know that $\config{\cvec{\pr_1}{\st_1}[\mu]}= \config{\cvec{\pr_2}{\st'}[\mu]}$. 
	Let $Q'$ be the quorum that replies to the $\act{read-next-config}$ occurred in $\pr_2$, on configuration  $\config{\cvec{\pr_2}{\st'}[\mu]}$.
	By definition $Q\cap Q'\neq \emptyset$, thus there is a server $s\in Q\cap Q'$ that sends $s.nextC = \cvec{\pr_1}{\st_1}[\mu+1]$
	to $\pr_2$ during $\op_2$. Since $\cvec{\pr_1}{\st_1}[\mu+1]\neq\bot$ then $\pr_2$ assigns $\cvec{\pr_2}{*}[\mu+1]=\cvec{\pr_1}{\st_1}[\mu+1]$, and 
	repeats the process in the configuration $\config{\cvec{\pr_2}{*}[\mu+1]}$. Since every configuration $\config{\cvec{\pr_1}{\st_1}[j]}$, 
		for $\mu\leq j<\nu_1$, has a quorum of servers with $s.nextC$, then by a simple induction it can be shown that the process will 
		be repeated for at least $\nu_1-\mu$ iterations, and  every configuration
		$\cvec{\pr_2}{\st''}[j]=\cvec{\pr_1}{\st_1}[j]$, at some state $\st''$ before $\st_2$. 
		Thus, $\cvec{\pr_2}{\st_2}[j]=\cvec{\pr_1}{\st_1}[j]$, for $0\leq j\leq \nu_1$. Hence $\nu_1\leq\nu_2$ and the lemma follows in this case as well. 
\end{proof}

Thus far we focused on the configuration member of each element in $cseq$. As operations do get in account
the \emph{status} of a configuration, i.e. $P$ or $F$, in the next lemma we will examine the relationship of 
the last finalized configuration as detected by two operations. First we present a lemma that shows the 
monotonicity of the finalized configurations.

\begin{lemma}
	\label{lem:final:monotonic}
	Let $\st$ and $\st'$ two states in an execution $\EX$ such that $\st$ appears before $\st'$ in $\EX$.
 	Then for any process $\pr$ must hold that $\mu(\cvec{\pr}{\st})\leq \mu(\cvec{\pr}{\st'})$.  %\red{must refer to a read-config}
\end{lemma}

\begin{proof}
	This lemma follows from the fact that if a configuration $k$ is such that 
	$\status{\cvec{\pr}{\st}[k]}=F$ at a state $\st$, then $\pr$ will start any 
	future $\act{read-config}$ action from a configuration $\config{\cvec{\pr}{\st'}[j]}$
	such that $j\geq k$. But $\config{\cvec{\pr}{\st'}[j]}$ is the last finalized configuration 
	at $\st'$ and hence $\mu(\cvec{\pr}{\st'})\geq \mu(\cvec{\pr}{\st})$.
\end{proof}

\begin{lemma}  [Sequence Progress]
	\label{lem:finalconf}
	Let $\op_1$ and $\op_2$ two 
	%read/write/install operations 
	completed \act{read-config} actions invoked by processes $\pr_1, \pr_2\in\idSet$ 
	respectively, such that $\op_1\bef\op_2$ in an execution $\EX$. 
	Let $\st_1$ be the state after the response 
	step of $\op_1$ and $\st_2$ the state after the response step 
	%after the completion of $\op_1$ and $\st_2$ the state after the termination of the first $\act{read-config}$ 
	of $\op_2$. Then 
	$\mu(\cvec{\pr_1}{\st_1})\leq\mu(\cvec{\pr_2}{\st_2})$.
\end{lemma}

\begin{proof}
	By Lemma \ref{lem:prefix} it follows that $\cvec{\pr_1}{\st_1}$ is a prefix of $\cvec{\pr_2}{\st_2}$.
	Thus, if $\nu_1 = \nu(\cvec{\pr_1}{\st_1})$ and $\nu_2 = \nu(\cvec{\pr_2}{\st_2})$, $\nu_1\leq\nu_2$.
	Let $\mu_1=\mu(\cvec{\pr_1}{\st_1})$, such that $\mu_1\leq\nu_1$, be the last element in $\cvec{\pr_1}{\st_1}$,
	where $\status{\cvec{\pr_1}{\st_1}[\mu_1]} = F$. Let now $\mu_2=\mu(\cvec{\pr_2}{\st'})$, 
	be the last element which $\pr_2$ obtained in Line A\ref{algo:parser}:\ref{line:readconfig:final} 
	during $\op_2$ 
	%of the  $\act{read-config}$ action 
	such that $\status{\cvec{\pr_2}{\st'}[\mu_2]} = F$ in some state $\st'$ before $\st_2$. 
	If $\mu_2\geq\mu_1$, and since $\st_2$ is after $\st'$, then by Lemma \ref{lem:final:monotonic} 
	$\mu_2\leq \mu(\cvec{\pr_2}{\st_2})$ and hence $\mu_1\leq \mu(\cvec{\pr_2}{\st_2})$ as well. 
	
	It remains to examine the case where $\mu_2<\mu_1$. Process  $\pr_1$ 
	sets the status of $\cvec{\pr_1}{\st_1}[\mu_1]$ to $F$ in two cases: (i) either when finalizing 
	a reconfiguration, or (ii) when receiving an $s.nextC = \tup{\config{\cvec{\pr_1}{\st_1}[\mu_1]}, F}$ %with status $F$ 
	from some server $s$ during a $\act{read-config}$ action. In both cases $\pr_1$ propagates the 
	$\tup{\config{\cvec{\pr_1}{\st_1}[\mu_1]}, F}$ to a quorum of servers in  
	$\config{\cvec{\pr_1}{\st_1}[\mu_1-1]}$ before completing. We know by Lemma
	\ref{lem:prefix} that since $\op_1\bef\op_2$ then $\cvec{\pr_1}{\st_1}$ is a prefix 
	in terms of configurations of the $\cvec{\pr_2}{\st_2}$. So it must be the case 
	that $\mu_2 < \mu_1 \leq \nu(\cvec{\pr_2}{\st_2})$. Thus, during $\op_2$, %the $\act{read-config}$ action, 
	$\pr_2$ starts from the configuration at index $\mu_2$ and in some iteration 
	performs $\act{get-next-config}$ in configuration $\cvec{\pr_2}{\st_2}[\mu_1-1]$. 
	According to Lemma \ref{lem:unique}, $\config{\cvec{\pr_1}{\st_1}[\mu_1-1]} = \config{\cvec{\pr_2}{\st_2}[\mu_1-1]}$.
	Since $\op_1$ completed before $\op_2$, then it must be the case that $\st_1$ appears before 
	$\st'$ in $\EX$. However, $\pr_2$ invokes the $\act{get-next-config}$ operation in a state $\st''$
	which is either equal to $\st'$ or appears after $\st'$ in $\EX$. Thus, $\st''$ must appear after $\st_1$ in $\EX$.
	From that it follows that when the $\act{get-next-config}$ is executed by $\pr_2$ there is already 
	a quorum of servers in $\config{\cvec{\pr_2}{\st_2}[\mu_1-1]}$, say $Q_1$, that received 
	$\tup{\config{\cvec{\pr_1}{\st_1}[\mu_1]}, F}$from $\pr_1$. 
	Since, $\pr_2$ waits from replies from a quorum of servers from the same configuration, say $Q_2$, and since the 
	$nextC$ variable at each server is monotonic (Lemma \ref{lem:server:monotonic}), then there is a server $s\in \quo{1}\cap \quo{2}$, 
	such that $s$ replies to $\pr_2$ with $s.nextC = \tup{\config{\cvec{\pr_1}{\st_1}[\mu_1]}, F}$. So, 
	$\cvec{\pr_2}{\st_2}[\mu_1]$ gets $\tup{\config{\cvec{\pr_1}{\st_1}[\mu_1]}, F}$, and 
	hence $\mu(\cvec{\pr_2}{\st_2})\geq \mu_1$ in this case as well. This completes our proof.
\end{proof}

Using the previous Lemmas we can conclude to the main result of this section.

\begin{theorem}
\label{thm:recon:properties}
	Let $\op_1$ and $\op_2$ two 
%read/write/install operations 
completed \act{read-config} actions invoked by processes $\pr_1, \pr_2\in\idSet$ 
respectively, such that $\op_1\bef\op_2$ in an execution $\EX$. 
Let $\st_1$ be the state after the response 
step of $\op_1$ and $\st_2$ the state after the response step 
%after the completion of $\op_1$ and $\st_2$ the state after the termination of the first $\act{read-config}$ 
of $\op_2$.
%$\cvec{\pr_1}{\st_1}$ and $\cvec{\pr_2}{\st_2}$ at two states $\st_1$ and $\st_2$ 
%in an execution $\EX$ of the algorithm  such that $\st_1$ appears before $\st_2$ in $\EX$. 
Then the following properties hold: 
\begin{enumerate}
\item [$(a)$] 
\textbf{Configuration Consistency}: $\cvec{\pr_2}{\st_2}[i].cfg = \cvec{\pr_1}{\st_1}[i].cfg$,  for $ 1 \leq i \leq \nu(\cvec{\pr_1}{\st_1})$,
\item [$(b)$]
 \textbf{Sequence Prefix}: 
 $\cvec{\pr_1}{\st_1}  \preceq_p \cvec{\pr_2}{\st_2}$, and
\item [$(c)$] 
\textbf{Sequence Progress}:
%$(c)$  
  $\mu(\cvec{\pr_1}{\st_1}) \leq \mu(\cvec{\pr_2}{\st_2})$
  %; and 
%\item [$(d)$]  
%\nn{????$(d)$  $\cvec{\pr_2}{\st_2}[i]   = \cvec{\pr_1}{\st_1}[i]$,  for  $ 1 \leq i \leq \mu(\cvec{\pr_1}{\st_1})$????} 
\end{enumerate}
\end{theorem}

\begin{proof}
Statements $(a)$, $(b)$ and $(c)$ follow from Lemmas \ref{lem:unique}, \ref{lem:prefix}, and 
\ref{lem:final:monotonic}.
%are from  the atomic read-modify-write property of ${\recBox}.\act{add-config}(\cdot)$ 
%and $(d)$ is due to the recon client.
\end{proof}

\subsection{Atomicity  Property of \ares{}}
\label{sec:safety:atomic}

Given the properties satisfied by the reconfiguration algorithm of \ares{} 
% and assuming that the DAP used satisfy Property~\ref{property:dap}, as presented
% in Section \ref{ssec:dap}, then  we have the following result. 
in any execution, we can now proceed to examine whether our algorithm satisfies the safety (atomicity) conditions. The
propagation of the information of the distributed object is achieved using the get-tag, get-data, and
put-data actions. We assume that the DAP used satisfy  Property~\ref{property:dap} as presented in
Section \ref{ssec:dap}, and we will show that, given such assumption, \ares{} satisfies atomicity.

We begin with a lemma stating that if a reconfiguration operation retrieves a configuration
sequence of length $k$ during its $\act{read-config}$ action, then it installs/finalizes the $k + 1$ configuration in the global configuration sequence.

\begin{lemma}%{\bf 
	\label{lem:recon:incremental}
	Let $\op$ be a complete reconfiguration operation by a reconfigurer $\rec$ in an execution $\EX$ of \ares{}.
	if $\state_1$ is the state in $\EX$ following the termination of the $\act{read-config}$ action during $\op$,
	then $\op$ invokes a $\act{finalize-config}(\cvec{\rec}{\state_2})$ at a state $\state_2$ in $\EX$, 
	with $\nu(\cvec{\rec}{\state_2}) = \nu(\cvec{\rec}{\state_1}) + 1$.
	%		
	%Let $\rho$ denote a complete reconfiguration operation in an execution $\EX$ of \ares{} and let $\mathbf{x}$ be the 
	%configuration sequence returned by  the $\act{read-config}$ step, $a_1$  in $\rho$. Then if $\mathbf{y}$ is a reconfiguration 
	%sequence returned by another $\act{read-config}$ step $a_2$ that appears in $\EX$ after $a_1$ completes then during $\rho$ the call $\act{finalize-config}( \mathbf{y}( \nu(\mathbf{x}) + 1)$ is executed.
\end{lemma}

\begin{proof}
	This lemma follows directly from the implementation of the $\act{reconfig}$ operation. 
	Let $\op$ be a reconfiguration operation $\act{reconfig}(c)$. At first, $\op$ invokes a
	$\act{read-config}$ to retrieve a latest value of the global configuration sequence, $\cvec{\rec}{\state_1}$, 
	in the state $\state_1$ in $\EX$. During the $\act{add-config}$ action, $\op$ proposes the addition of $c$, 
	and appends at the end of $\cvec{\rec}{\state_1}$ the decision $d$ of the consensus protocol. 
	Therefore, if $\cvec{\rec}{\state_1}$ is extended by $\tup{d, P}$ (Line A~\ref{algo:reconfigurer}:\ref{line:addconfig:assign}), and hence the 
	$\act{add-config}$ action returns a configuration sequence $\cvec{\rec}{\state_1'}$ with length 
	$\nu(\cvec{\rec}{\state_1'})= \nu(\cvec{\rec}{\state_1}) + 1$. As $\nu(\cvec{\rec}{\state_1'}$ does not change 
	during the $\act{update-config}$ action, then $\cvec{\rec}{\state_1'}$ is passed to the $\act{finalize-config}$ action at state $\state_2$,
	and hence $\cvec{\rec}{\state_2}=\cvec{\rec}{\state_1'}$. Thus, $\nu(\cvec{\rec}{\state_2})=\nu(\cvec{\rec}{\state_1'})= \nu(\cvec{\rec}{\state_1}) + 1$ and the lemma follows.
\end{proof}

The next lemma states that only some reconfiguration operation $\op$ may finalize a configuration $c$ at index $j$ in a configuration sequence $\pr.cseq$ at any process $\pr$.
%, is finalized by some . 
To finalize $c$, the lemma shows that $\op$ must witness a configuration
sequence such that its last finalized configuration appears at an index $i<j$ in the configuration sequence $\pr.cseq$. In other words,  reconfigurations always finalize configurations that are ahead from their
latest observed final configuration, and it seems like ``jumping'' from one final configuration to the next.

\begin{lemma} %{\bf 
\label{lem:recon:jump}
Suppose $\EX$ is an execution of \ares{}. 
%Let $\cvec{*}{\state}$ be a configuration sequence returned by any $\act{read-config}$ action at state $\state$ in $\EX$,
 %and let $\cvec{*}{\state}[j].status = F$, for some $0 < j \leq \nu( \cvec{*}{\state})$. 
 For any state $\state$ in $\EX$, if $\status{\cvec{\pr}{\state}[j]} = F$ for some process $\pr\in\idSet$, 
 then there exists a $\act{reconfig}$ operation $\op$ by a reconfigurer $\rec\in\recSet$, such that
 (i) $\rec$ invokes $\act{finalize-config}(\cvec{\rec}{\state'})$ during $\op$ at some state $\state'$ in $\EX$, 
 (ii) $\nu(\cvec{\rec}{\state'}) = j$, and (iii) $\mu(\cvec{\rec}{\state'}) < j$. 
% 
% there exists a $\act{recon}$ operation $\rho$ that introduces $\cvec{}{}[j]$ 
% such that there exists  a configuration in $\cvec{rc}{}$ returned by a $\act{read-config}$ of $\rho$,  such that, $\cvec{rc}{}[k]$  $ k < j$ and  $\cvec{rc}{}[k].status = F$.
\end{lemma}
%
%\begin{lemma} \label{lem:monotonicity:cseq}
%If $cseq_1$ and $cseq_2$ are the values returned by two executions of  $\act{read-config}(c)$, where $c \in \mathcal{C}$ in an execution $\EX$ of \ares{}; and $cseq_1$ is returned  before the procedure call that returned $cseq_2$ is invoked in $\EX$  then $cseq_1 \preceq_p cseq_2$.  \blue{use this template}
%\end{lemma}

\begin{proof}
	A process sets the status of a configuration $c$ to $F$ in two cases: (i) either during 
	a $\act{finalize-config}(seq)$ action such that $\nu(seq) = \tup{c,P}$ (Line A\ref{algo:reconfigurer}:\ref{line:status:finalize}), or 
	(ii) when it receives $\tup{c,F}$ from a server $s$ during a $\act{read-next-config}$ action. 
	Server $s$ sets the status of a configuration $c$ to $F$ only if it receives a message that contains $\tup{c,F}$
	(Line A\ref{algo:server}:\ref{line:server:finalize}). So, (ii) is possible only if $c$ is finalized during a $\act{reconfig}$ operation. 
	
	Let, w.l.o.g., $\op$ be the first reconfiguration operation that finalizes $\config{\cvec{\pr}{\state}[j]}$. 
	To do so, process $\rec$
	invokes $\act{finalize-config}(\cvec{\rec}{\state_1'})$ during $\op$, at some state $\state'$ that appears
	before $\state$ in $\EX$. By Lemma \ref{lem:unique}, $\config{\cvec{\pr}{\state}[j]} = \config{\cvec{\rec}{\state'}[j]}$.
	Since, $\rec$ finalizes $\cvec{\rec}{\state'}[j]$, then  this is the last entry of $\cvec{\rec}{\state'}$ and 
	hence $\nu(\cvec{\rec}{\state'}) = j$. 
	Also, by Lemma \ref{lem:recon:jump} it follows that the 
	$\act{read-config}$ action of $\op$ returned a configuration $\cvec{\rec}{\state''}$ in some state 
	$\state''$ that appeared before $\state'$ in $\EX$, such that $\nu(\cvec{\rec}{\state''}) < \nu(\cvec{\rec}{\state'})$.
	Since by definition, $\mu(\cvec{\rec}{\state''})\leq\nu(\cvec{\rec}{\state''})$, then $\mu(\cvec{\rec}{\state''}) < j$.
	However, since only $\tup{c, P}$ is added to $\cvec{\rec}{\state''}$ to result in $\cvec{\rec}{\state'}$,
	then $\mu(\cvec{\rec}{\state''}) = \mu(\cvec{\rec}{\state'})$. Therefore, $\mu(\cvec{\rec}{\state'}) < j$ as well 
	and the lemma follows.
\end{proof}

We now reach an important lemma of this section. By \ares{}, before a read/write/reconfig operation completes 
it propagates the maximum tag it discovered by executing the $\act{put-data}$ action in the last 
configuration of its local configuration sequence (Lines A\ref{algo:reconfigurer}:\ref{line:addconfig:put}, A\ref{algo:writer}:\ref{line:writer:prop}, A\ref{algo:writer}:\ref{line:reader:prop}). 
When a subsequent operation is invoked, it reads the latest configuration 
sequence by beginning from the last finalized configuration in its local sequence and invoking $\act{read-data}$ to
all the configurations until the end of that sequence. The lemma shows that the latter operation 
will retrieve a tag which is higher than the tag used in the $\act{put-data}$ action of any preceding operation.

% For the following proof we use the notation $\nu^c(\cvec{\pr}{\st}) = \config{\cvec{\pr}{\st}[\nu(\cvec{\pr}{\st})]}$. In other words, $\nu^c(\cvec{\pr}{\st})$ denotes 
% the last configuration in the sequence $\cvec{\pr}{\st}$.

\begin{lemma}%{\bf  
	\label{lem6}
	Let $\pi_1$ and $\pi_2$ be two completed read/write/reconfig  operations invoked by processes
	$p_1$ and $p_2$ in $\idSet$,  in an execution  $\EX$ of \ares{}, such that, $\pi_1\bef\pi_2$. 
	If $c_1.\act{put-data}(\tup{\tg{\op_1}, v_{\op_1}})$ is the last \act{put-data} action of $\pi_1$
	and $\state_2$ is the state in $\EX$ after the completion of the first $\act{read-config}$ action of $\pi_2$, 
	then there exists a $c_2.\act{put-data}(\tup{\tg{},v})$ action in some configuration 
	$c_2 = \config{\cvec{\pr_2}{\state_2}[k]}$, for $\mu(\cvec{\pr_2}{\state_2}) \leq k \leq
	\nu(\cvec{\pr_2}{\state_2})$, such that (i) it completes in a state $\state'$ before $\state_2$ in $\EX$, and (ii) $\tg{}\geq\tg{\op_1}$.
\end{lemma}

\begin{proof}
Note that from the definitions of $\nu(\cdot)$ and $\mu(\cdot)$, we have  
$\mu(\cvec{\pr_2}{\state_2}) \leq \nu(\cvec{\pr_2}{\state_2})$.  Let $\state_1$ be the state in $\EX$ 
after the completion of $c_1.\act{put-data}(\tup{\tg{\op_1}, v_{\op_1}})$ and 
$\state'_1$ be the state in $\EX$ 
%immediately 
following the response step of $\pi_1$. Since any operation executes $\act{put-data}$ 
on the last discovered configuration then $c_1$ is the last configuration found in 
$\cvec{\pr_1}{\state_1}$, and hence $c_1 =  \config{\cvec{\pr_1}{\st_1}[\nu(\cvec{\pr_1}{\st_1})]}$. %\nu^c(\cvec{\pr_1}{\state_1})$.  
By Lemma~\ref{lem:final:monotonic}   we have
$\mu(\cvec{\pr_1}{\state_1}) \leq \mu(\cvec{\pr_1}{{\state}'_1})$ 
and by Lemma~\ref{lem:finalconf} we have 
$\mu(\cvec{\pr_1}{{\state}'_1}) \leq \mu(\cvec{\pr_2}{{\state}_2})$, since $\pi_2$ (and thus its first \act{read-config}
action) is invoked after ${\state}'_1$ (and thus after the last \act{read-config} action during $\op_1$). %, i.e., after $\pi_1$ completes  and 
Hence, combining the two implies that
$\mu(\cvec{\pr_1}{\state_1}) \leq \mu(\cvec{\pr_2}{{\state}_2})$.   Now from the last implication and the first statement we have $\mu(\cvec{\pr_1}{{\state}_1}) \leq \nu(\cvec{\pr_2}{{\state}_2})$.
Therefore,  it remains to examine whether the last finalized configuration witnessed by $\pr_2$ appears before or after $c_1$, i.e.:
$(a)$ $\mu(\cvec{\pr_2}{\state_2}) \leq \nu(\cvec{\pr_1}{{\state}_1})$ and 
$(b)$ $\mu(\cvec{\pr_2}{\state_2}) > \nu(\cvec{\pr_1}{{\state}_1})$.
% Let for appreviation denote by $\mu_i = \mu(\cvec{\pr_i}{\state_i})$, 
% and $\nu_i = \nu(\cvec{\pr_i}{\state_i})$.

\vspace{1em}

\noindent{\textbf{Case $(a)$:}} Since $\op_1\bef\op_2$
%operation $\pi_2$ is invoked after $\pi_1$ completes 
then, by Theorem \ref{thm:recon:properties}, $\cvec{\pr_2}{\state_2}$ value returned
by $\act{read-config}$ at $p_2$ during the execution of $\pi_2$ satisfies $\cvec{\pr_1}{\state_1}  \preceq_p \cvec{\pr_2}{\state_2}$. 
Therefore, $\nu(\cvec{\pr_1}{{\state}_1})  \leq \nu(\cvec{\pr_2}{{\state}_2})$, and hence in this case 
$\mu(\cvec{\pr_2}{\state_2}) \leq \nu(\cvec{\pr_1}{{\state}_1})  \leq \nu(\cvec{\pr_2}{{\state}_2})$. 
Since $c_1$ is the last configuration in $\cvec{\pr_{1}}{\state_1}$, then it has index $\nu(\cvec{\pr_1}{{\state}_1})$.
So if we take $c_2 = c_1$ then the $c_1.\act{put-data}(\tup{\tg{\op_1}, v_{\op_1}})$ action trivially 
satisfies both conditions of the lemma as: (i) it completes in state $\state_1$ which appears before $\state_2$, and
(ii) it puts a pair $\tup{\tg{}, v}$ such that $\tg{}=\tg{\op_1}$. 
%if $\pi_2$ is either a  write or a read operation then   $\act{get-tag}$  or $\act{get-data}$   is called on each configuration 
%  $\cvec{p_2}{\sigma_2}[i]$ for $ \mu(\cvec{p_2}{\sigma_2}) \leq i \leq \nu(\cvec{p_2}{\sigma_2})$ .  Note that 
%  $\mu(\cvec{\pr_2}{\state_2}) \leq \nu(\cvec{\pr_1}{{\state}_1})  \leq \nu(\cvec{\pr_2}{{\state}_2}) $ 
%  then by Definition~\ref{def:consistency} and because $\act{put-data}$ corresponding to  $\sigma_1$ at $p_1$ completes before $\act{get-tag}$ or $\act{get-data}$ the claim of the lemma holds.  
  \vspace{1em}
  
\noindent{\textbf{Case $(b)$:}} This case is possible if there exists a reconfiguration client $\rec$ that invokes $\act{reconfig}$ operation $\rho$, 
%say corresponding to some reconfiguration operation $\rho$,  that 
during which it executes the 
$\act{finalize-config}( \cvec{\rec}{*})$ that finalized configuration with index 
$\nu(\cvec{\rec}{*}) = \mu(\cvec{\pr_2}{\state_2})$. 
Let $\state$ be the state immediately after the $\act{read-config}$ of $\rho$. Now, we consider two sub-cases: $(i)$ $\state$ appears before $\state_1$ in $\EX$, or  $(ii)$ $\state$ appears after $\state_1$ in $\EX$.
  \vspace{1em}
  
 \noindent{\emph{Subcase $(b)(i)$:}} Since $\act{read-config}$ at $\state$ completes before the invocation of last $\act{read-config}$ of operation  $\pi_1$ then, either
   $\cvec{\rec}{\state}  \prec_p \cvec{\pr_1}{\state_1}$, or $\cvec{\rec}{\state}  = \cvec{\pr_1}{\state_1}$ due to Lemma~\ref{lem:prefix}.  
  Suppose  $\cvec{\rec}{\state}  \prec_p \cvec{\pr_1}{\state_1}$,  then according to Lemma~\ref{lem:recon:incremental} $\rec$  executes $\act{finalize-config}$  on configuration sequence  $\cvec{\rec}{*}$ with  $\nu(\cvec{\rec}{*}) = \nu(\cvec{\rec}{\state}) + 1$. 
  %configuration as argument  but since
 %$\rho_0$ installs $\mu^c(\cvec{p_2}{\sigma_2})$ then 
  Since $\nu(\cvec{\rec}{*}) = \mu(\cvec{\pr_2}{\state_2})$, then
  $\mu(\cvec{p_2}{\sigma_2}) = \nu(\cvec{rc}{\sigma}) + 1$. 
  If however, $\cvec{\rec}{\state}  \prec_p \cvec{\pr_1}{\state_1}$, then 
  $\nu(\cvec{\rec}{\state}) < \nu(\cvec{\pr_1}{\state_1})$ and thus 
  $\nu(\cvec{\rec}{\state})+1 \leq \nu(\cvec{\pr_1}{\state_1})$. 
  This implies that
%  But this implies 
 %  $\nu(\cvec{rc}{\sigma})  + 1 \leq  \nu(\cvec{p_1}{\sigma_1}) $ which
      $\mu(\cvec{p_2}{\sigma_2})  \leq  \nu(\cvec{p_1}{\sigma_1})$  which 
      contradicts our initial assumption for this case that $\mu(\cvec{p_2}{\sigma_2}) >  \nu(\cvec{p_1}{\sigma_1})$. So this sub-case is impossible.
      %a contradiction. 

  Now suppose,  that $\cvec{\rec}{\state}  = \cvec{\pr_1}{\state_1}$. Then it follows that 
  $\nu(\cvec{\rec}{\state})  = \nu(\cvec{\pr_1}{\state_1})$, and that  
  $\mu(\cvec{p_2}{\sigma_2}) = \nu(\cvec{p_1}{\sigma_1}) + 1$ in this case.
  Since $\state_1$ is the state after the last $\act{put-data}$ during $\op_1$, then if $\state_1'$ is the state after the completion of the last $\act{read-config}$ of $\op_1$ (which follows the \act{put-data}), it must be the case that $\cvec{\pr_1}{\state_1} = \cvec{\pr_1}{\state_1'}$. So, during its last $\act{read-config}$ process  $p_1$ does not read the configuration indexed at $ \nu(\cvec{p_1}{\sigma_1}) + 1$. This means that the $\act{put-config}$ completes in $\rho$ at state ${\state}_{\rho}$ after ${\state}'_1$ and the 
  $\act{update-config}$ operation is invoked at state $\state_{\rho}'$ after $\state_{\rho}$ with 
  a configuration sequence $\cvec{\rec}{\state_{\rho}'}$.
  During the update operation $\rho$ invokes $\act{get-data}$ operation in every configuration 
  $\config{\cvec{\rec}{\state_{\rho}'}[i]}$, for $\mu(\cvec{\rec}{\state_{\rho}'})\leq i\leq\nu(\cvec{\rec}{\state_{\rho}'})$.
  Notice that $\nu(\cvec{\rec}{\state_{\rho}'})= \mu(\cvec{p_2}{\sigma_2}) = \nu(\cvec{p_1}{\sigma_1}) + 1$
  and moreover the last configuration of $\cvec{\rec}{\state_{\rho}'}$ was just added by $\rho$ and it is 
  not finalized. From this it follows that $\mu(\cvec{\rec}{\state_{\rho}'})<\nu(\cvec{\rec}{\state_{\rho}'})$,
  and hence $\mu(\cvec{\rec}{\state_{\rho}'})\leq\nu(\cvec{p_1}{\sigma_1}) $.
  Therefore,  $\rho$ executes $\act{get-data}$ in configuration $\config{\cvec{\rec}{\state_{\rho}'}[j]}$ for 
  $j = \nu(\cvec{p_1}{\state_1}) $. Since $\pr_1$ invoked $\act{put-data}(\tup{\tg{\op_1},v_{\op_1}})$ 
  at the same configuration $c_1$, 
  %$\nu^c(\cvec{p_1}{\state_1})$, 
  and completed in a state $\state_1$ before $\state_{\rho}'$, then by
  \textbf{C1} of Property~\ref{property:dap}, 
  it follows that the $\act{get-data}$ action will 
  return a tag $\tg{}\geq \tg{\op_1}$. Therefore, the maximum tag that $\rho$ discovers is 
  $\tg{max}\geq\tg{}\geq\tg{\op_1}$. Before invoking the $\act{finalize-config}$ action, $\rho$ 
  invokes %$\nu^c(\cvec{\rec}{\state_{\rho}'}).\act{put-data}(\tup{\tg{max}, v_{max})}$
  $c_1.\act{put-data}(\tup{\tg{max}, v_{max})}$. Since 
  $\nu(\cvec{\rec}{\state_{\rho}'})= \mu(\cvec{p_2}{\state_2})$, and since by Lemma \ref{lem:unique},
  then the action $\act{put-data}$ is invoked in a configuration $c_2 = \config{\cvec{p_2}{\sigma_2}[j]}$ 
  such that $j=\mu(\cvec{p_2}{\sigma_2})$. Since the $\act{read-config}$ action of $\op_2$ 
  observed configuration $\mu(\cvec{p_2}{\sigma_2})$, then it must be the case that $\state_2$ 
  appears after the state where the $\act{finalize-config}$ was invoked and therefore after the 
  state of the completion of the $\act{put-data}$ action during $\rho$. Thus, in this case both 
  properties are satisfied and the lemma follows. 
%  in state ${\state}''_1$ after ${\state}'_1$ and hence after $\state_1$. 
%  
%  
%  So by Definition~\ref{def:consistency} it returns a tag $t > tag(\pi_1)$. 
  \vspace{1em}

\noindent{\emph{Subcase $(b)(ii)$:}} Suppose 
%$\act{read-config}$ 
in this case that $\state$ occurs in $\EX$ after $\sigma_1$. In this case the 
   last $\act{put-data}$ in $\pi_1$ completes   before the invocation of the  $\act{read-config}$ in $\rho$ in execution $\EX$. 
   Now we can argue recursively, $\rho$ taking the place of operation $\pi_2$,  that $\mu(\cvec{\rec}{{\state}}) \leq \nu(\cvec{\rec}{{\state}})$ and 
therefore,  we consider  two cases:
$(a)$ $\mu(\cvec{\rec}{\state}) \leq \nu(\cvec{\pr_1}{{\state}_1})$ and 
$(b)$ $\mu(\cvec{\rec}{\state}) > \nu(\cvec{\pr_1}{{\state}_1})$. 
Note that there are finite number of operations  invoked in $\EX$  before $\pi_2$ is invoked, and hence  the statement of the lemma can be shown to hold by a sequence of inequalities.
\end{proof}

The following lemma shows the consistency of operations as long as the DAP used satisfy 
Property~\ref{property:dap}. %properties \textbf{C1} and \textbf{C2}.

\begin{lemma} %{\bf \ref{dap:C1:C2}} 
	 Let $\op_1$ and $\op_2$ denote completed read/write operations in an execution $\EX$, from processes $\pr_1,\pr_2\in\idSet$ respectively, such that $\op_1\bef\op_2$. If $\tg{\op_1}$ and $\tg{\op_2}$ are the 
	local tags at $\pr_1$ and $\pr_2$ after the completion of $\op_1$ and $\op_2$ respectively, then $\tg{\op_1} \leq \tg{\op_2}$; if $\op_1$ is a write operation then $\tg{\op_1}  < \tg{\op_2}$.
\end{lemma}

\begin{proof}
		Let $\tup{\tg{\op_1}, v_{\op_1}}$ be the pair passed to the last $\act{put-data}$ action of $\op_1$.
		Also, let $\state_2$ be the state in $\EX$ that follows the completion of the first $\act{read-config}$ action 
		during $\op_2$.
		Notice that $\op_2$ executes a loop after the first $\act{read-config}$ operation 
		and performs $c.\act{get-data}$ (if $\op_2$ is a read) or $c.\act{get-tag}$ (if $\op_2$
		is a write) from all $c =\config{\cvec{\pr_2}{\state_2}[i]}$, for $\mu(\cvec{\pr_2}{\state_2})\leq i \leq \nu(\cvec{\pr_2}{\state_2})$. By Lemma \ref{lem6}, there exists a $c'.\act{put-data}(\tup{\tg{},v})$ action by 
		some operation $\op'$ on some configuration $c'=\config{\cvec{\pr_2}{\state_2}[j]}$, for 
		$\mu(\cvec{\pr_2}{\state_2})\leq j \leq \nu(\cvec{\pr_2}{\state_2})$, that completes in some state 
		$\state'$ that appears before $\state_2$ in $\EX$. Thus, the $\act{get-data}$ or $\act{get-tag}$ 
		invoked by $\pr_2$ on $\config{\cvec{\pr_2}{\state_2}[j]}$, occurs after state $\state_2$ and thus
		after $\state'$. Since the DAP primitives used satisfy \textbf{C1} and \textbf{C2} of 
		Property \ref{property:dap}, then the $\act{get-tag}$ action will return a tag $\tg{\op_2}'$ 
		or a $\act{get-data}$ action will return a pair $\tup{\tg{\op_2}', v_{\op_2}'}$, with $\tg{\op_2}'\geq\tg{}$.
		As $\pr_2$ gets the maximum of all the tags returned, then by the end of the loop 
		$\pr_2$ will retrieve a tag $\tg{max}\geq\tg{\op_2}'\geq\tg{}\geq\tg{\op_1}$.
		
		If now $\op_2$ is a read, it returns $\tup{\tg{max}, v_{max}}$ after propagating that value to 
		the last discovered configuration. Thus, $\tg{\op_2}\geq \tg{\op_1}$. If however $\op_2$ is 
		a write, then before propagating the new value the writer increments the maximum timestamp
		discovered (Line A\ref{algo:writer}:\ref{line:writer:increment}) generating a tag $\tg{\op_2}>\tg{max}$.
		Therefore the operation $\op_2$ propagates a tag $\tg{\op_2}>\tg{\op_1}$ in this case.
\end{proof}

%\subsection*{ \ref{ssec:liveness} \ares{} Liveness}
%\input{ssec-correct-liveness.tex}
And the main result of this section follows:

\begin{theorem}[Atomicity]
	In  any execution $\EX$ of \ares{}, if in every configuration $c\in\gseq$,
	$\dagetdata{c}$, $\daputdata{c}{}$, and $\dagettag{c}$
	 %the DAP primitives  
	 satisfy Property~\ref{property:dap}, then ~\ares{} satisfy atomicity.
	%, given that the 
	%$\act{get-data}$, $\act{get-tag}$, and $\act{put-data}$ primitives used satisfy properties
	%\textbf{C1} and \textbf{C2} of Definition \ref{def:consistency}.
\end{theorem}

%In  \ares{},  each configuration 
%may implement the DAPs in a different way as stated below:
As algorithm \ares{} handles each configuration separately, then we can observe that the algorithm
may utilize a different mechanism for the put and get primitives in each configuration. So the 
following remark:

\begin{remark}
	Algorithm \ares{} satisfies atomicity even when the implementaton of the  DAPs in two 
	different configurations $c_1$ and $c_2$ are not the same, given that the $c_i.\act{get-tag}$,
	$c_i.\act{get-data}$, and the $c_i.\act{put-data}$ primitives 
	in each $c_i$ satisfy Property~\ref{property:dap}.  
\end{remark}

\section{Performance Analysis of \ares{}}
\label{sec:performance}
A major challenge in reconfigurable atomic services is to examine the latency of terminating read and write operations, especially when those are invoked concurrently with reconfiguration operations. 
In this section we provide an in depth analysis of the latency of operations in \ares{}. Additionally, a storage and communication analysis is shown when \ares{} utilizes 
the erasure-coding algorithm presented in Section \ref{ssec:dap:impl}, in each configuration. 

% of the storage and communication costs of \ares{}, 
% and the latency of read and write operations. 

\subsection{Latency Analysis}
\label{sec:safety:d}
Liveness (termination) properties cannot be specified for ~\ares{}, without restricting asynchrony  or the
rate of arrival of \act{reconfig} operations, or if the consensus protocol never terminates.
Here,  we provide some conditional performance analysis of the operation, based on 
%assumptions on the 
latency bounds on the message \nnrev{delay}{delivery}. %in the network.
 We assume that local computations take negligible time and the latency of an 
operation is  due to the delays in the messages exchanged during the execution. 
%Before proceeding with our 
%analysis we define an upper and lower communication bounds. 
We measure delays in \myemph{time units} of some global clock, which is visible only to an external viewer.
No process has access to the clock.
Let $\smdelay$ and $\lgdelay$ be the minimum and maximum durations taken by 
 messages, sent during an execution  of~\ares,  to reach their destinations.
%denote the minimum message delivery delay 
%between any two processes in the service; let $\lgdelay$ be the 
%maximum delivery delay. 
 Also, let $\opdelay{\op}$ denote the duration 
 of an operation (or action) $\op$. In the statements that follow, 
 we consider any execution $\EX$ of \ares, which contains $k$ \act{reconfig} operations.
 \kmk{For any configuration $c$ in an execution of~\ares{},  we assume that any 
 	$\consensus{c}.\act{propose}$ operation, takes at least $\opdelaymin{CN}$ time units.}

\remove{
\begin{lemma}
\label{lem:opdelays}
Suppose $\pi$ and $\phi$ are operations of the type \act{put-config}, \act{read-next-config}, respectively, invoked by some non-faulty reconfiguration clients,  then the latency of these operations are bounded as follows: 
%\begin{itemize}
	%\item 
	$(i)$ $2\smdelay\leq \opdelay{\pi}\leq 2\lgdelay$
	%\item 
	and 
	$(ii)$ $2\smdelay\leq \opdelay{\phi}\leq 2\lgdelay$.
%\end{itemize}
\end{lemma}
}
Let us first examine what is the action delays based on the boundaries we assume. 
It is easy to see that actions \act{put-config}, \act{read-next-config} perform two message exchanges thus take time $2\smdelay\leq \opdelay{\phi}\leq 2\lgdelay$. 
From this we can derive the delay of  a \act{read-config} action.

\begin{lemma}
	\label{lem:rcdelay}
	Let $\phi$ be a $\act{read-config}$ operation invoked by a non-faulty reconfiguration client $\rec$, 
	with the input argument and returned values of $\phi$ as  $\cvec{\rec}{\st}$ and  $\cvec{\rec}{\st'}$ respectively. Then the delay of $\phi$ is:	$4\smdelay(\nu(\cvec{\rec}{\st'})-\mu(\cvec{\rec}{\st})+1)\leq \opdelay{\phi}\leq4\lgdelay(\nu(\cvec{\rec}{\st'})-\mu(\cvec{\rec}{\st})+1)$.
    %$4\smdelay(\nu-\mu+1)\leq \opdelay{\phi}\leq 4\lgdelay(\nu-\mu+1)$.
\end{lemma}

From Lemma \ref{lem:rcdelay} it is clear that the latency of a $\act{read-config}$ action 
depends on the number of configurations installed since the last  finalized configuration known to the recon client.

% \remove{
% Let $\seqlen = \nu-\mu$ denote the number of newly installed configurations.
% %Now let us examine when a new configuration gets inserted in the configuration 
% %sequence by a \act{reconfig} operation.
Given the latency of a \act{read-config}, we can compute the minimum amount 
of time it takes for $k$ configurations to be installed.

The following lemma shows the maximum latency of a read or a write operation, invoked by any non-faulty client. 
From~\ares{} algorithm,  the latency of a read/write operation depends on the delays of the  DAPs  operations. 
%influence  the delay of a read and write operation. 
For our analysis we assume 
that all $\act{get-data}$, $\act{get-tag}$ and $\act{put-data}$ primitives use 
two phases of communication.  Each phase consists of a communication between the client and the servers.
%\kmkremove{Such an assumption is justified by atomic algorithms like ~\treas{} and ABD. }

\begin{lemma}
	\label{lem:dapdelays}
Suppose $\pi$,  $\phi$ and $\psi$ are operations of the type \act{put-data}, \act{get-tag} and  \act{get-data}, respectively, invoked by some non-faulty reconfiguration clients,  then the latency of these operations are bounded as follows: 
	$(i)$ $2\smdelay\leq \opdelay{\pi}\leq 2\lgdelay$; $(ii)$
	 $2\smdelay\leq \opdelay{\phi}\leq 2\lgdelay$; and $(iii)$
	  $2\smdelay\leq \opdelay{\psi}\leq 2\lgdelay$.
\end{lemma}
%Now we show the delay of a read or a write operation $\op$.

In the following lemma, we estimate the time taken for a read or a write operation to complete,
	when it discovers $k$ configurations between its invocation and response steps.
%
%show that in a fair execution of ~\ares{} \nnrev{where there are at}{that contains} $k$ reconfiguration 
%operations, any read or write operation takes at most  $6\lgdelay\left(k+2\right)$.

\begin{lemma}
	\label{lem:rwdelay}
%Consider a execution of ~\ares{} where there are $k$ reconfiguration and a read or a write operation $\pi$ invoked by a non-faulty then   $6\lgdelay(k+1)$.
Consider any  execution of ~\ares{} where at most  $k$ reconfiguration operations are invoked.
Let $\sigma_s$ and $\sigma_e$ be the states before the invocation 
and after the completion step of a read/write operation $\op$,
in some fair execution $\EX$ of \ares{}. 
%If $k=\nu(\cvec{\pr}{\sigma_e}) - \mu(\cvec{\pr}{\sigma_s})$, 
Then we have 
$\opdelay{\op}\leq 6\lgdelay\left(k+2\right)$ to complete. 
\end{lemma}

\begin{proof}
	Let $\state_s$ and $\state_e$ be the states before the invocation 
	and after the completion step of a read/write operation $\op$ by $\pr$ respectively,
	in some execution $\EX$ of \ares. 
	%Then $\op$ takes time at most:
%	\[
%		\opdelay{\op}\leq 6\lgdelay\left[\nu(\cvec{\pr}{\state_e}) - \mu(\cvec{\pr}{\state_s})+2\right]
%	\] 
	By algorithm examination we can 
	see that any read/write operation performs the following actions in this order:
	$(i)$  \act{read-config}, $(ii)$ \act{get-data} (or \act{get-tag}), $(iii)$ \act{put-data},
	and $(iv)$ \act{read-config}. Let $\state_1$ be the state when the first \act{read-config}, denoted by $\act{read-config}_1$, 
	action terminates. By Lemma \ref{lem:rcdelay} the action will take time:
	\[
		\opdelay{\act{read-config}_1} \leq 4\lgdelay(\nu(\cvec{\pr}{\state_1})-\mu(\cvec{\pr}{\state_s})+1)
	\]
	The $\act{get-data}$ action that follows the \act{read-config} (Lines Alg.~\ref{algo:writer}:\ref{line:rw:getdata:start}-\ref{line:rw:getdata:end}) also took at most $(\nu(\cvec{\pr}{\state_1})-\mu(\cvec{\pr}{\state_s})+1)$ time units,
	given that no new finalized configuration was discovered by the \act{read-config} action. 
	Finally, the \act{put-data}  and the second \act{read-config} actions of $\op$ may be invoked at most
	$(\nu(\cvec{\pr}{\state_e})-\nu(\cvec{\pr}{\state_1})+1)$ times, given that the \act{read-config} action discovers 
	one new configuration every time it runs. Merging all the outcomes, the total time of $\op$ can be at most:
	\begin{eqnarray*}
	\opdelay{\op} & \leq & 4\lgdelay(\nu(\cvec{\pr}{\state_1})-\mu(\cvec{\pr}{\state_s})+1) + 2\lgdelay(\nu(\cvec{\pr}{\state_1})-\mu(\cvec{\pr}{\state_s})+1) + (4\lgdelay+2\lgdelay)(\nu(\cvec{\pr}{\state_e})-\nu(\cvec{\pr}{\state_1})+1) \\
%	 & \leq & 6\lgdelay\nu(\cvec{\pr}{\state_1})-6\lgdelay\mu(\cvec{\pr}{\state_s})+6\lgdelay\nu(\cvec{\pr}{\state_e})-6\lgdelay\nu(\cvec{\pr}{\state_1})+12\lgdelay \\
	 & \leq & 6\lgdelay\left[\nu(\cvec{\pr}{\state_e}) - \mu(\cvec{\pr}{\state_s})+2\right] \leq 6D(k+1)
	\end{eqnarray*}
where  $\nu(\cvec{\pr}{\state_e}) - \mu(\cvec{\pr}{\state_s})\leq k + 1$ since there can be at most $k$ new configurations installed. and the result of the lemma follows.
\end{proof}

It remains now to examine the conditions under which a read/write operation may “catch up” with an infinite number of reconfiguration operations.
For the sake of a worst case analysis we will assume that reconfiguration operations suffer 
the minimum delay $d$, whereas read and write operations suffer the maximum
delay $D$ in each message exchange. 
% Also, we assume that any consensus operation takes the least amount of time to complete $\opdelaymin{CN}$.
We first show how long it takes for $k$ configurations to be installed.

\begin{lemma}
	\label{lem:configdelay}
	Let $\sigma$ be the last state of a fair execution of \ares{}, $\EX$. 
	Then $k$ configurations can be installed to $\cvec{}{\sigma}$, in time
	$\opdelay{k} \geq 4\smdelay\sum_{i=1}^{k}i+ k\left(\opdelaymin{CN}+2\smdelay\right)$ time units.
%	\begin{equation}
	%$ \left(\opdelaymin{CN}+6\smdelay\right)k$.
%	\end{equation}
	%by the completion of its $\act{add-config}$ action 
%	in our execution construction.
\end{lemma}

\begin{proof}
    In \ares{} a \act{reconfig} operation has 
four phases: $(i)$ $\act{read-config}(cseq)$,  reads the latest configuration sequence, 
$(ii)$ $\text{\act{add-config}}(cseq, c)$,  attempts to add  the new configuration 
at the end of the global sequence $\mathcal{G}_L$, 
$(iii)$ $\text{\act{update-config}}(cseq)$,   transfers the knowledge to the added configuration,
and 
$(iv)$  $\text{\act{finalize-config}}(cseq)$ finalizes the added configuration. So, a new configuration is appended to the 
end of the configuration sequence (and it becomes visible to any operation) during the 
\act{add-config} action.  In turn, the \act{add-config} action, runs a consensus algorithm
to decide on the added configuration and then invokes a \act{put-config} action to add
the decided configuration. Any operation that is invoked after the \act{put-config} action 
observes the newly added configuration. 

Notice that when multiple reconfigurations are invoked concurrently, then it might be the case 
that all participate to the same consensus instance and the configuration sequence is appended 
by a single configuration. The worst case scenario happens when all concurrent reconfigurations
manage to append the configuration sequence by their configuration. In brief, this is possible when 
the \act{read-config} action of each \act{reconfig} operation appears after the \act{put-config}
action of another \act{reconfig} operation. 

\begin{figure}[ht]
	\begin{center}
		\includegraphics[width=0.60\textwidth]{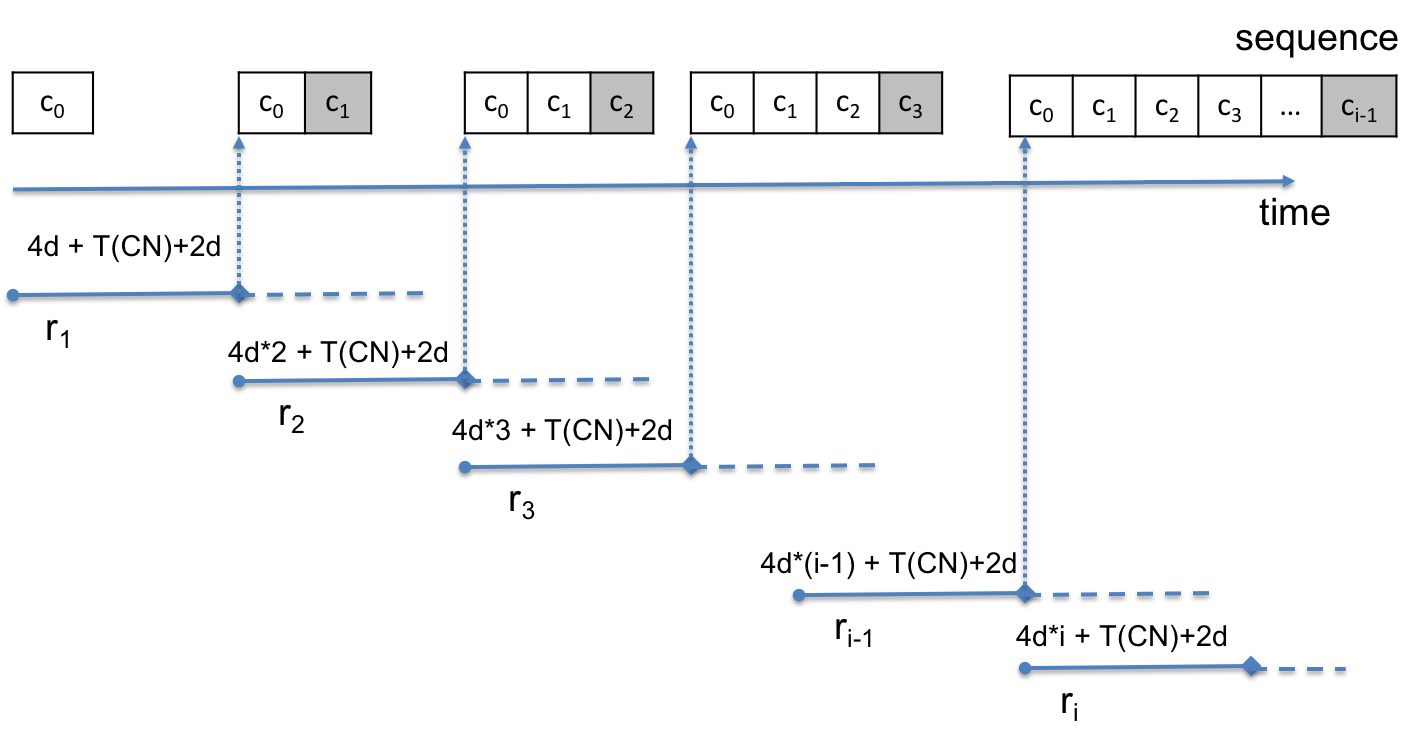}
		\caption{Successful \act{reconfig} operations.}
		\label{fig:reconfigExec}
	\end{center}
\end{figure}

More formally we can build an execution where all \act{reconfig} operations append their configuration
in the configuration sequence. Consider the partial execution $\EX$ that ends in a state $\state$. Suppose that 
every process $\pr\in\idSet$ knows the same configuration sequence, $\cvec{\pr}{\state}=\cvec{}{\state}$. Also let 
the last finalized operation in $\cvec{}{\state}$ be the last configuration of the sequence, e.g. $\mu(\cvec{}{\state}) = \nu(\cvec{}{\state})$.  
Notice that $\cvec{}{\state}$ can also be the initial configuration sequence $\cvec{\pr}{\state_0}$. 
We extend $\EX_0$ by a series of \act{reconfig} operations, such that each reconfiguration 
$\rec_i$ is invoked by a reconfigurer $r_i$ and attempts to add a configuration $c_i$. 
Let $\rec_1$ be the first reconfiguration that performs the following actions 
without being concurrent with any other \act{reconfig} operation: 
\begin{itemize}
	\item \act{read-config} starting from $\mu(\cvec{}{\state})$
	\item \act{add-config} completing both the consensus proposing $c_1$ and 
	the $\act{put-config}$ action writing the decided configuration
\end{itemize}
%\act{read-config} and \act{add-config} actions without being concurrent 
%with any other operation. 
Since  $\rec_1$ its not concurrent with any other $\act{reconfig}$ operation, then is the only process to propose
a configuration in $\mu(\cvec{}{\state})$,  and hence by the consensus algorithm properties,
$c_1$ is decided. Thus, $\cvec{}{\state}$ is appended by a tuple $\tup{c_1,P}$.

Let now reconfiguration $\rec_2$ be invoked immediately after the completion of the 
$\act{add-config}$ action from $\rec_1$. Since the local sequence at the beginning 
of $\rec_2$ is equal to $\cvec{}{\state}$, then the $\act{read-config}$ action of $\rec_2$
will also start from $\mu(\cvec{}{\state})$.  Since, $\rec_1$ already propagated $c_1$ 
to $\mu(\cvec{}{\state})$ during is $\act{put-config}$ action, then $\rec_2$ will discover 
$c_1$ during the first iteration of its $\act{read-config}$ action, and thus it will 
repeat the iteration on $c_1$. Configuration $c_1$ is the last in the sequence and 
thus the $\act{read-config}$ action of $\rec_2$ will terminate after the second 
iteration.  Following the $\act{read-config}$ action, $\rec_2$ attempts to 
add $c_2$ in the sequence. Since $\rec_1$ is the only reconfiguration that might 
be concurrent with $\rec_2$, and since $\rec_1$ already completed consensus 
in $\mu(\cvec{}{\state})$, then $\rec_2$ is the only operation to run consensus in $c_1$.
Therefore, $c_2$ is accepted and $\rec_2$ propagates $c_2$ in $c_1$ using a 
$\act{put-config}$ action. 

So in general we let configuration $\rec_i$ to be invoked after the completion of 
the $\act{add-config}$ action from $\rec_{i-1}$. As a result, the $\act{read-config}$
action of $\rec_i$ performs $i$ iterations, and the configuration $c_i$ is added 
immediately after configuration $c_{i-1}$ in the sequence. Figure \ref{fig:reconfigExec}
illustrates our execution construction for the reconfiguration operations. 

It is easy to notice that such execution results in the worst case latency for all the 
reconfiguration operations $\rec_1, \rec_2,\ldots, \rec_i$. As by Lemma \ref{lem:rcdelay}
a \act{read-config} action takes at least $4d$ time to complete, then as also 
seen in Figure \ref{fig:reconfigExec}, $k$ reconfigs may take time 
$\opdelay{k} \geq \sum_{i=1}^{k}\left[4\smdelay*i+ \left(\opdelaymin{CN}+2\smdelay\right)\right]$. Therefore, it will take time
$\opdelay{k} \geq 4\smdelay\sum_{i=1}^{k}i+ k\left(\opdelaymin{CN}+2\smdelay\right)$ and the lemma follows.
%
% We can now compute 
% the minimum latency we need to add $k$ new configurations in the configuration 
% sequence starting from the state $\state$ of execution $\EX$.
\end{proof}

The following theorem is the main result of this section, in which we define the relation between $\opdelaymin{CN}$, $d$ and $D$
%show that in any fair execution of ~\ares{}, if the  consensus operations are ``slow enough'' then  
so to guarantee that any read or write issued by a non-faulty client always terminates.

\begin{theorem}
%Consider a  read or  write operation $\pi$, invoked by a non-faulty client, in a fair and well-formed execution of ~\ares{}. Suppose at the point of invocation of $\pi$ the client has $|cseq| = p$. Then if $\opdelaymin{CN} \geq 6D(p + 2) -5d$ the operation $\pi$ completes.
 Suppose  $\opdelaymin{CN} \geq 3(6D-\smdelay)$, then  any  read or write operation $\op$ completes in any execution  $\EX$ of 
\ares{}  for any number of reconfiguration operations in $\EX$.
%$\smdelay \geq \frac{3\lgdelay}{k}-\frac{\opdelay{CN}}{2(k+2)}$
\end{theorem}

\begin{proof}
	By Lemma \ref{lem:configdelay}, 
	$k$ configurations may be installed in:
		$\opdelay{k} \geq 4\smdelay\sum_{i=1}^{k}i+ k\left(\opdelaymin{CN}+2\smdelay\right)$.	
	Also by Lemma \ref{lem:rwdelay}, we know that operation $\op$ takes at most 
$	\opdelay{\op}\leq 6\lgdelay\left(\nu(\cvec{\pr}{\state_e}) - \mu(\cvec{\pr}{\state_s})+2\right)$.
	Assuming that $k=\nu(\cvec{\pr}{\state_e}) - \mu(\cvec{\pr}{\state_s})$, the total number of 
	configurations observed during $\op$, then $\op$ may terminate before a $k+1$ configuration 
	is added in the configuration sequence if 
	  $6\lgdelay(k+2) \leq  4\smdelay\sum_{i=1}^{k}i+ k\left(\opdelaymin{CN}+2\smdelay\right)$ then we have
	  $d \geq \frac{3\lgdelay}{k}-\frac{\opdelaymin{CN}}{2(k+2)}$.
	And that completes the lemma. 
\end{proof}

\subsection{Storage and Communication Costs for \ares{}.}\label{sec:safety:c}
Storage and Communication costs for \ares{} highly depends on the DAP that we use 
in each configuration. For our analysis we assume that each configuration utilizes the 
algorithms and the DAPs presented in Section \ref{ssec:dap:impl}.
% We now briefly present the storage and communication costs associated with the presented DAPs.
%{\treas{}  %Due to space limitations the proofs appear in \cite{}.

Recall that by our assumption, the storage cost counts the size (in bits) of the coded elements 
%that are locally stored, which are 
stored in variable $List$  at each server. We ignore the storage cost due to meta-data.
For  communication cost we measure the bits sent on the wire between the nodes.

\begin{lemma}\label {thm:storage_TREAS}
	The worst-case total storage cost of Algorithm \ref{fig:casopt} is $(\delta +1 )\frac{n}{k}$.
\end{lemma}
\begin{proof}
  The maximum number of  (tag, coded-element) pair in the $List$ is $\delta+1$, and the size of each coded element is 
  $\frac{1}{k}$ while the tag variable is a metadata and therefore, not counted. So, the total storage cost is $(\delta +1)\frac{n}{k}$.
\end{proof}

We next state  the communication cost for the write and read operations in  Aglorithm \ref{fig:casopt}. Once again, note that we ignore the communication cost arising from exchange of meta-data.

\begin{lemma} \label{treas:write_cost}
	The communication cost associated with a successful  write operation in Algorithm \ref{fig:casopt} is at most $\frac{n}{k}$. 
\end{lemma}

\begin{proof}
  During read operation, in the $\act{get-tag}$ phase the servers respond with their highest tags variables, which are metadata. However, in the $\act{put-data}$ phase, the reader sends each server the  coded elements of size  $\frac{1}{k}$ each, and hence the total cost of communication for this is $\frac{n}{k}$. Therefore, we have the worst case communication cost of a write operation is $ \frac{n}{k}$.
\end{proof}

\begin{lemma} \label{radonc:read_cost}
	The communication cost associated with a successful read operation in Algorithm \ref{fig:casopt} is at most $(\delta +2)\frac{n}{k}$. 
\end{lemma}
\begin{proof}
  During read operation, in the $\act{get-data}$ phase the servers respond with their $List$ variables and hence each such list 
  is of size at most $(\delta +1)\frac{1}{k}$, and then counting all such responses give us $(\delta +1)\frac{n}{k}$.  In the $\act{put-data}$ phase, the reader sends each server the  coded elements of size  $\frac{1}{k}$ each, and hence the total cost of communication for this is $\frac{n}{k}$. Therefore, we have the worst case communication cost of a read operation is 
  $(\delta+2) \frac{n}{k}$.
\end{proof}

From the above Lemmas we get.

\begin{theorem}\label{treas:performance}
 The \ares{} algorithm has: (i) storage cost $(\delta +1 )\frac{n}{k}$, (ii) communication 
cost for each write at most to $\frac{n}{k}$, and (iii) communication 
cost for each read at most $(\delta +2)\frac{n}{k}$.
\end{theorem}

%\section{\treas: A new two-round erasure-code based algorithm}\label{sec:treas}
%\input{algorithm_atomic_v3.tex}

\section{Flexibility of DAPs}\label{sec:dap:flexible}
 %For a read/write algorithm
%that uses the presented data-primitives, it can provide atomic guarantees if the data primitives 
%satisfy the following consistency properties: 
%To implement an algorithm, which is based on the DAP primitives, the DAP primitives must satisfy the following
%consistency properties.

In this section, we argue that  various implementations of DAPs  can be used in \ares{}. In fact, via \act{reconfig} 
operations, one can implement a highly adaptive 
%strongly consistent 
atomic DSS: 
replication-based can be transformed into erasure-code based DSS; increase or decrease the number of storage servers; 
 study 
the performance of the DSS under various code parameters, etc. 
The insight to implementing various DAPs comes from the observation that the  simple  algorithmic 
	template $A$ (see Alg.~\ref{algo:atomicity:generic1}) for reads and writes protocol combined with 
	any implementation of DAPs, satisfying Property~\ref{property:dap} gives rise to a MWMR atomic memory service.
	Moreover, the read and writes operations terminate as long as the implemented DAPs complete. 
%	implements an atomic  storage if the implementation of the  DAPs satisfy Property \ref{property:dap}, and $A$ preserves liveness (termination),
%	if every invocation of the used DAPs terminate.
\begin{algorithm}[!ht]
	\begin{algorithmic}[2]
		\begin{multicols}{2}
			{\scriptsize
				%\Part{Generic Algorithm $A_1$}
				\Operation{read}{} 
				%\State $wCounter\gets wCounter+1$
				\State $\tup{t, v} \gets \dagetdata{c}$
				\State $\daputdata{c}{ \tup{t,v}}$
				\State return $ \tup{t,v}$
				\EndOperation
				%\Statex
				\Operation{write}{$v$} 
				%\State $wCounter\gets wCounter+1$
				\State $t \gets \dagettag{c}$
				\State $t_w \gets inc(t)$
				\State $\daputdata{c}{\tup{t_w,v}}$
				\EndOperation
				%\EndPart
			}
		\end{multicols}
	\end{algorithmic}
	\caption{Template $A$ for the client-side read/write steps.}
	\label{algo:atomicity:generic1}
	\vspace{-1em}
\end{algorithm}

	A read operation in $A$ performs $\dagetdata{c}$ to retrieve a tag-value pair, $\tup{\tg{},v}$ from a configuration $c$, and then 
	it performs a $c.\act{put-data}(\tup{\tg{},v})$ to propagate that pair to the configuration $c$. A write operation is similar to the read but before 
	performing the $\act{put-data}$ action it generates a new tag which associates with the value to be written.
The following result shows 
	%\cite{ARES:Arxiv:2018} 
	that $A$ is atomic and live,
	if the DAPs satisfy Property~\ref{property:dap} and live.
	 
 \begin{theorem}[\nn{Atomicity} of template $A$]\label{atomicity:A1}
 Suppose the DAP implementation satisfies the consistency properties $C1$ and $C2$ of  Property \ref{def:consistency}
 for a configuration $c\in\confSet$. 
 Then any execution $\EX$  of \nn{algorithm $A$ in configuration $c$}
 %the  atomicity protocols $A_1$  on a 
 %fixed set of servers $S$, of some 
 %configuration $c\in\confSet$,  
 %as in Fig. ~\ref{algo:atomicity:generic1} satisfies atomic read and write and is live if the  primitive functions are live in $\xi$.
 is atomic and live if each DAP \nn{invocation terminates} in $\EX$ \nn{under the failure model $c.\mathcal{F}$}.
 \end{theorem}
 \begin{proof}
 We  prove the atomicity by proving properties $A1$, $A2$ and $A3$ presented in Section \ref{model} for any execution of the algorithm.
					
	\emph{Property $A1$}: Consider two operations $\phi$ and $\pi$ such that $\phi$ completes before $\pi$ is invoked. 
	We need to show that it cannot be  the case that $\pi \prec \phi$. We break our analysis into the following four cases:

	Case $(a)$: {\em Both $\phi$ and $\pi$ are writes}. The $\daputdata{c}{*}$ of $\phi$ completes before 
	$\pi$ is invoked. 
	%which implies that by well-formedness 
	By property {\bf C1} the tag $\tg{\pi}$ returned by the $\dagetdata{c}$ at $\pi$ is 
	at least as large as $\tg{\phi}$. Now, 
	%since $\tg{\pi}$ is larger than $t_{\phi}$, by the steps of 
	since $\tg{\pi}$ is incremented by the write operation then $\pi$ puts a tag $\tg{\pi}'$ such that
	$\tg{\phi} < \tg{\pi}'$ and hence we cannot have $\pi \prec \phi$.
	
	Case $(b)$: {\em $\phi$ is a write and  $\pi$ is a read}. In execution $\EX$ since 
$\daputdata{c} {\tup{t_{\phi}, *}}$ of $\phi$ completes 
	before the $\dagetdata{c}$ of $\pi$ is invoked, by 
	%the well-formedness 
	property {\bf C1} the tag $\tg{\pi}$ obtained from the above
	$\dagetdata{c}$ is at least as large as $\tg{\phi}$. Now $\tg{\phi} \leq \tg{\pi}$ implies that we cannot have $\pi \prec \phi$.
	
	Case $(c)$: {\em $\phi$ is a read and  $\pi$ is a write}.  Let the id of the writer that invokes $\pi$ we $w_{\pi}$.  
	The 
$\daputdata{c}{\tup{\tg{\phi}, *}}$  call of $\phi$ completes
	before  $\dagettag{c}$ of $\pi$ is initiated. Therefore, by 
	%the well-formedness 
	property {\bf C1} %of data-primitives the above 
	$\act{get-tag}(c)$ returns $\tg{}$ such that, $\tg{\phi} \leq \tg{}$. Since $\tg{\pi}$ is equal to $inc(\tg{})$ 
	by design of the algorithm, hence $\tg{\pi} > \tg{\phi}$ and we cannot have $\pi \prec \phi$.
	
	Case $(d)$: {\em Both $\phi$ and $\pi$ are reads}. In execution $\EX$  
the $\daputdata{c}{\tup{t_{\phi}, *}}$ is executed as a part of $\phi$ and 
	completes before $\dagetdata{c}$ is called in $\pi$. By 
	%the well-formedness
	 property {\bf C1} of the data-primitives, 
	we have $\tg{\phi} \leq \tg{\pi}$ and hence we cannot have $\pi \prec \phi$.
	
	\emph{Property $A2$}: Note that because the tag set $\tsSet$ is well-ordered we can show that A2 holds by first showing that
	every write has a unique tag. This means that any two pair of writes can be ordered. 
	Note that 
	a read can be ordered w.r.t. any write operation trivially if the respective tags are different, and by definition, if the 
	tags are equal the write is ordered before the read.
	
	Observe that two tags generated from different writers are necessarily distinct because of the 
	id component of the tag. Now if the operations, say $\phi$ and $\pi$ are writes  from the same writer then, by 
	well-formedness property, the second operation will witness a higher integer part in the tag 
	%is invoked after the first completes, say without loss of generality $\phi$ completes before 
	%$\pi$ is invoked.   In that case the integer part of the tag of $\pi$ is higher 
	%because the well-formedness 
	by property  {\bf C1}, and since the $\dagettag{c}$  is followed by $\daputdata{c}{*}$. Hence $\pi$ is ordered after $\phi$. 
	
	\emph{Property $A3$}:  By {\bf C2} the $\dagetdata{c}$ may return a tag $\tg{}$, only 
	when there exists an operation $\op$ that invoked a $\daputdata{c}{\tup{\tg{},*}}$. Otherwise it returns the initial value. Since a write is the only operation to put a new tag $\tg{}$ in the system then Property $A3$ follows from {\bf C2}.
 \end{proof}
	
	\subsection{Representing  Known Algorithms in terms of data-access primitives}
\label{ssec:generic}

A number of known tag-based algorithms that implement atomic read/write objects 
	(e.g., ABD \cite{ABD96}, \fast \cite{CDGL04} ), can be expressed 
	in terms of DAP. In this subsection we demonstrate how we can transform the very celebrated ABD algorithm \cite{ABD96}.

\paragraph{\mwABD{} Algorithm.}
The multi-writer version of the ABD can be transformed to the generic algorithm Template $A$. Algorithm \ref{algo:dap:abd}
illustrates the three DAP for the ABD algorithm. The $\act{get-data}$ primitive encapsulates the query phase 
of \mwABD{}, while the $\act{put-data}$ primitive encapsulates the propagation phase of the algorithm. 
%Thus, given a configuration $c\in\confSet$ the DAPs in the transformation of ABD are implemented as follows:

\begin{algorithm}[!h]
	%\hrule \F
	\begin{algorithmic}[2]
		\begin{multicols}{2}{\scriptsize
				\Part{Data-Access Primitives at process $\pr$}
				\Procedure{c.put-data}{$\tup{\tg{},v})$}
				\State {\bf send} $(\text{{\sc write}}, \tup{\tg{},v})$ to each $s \in \servers{c}$
				\State {\bf until} $\exists \quo,  \quo \in \quorums{c}$ s.t. $\pr$ receives {\sc ack} from $\forall s\in\quo{}$
				\EndProcedure
				
				\Statex				
				
				\Procedure{c.get-tag}{}
				%	\State {\bf send} $(\text{\act{query}},\rdr)$ to every server $s\in \bigcup_{cseq[i]}members(\qs_{cseq[i].conf})$
				\State {\bf send} $(\text{{\sc query-tag}})$ to each  $s\in \servers{c}$
				\State {\bf until}    $\exists \quo{}, \quo{}\in\quorums{c}$ s.t. 
				\State\TT$\pr$ receives $\tup{\tg{s},v_s}$ from $\forall s\in\quo{}$ 
				\State $\tg{max} \gets \max(\{\tg{s} : \pr \text{ received } \tup{\tg{s},v_s} \text{ from } s \})$
				\State {\bf return} $\tg{max}$
				\EndProcedure
				
				\Statex
				
				\Procedure{c.get-data}{}
				%	\State {\bf send} $(\text{{\sc query}},\rdr)$ to every server $s\in \bigcup_{cseq[i]}members(\qs_{cseq[i].conf})$
				\State {\bf send} $(\text{{\sc query}})$ to each  $s\in \servers{c}$
				\State {\bf until}    $\exists \quo{}, \quo{}\in\quorums{c}$ s.t. 
				\State\TT $\pr$ receives $\tup{\tg{s},v_s}$ from $\forall s\in\quo{}$ 
				\State $\tg{max} \gets \max(\{\tg{s} : \rdr_i \text{ received } \tup{\tg{s},v_s} \text{ from } s \})$
				\State {\bf return} $\{\tup{\tg{s},v_s}:\tg{s}=\tg{max} \wedge ~\pr \text{ received } \tup{\tg{s},v_s} \text{ from } s\}$
				\EndProcedure
				\EndPart
			}\end{multicols}
				
			\\\hrulefill 
			
			\begin{multicols}{2}{\scriptsize
				\Part{Primitive Handlers at server $s_i$ in configuration $c$}
				\Receive{{\sc query-tag}}{} %\Comment{Called upon reception of a message}
				%\State $\act{handle-get-tag(c_k)}$
				\State send $\tg{}$ to $q$
				\EndReceive
				
				\Statex
				
				\Receive{{\sc query}}{}
				%\State $\act{handle-get-data(c_k)}$
					\State send $\tup{\tg{}, v}$ to $q$
				\EndReceive
				
				\Statex
				\Statex
				
				\Receive{{\sc write}, $\tup{\tg{in}, v_{in}}$}{}
				%\State $\act{handle-put-data(c_k)}$
					\If {$\tg{in}> \tg{}$} 	\label{line:server:tg-comparison}
						\State  $\tup{\tg{},v}\gets \tup{\tg{in},v_{in}}$ \label{line:server:update}
					\EndIf
					\State  send  {\sc ack} to $q$ 	\label{line:server:reply}
				\EndReceive
				\EndPart
				
		}\end{multicols}	
	\end{algorithmic}
	%\hrule \B
	\caption{Implementation of DAP for ABD at each process $\pr$ using configuration $c$}
	\label{algo:dap:abd}
\end{algorithm} 

Let us now examine if the primitives satisfy properties \textbf{C1} and \textbf{C2} of Property \ref{property:dap}.
We begin with a lemma that shows the monotonicity of the tags at each server.

\begin{lemma} 
	\label{lem:server:tag:monotonic}
	Let $\st$ and $\st'$ two states in an execution $\EX$ such that $\st$ appears before $\st'$ in $\EX$.
	Then for any server $s\in\srvSet$ it must hold that $\atT{s.tag}{\st}\leq\atT{s.tag}{\st'}$.
\end{lemma}

\begin{proof}
	According to the algorithm, a server $s$ updates its local tag-value pairs when it receives a message
	with a higher tag. So if $\atT{s.tag}{\st}=\tau$ then in a state $\st'$ that appears after $\st$ in $\EX$,
	$\atT{s.tag}{\st'}\geq \tau$.
\end{proof}

In the following two lemmas we show that property \textbf{C1} is satisfied, that is if a $\act{put-data}$ action completes,
then any subsequent $\act{get-data}$ and $\act{get-tag}$ actions will discover a higher tag than the one propagated 
by that $\act{put-data}$ action.

\begin{lemma}
	\label{lem:putdata:gettag}
	Let $\phi$ be a $\daputdata{c}{\tup{\tau,v}}$ action invoked by $\pr_1$ and $\gamma$ be a $\dagettag{c}$ action
	invoked by $\pr_2$ in a configuration $c$, such that 
	$\phi\bef\gamma$ in an execution $\EX$ of the algorithm. Then $\gamma$ returns a tag $\tau_\gamma \geq \tau$. 
\end{lemma}

\begin{proof}
	The lemma follows from the intersection property of quorums. In particular, during the $\daputdata{c}{\tup{\tau,v}}$ 
	action, $\pr_1$ sends the pair $\tup{\tau,v}$ to all the servers in $\servers{c}$ and waits until all the servers in 
	a quorum $Q_i\in\quorums{c}$ reply. When those replies are received then the action completes. 
	
	During a $\dagetdata{c}$ action on the other hand, $\pr_2$ sends query messages to all the servers in $\servers{c}$ 
	and waits until all servers in a quorum $Q_j\in\quorums{c}$ (not necessarily different than $Q_i$) reply. By definition 
	$Q_i \cap Q_j\neq\emptyset$, thus any server $s\in Q_i\cap Q_j$ reply to both $\phi$ and $\gamma$ actions.  
	By Lemma \ref{lem:server:tag:monotonic} and since $s$ received a tag $\tau$, then $s$ replies to $\pr_2$ with a tag
	$\tau_s\geq\tau$. Since $\gamma$ returns the maximum tag it discovers then $\tau_\gamma \geq \tau_s$. Therefore
	$\tau_\gamma \geq \tau$ and this completes the proof.
\end{proof}

With similar arguments and given that each value is associated with a unique tag then we can show the following lemma.

\begin{lemma}
	\label{lem:putdata:getdata}
	Let $\pi$ be a $\daputdata{c}{\tup{\tau,v}}$ action invoked by $\pr_1$ and $\phi$ be a $\dagetdata{c}$ action
	invoked by $\pr_2$ in a configuration $c$, such that $\pi\bef\phi$ in an execution $\EX$ of the algorithm. 
	Then $\phi$ returns a tag-value  $\tup{\tau_\phi, v_\phi}$
	such that $\tau_\phi \geq \tau$. 
\end{lemma}

Finally we can now show that property \textbf{C2} also holds. 

\begin{lemma}
	If $\phi$ is a $\dagetdata{c}$ that returns $\tup{\tg{\pi}, v_\pi } \in \tsSet \times \valSet$, 
	then there exists $\pi$ such that $\pi$ is a $\daputdata{c}{\tup{\tg{\pi}, v_{\pi}}}$ and $\phi \not \rightarrow \pi$.
\end{lemma}

\begin{proof}
	This follows from the facts that (i) servers set their tag-value pair to a pair received by a $\act{put-data}$ action, and (ii) 
	a $\act{get-data}$ action returns a tag-value pair that it received from a server. So if a $\dagetdata{c}$  operation $\phi$ 
	returns a tag-value pair $\tup{\tg{\pi}, v_\pi }$, there should be a server $s$  that replied to that operation with 
	$\tup{\tg{\pi}, v_\pi }$,  and $s$ received $\tup{\tg{\pi}, v_\pi }$ from some $\daputdata{c}{\tup{\tg{\pi}, v_\pi }}$
	action, $\pi$. Thus, $\pi$ can proceed or be concurrent with $\phi$, and hence $\phi\not\bef \pi$.
\end{proof}

% NN: we decided to not include the following optimization for now
%\section{Efficient state transfer during reconfiguration}
%\label{sec:transfer}
%\input{transferstate_v1.tex}

\section{Experimental Evaluation}
\label{sec:evaluation}
The theoretical findings suggest that \ares{} is an algorithm
to provide robustness and flexibility on shared memory 
implementations,  without sacrificing strong consistency. In this section 
we present a proof-of-concept implementation of \ares{} and we run 
preliminary experiments to get better insight on the 
efficiency and adaptiveness of \ares{}. In particular,
our experiments measure the latency of each \act{read}, \act{write}, and 
\act{reconfig} operations, and examine the persistence of consistency 
even when the service is reconfigured between configurations that utilize
different shared memory algorithms.  

\subsection{Experimental Testbed}
\label{ssec:testbed}

We ran experiments on two different setups: (i) simulated locally on a single machine, and (ii) on a LAN. Both type of experiments run on \emph{Emulab}~\cite{emulab}, an emulated WAN environment testbed used for developing, debugging, and evaluating the systems. We used nodes with two 2.4 GHz 64-bit 8-Core E5-2630 "Haswell" processors and \SI{64}{\giga\byte} RAM. In both setups we used an external implementation of Raft\cite{} consensus algorithms, which was used for the service reconfiguration and was deployed on top of small RPi devices. Small devices introduced further delays in the system, reducing the speed of reconfigurations and creating harsh conditions for longer periods in the service.

\noindent\textbf{Local Experimental Setup:}
The local setup was used 
to have access to a global synchronized clock (the clock of the local machine)
in order to examine whether our algorithm preserves global ordering and hence 
atomicity even when using different algorithms between configurations. 
%When testing the persistence of consistency, 
Therefore, all the instances are hosted on the same physical machine 
%instead of having one machine for each node. This preserves 
avoiding the skew between computer clocks in a
distributed system. Furthermore, the use of one clock guarantees that when an event occurs after another, it will assign a later time. 

\noindent\textbf{Distributed Experimental Setup:}
The distributed experiments in Emulab enabled the examination of the performance of the algorithm in a close to real environment.
For the deployment and remote execution of the experimental tasks on the Emulab, we used {\em Ansible Playbooks}~\cite{ansible}. 
All physical nodes were placed on a single LAN using a DropTail queue with the default traffic parameters, i.e. \SI{100}{Mb} bandwidth, and no delay or packet loss. Each physical machine runs one server or client process. This guarantees a fair communication delay between a client and a server node.

%To execute the tasks of the experiments on the Emulab remote nodes, we used {\em Ansible Playbooks}~\cite{ansible}, i.e. scripts written in YAML syntax, in which we defined the tasks. Running these Ansible Playbooks, ansible connects to the remote machines over ssh and starts execution of the tasks there.
%\vspace{0.5cm}
\noindent\textbf{Node Types:}
In all experiments, we use three distinct types of nodes, writers, readers and servers. Their main role is listed below:

\begin{itemize}
\item {\bf writer $w \in W \subseteq C$ :} 
a client that sends write requests to all servers and waits for a quorum of the servers to reply 

\item {\bf reader $r \in R \subseteq C$:} 
a client that sends read requests to servers and waits for a quorum of the servers to reply

\item {\bf reconfigurer $g \in G \subseteq C$:} 
a client that sends reconfiguration requests to servers and waits for a quorum of the servers to reply

\item {\bf server $s \in S$:} a server listens for read and write requests, it updates its object replica according to the atomic shared memory and replies to the process that originated the request. 

\end{itemize}

\noindent\textbf{Performance Metric:} 
The metric for evaluating the algorithms is the operational latency. This includes both communication and computational delays. The operation latency is computed as the average of all clients' average operation latencies. For better estimations, each experiment in every scenario was repeated 3 times.

% Computational latency encompasses the time necessary for the {FM} to fragment a file object and generate the respective hashes for its blocks. 
%t the read-config action discovers one new configuration every time it runs
\subsection{Experimental Scenarios}
\label{ssec:scenarios}
In this section, we describe the scenarios we constructed and the settings for each of them. In our scenarios we constructed the DAPs and used two different atomic storage algorithms in \ares{}: (i) the erasure coding based algorithm presented in Section \ref{ssec:dap:impl}, and (ii) the ABD algorithm (see Section \ref{ssec:generic}). 

\noindent\textbf{Erasure Coding:} The type of erasure coding we use is (n,k)-Reed-Solomon code, which guarantees that any k of n coded fragments is enough to reassemble the original object. The parameter k is the number of encoded data fragments, n is the total number of servers and m is the number of parity fragments, i.e. n-k. A high number of m and consequently a small number of k means less redundancy.   

\noindent\textbf{Fixed Parameters:} In all scenarios the number of servers is fixed to 10. The number of writers and the value of delta are set to 5; delta being the maximum number of concurrent put-data operations. The parity value of the EC algorithm is set to 2, in order to minimize the redundancy; leading to 8 data servers and 2 parity servers. 

It is worth mentioning that the quorum size of the EC algorithm is $\left\lceil \frac{10+8}{2} \right\rceil=9$, while the quorum size of ABD algorihtm is  $\left\lfloor \frac{10}{2} \right\rfloor+1=6$. In relation to the EC algorithm, we can conclude that the parameter k is directly proportional to the quorum size. But as the value of k and quorum size increase, the size of coded elements decreases.

\noindent\textbf{Distributed Experiments:} 
 For the distributed experiments we use a \emph{stochastic} invocation scheme in which readers, writers and reconfigurers pick a random time between intervals to invoke their next operations. Respectively the intervals are $[1...rInt]$, $[1..wInt]$ and  $[1..reconInt]$, where $rInt, wInt = 2sec$ and $reconInt = 15sec$. In total, each writer performs 60 writes, each reader 60 reads and the reconfigurer if any 6 reconfigurations. 

\paragraph{In particular, we present four types of scenarios:}
\begin{itemize}
 \item\textbf{File Size Scalability (Emulab):}
The first scenario is made to evaluate how the read and write latencies are affected by the size of the shared object. There are two separated runs, one for each examined storage algorithm. The number of readers is fixed to 5, without any reconfigurers. The file size is doubled from \SI{1}{\mega\byte} to \SI{128}{\mega\byte}.

\item\textbf{Reader Scalability (Emulab):}
This scenario is constructed to compare the read and write latency of the system with two different storage algorithms, while the readers increase. In particular, we execute two separate runs, one for each storage algorithm. We use only one reconfigurer which requests recon operations that lead to the same shared memory emulation and server nodes. The size of the file used is \SI{4}{\mega\byte}.

\item\textbf{Changing Reconfigurations (Emulab):}
In this scenario, we evaluate how the read and write latencies are affected when increasing the number of readers, while also changing the storage algorithm. We run two different runs which differ in the way the reconfigurer chooses the next storage algorithm: (i) the reconfigurer chooses randomly between the two storage algorithms, and (ii) the reconfigurer switches between the two storage algorithms. The size of the file used, in both scenarios, is \SI{4}{\mega\byte}.

\item\textbf{Consistency Persistence (Local):}
In this scenario we run multiple client operations in order to check if the data is consistent across servers. The number of readers is set to 5. The readers and writers invoke their next operations without any time delay, while the reconfigurer waits $15sec$ for the next invocation. In total, each writer performs 500 writes, each reader 500 reads and the reconfigurer 50 reconfigurations. The size of the file used is \SI{4}{\mega\byte}.

\end{itemize}

\subsection{Experimental Results}
\label{ssec:results}
% storage and communicational cost

In this section, we present and explain the evaluation results of each scenario. As a general observation, the \ares{} algorithm with the EC storage provides data redundancy with a lower communicational and storage cost compared to the ABD storage that uses a strict replication technique. 

\begin{figure}[t]
{
\begin{tabular}{cc}
	\includegraphics[scale=0.5,width=0.5\textwidth,height=50mm]{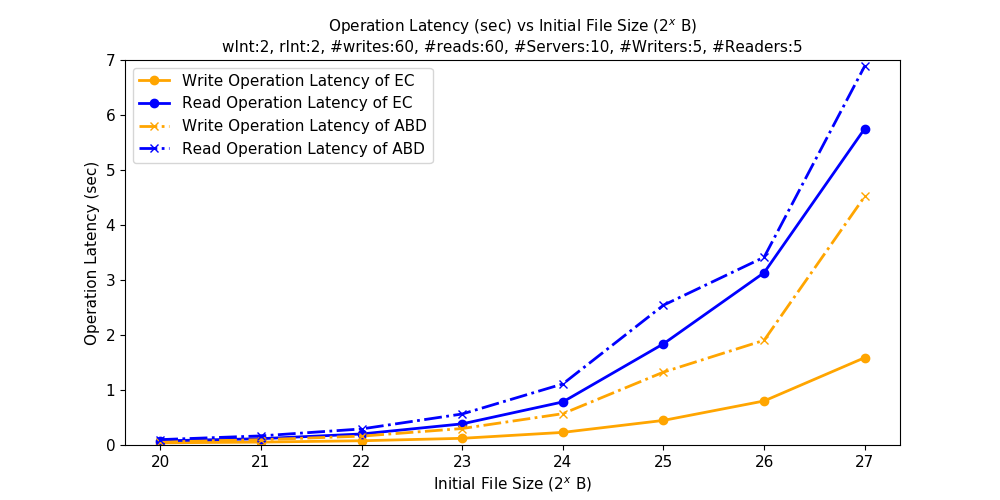}
	&
	\includegraphics[scale=0.5,width=0.5\textwidth,height=50mm]{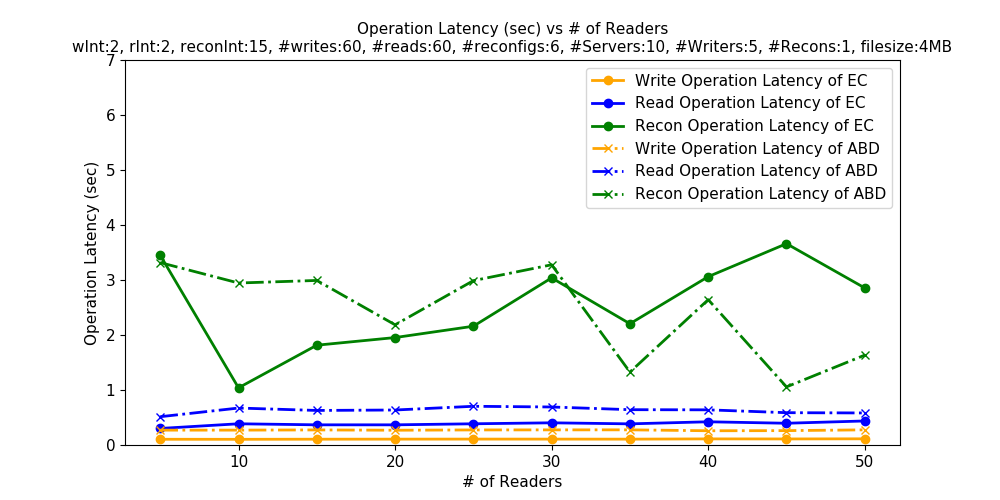}\\
	(a)  & (b) \\
	\includegraphics[scale=0.5,width=0.5\textwidth,height=50mm]{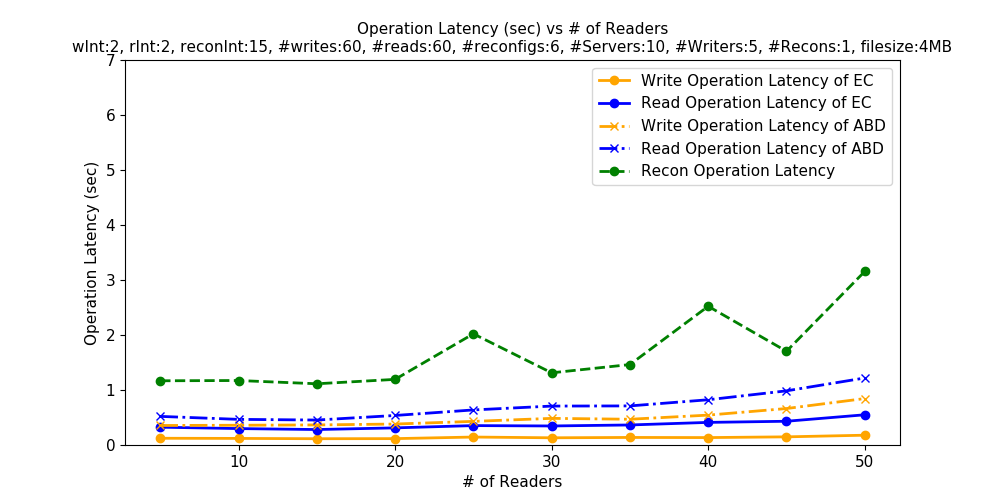}
	&		 
	\includegraphics[scale=0.5,width=0.5\textwidth,height=50mm]{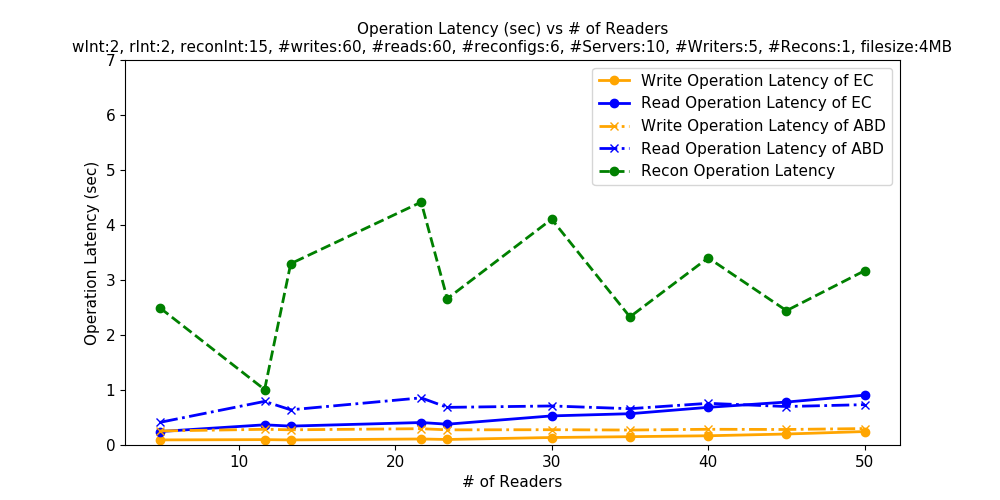}\\
	(c) & (d) \\ 
	
\end{tabular}
}
\vspace{-1em}
\caption{
Simulation results.
}
\label{fig:plots}
\end{figure}

\noindent\textbf{File Size Scalability Results:}
Fig. \ref{fig:plots}(a) shows the results of the file size scalability experiments. The read and write latencies of both storage algorithms remain in low levels until \SI{16}{\mega\byte}. In bigger sizes we observe the latancies
of all operations to grow significantly. It is worth noting that the fragmentation applied by the EC algorithm, benefits its write operations 
which follow a slower increasing curve than the rest of the operations. From the 
rest the reads seem to the one to suffer the worst delay hit, as they are 
engaged in more communication phases. Nevertheless, the larger messages 
sent by ABD result in slower read operations. 

% After that, the read and write latencies of ABD grow significantly. However the write latency of EC increases slightly and its read latency increases less than the one of ABD. The read operation is slow in both algorithms, since it transfers data twice, once to fetch the data from the servers, and once during the propagation phase. However, the read operation of ABD is slower than the EC, since it sends larger messages. 

\noindent\textbf{Reader Scalability Results:}
The results of reader scalability experiments can be found in Fig. \ref{fig:plots}(b). The read and write latencies of both algorithms remain almost unchanged, while the number of readers increases. This indicates that the system does not reach a state where it can not handle the concurrent read operations. Still, the reduced message size of read and write operation in EC keep their latencies lower than the coresponding latencies of ABD.
%, since EC uses coding which reduces the message size. 
On the other hand, the reconfiguration latency in both algorithms witnesses wild fluctuations between about 1 sec and 4 sec. This is probably due to the unstable connection in the external service which handles the reconfigurations.  

\noindent\textbf{Changing Reconfigurations Results:}
Fig. \ref{fig:plots}(c) illustrates the results of experiments with the random storage change. While, in Fig. \ref{fig:plots}(d), we can find the results of the experiments when the reconfigurer switches between storage algorithms. During both experiments, there are cases where a single read/write operation may access configurations that implement both ABD and EC algorithms, when concurrent with a recon operation. Thus, the latencies of such operations are accounted in both ABD and EC latencies. As we mentioned earlier, our choice of k minimizes the coded fragment size 
%and thus the amount of redundant information that is added to each fragment. However, this choice 
but introduces bigger quorums and thus larger communication overhead. 
%adversely affects the system since each client waits for more responses. 
As a result, in smaller file sizes, the \ares{} may not benefit so much from the coding, bringing the delays of the two storage algorithms closer to each other. It is again obvious that the reconfiguration delays is higher than the delays of all other operations. 

\noindent\textbf{Consistency Persistence Results:}
Though \ares{} protocol is provably strongly consistent, it is important ensure that our implementation is correct. Validating strong consistency of an execution requires precise clock synchronization across all processes, so that one can track operations with respect to a global time. This is impossible to achieve in a distributed system where clock drift is inevitable. To circumvent this, we deploy all the processes in a single beefy machine so that every process observers the same clock running in the same physical machine. 

Our checker gathers data regarding an execution, and this data includes start and end times of all the operations, as well as other parameters like logical timestamps used by the protocol. The checker logic is based on the conditions appearing in Lemma 13.16~\cite{Lynch1996}, which provide a set of sufficient conditions for guaranteeing strong consistency. The checker validates strong consistency property for every atomic object individually for the execution under consideration.

\section{Conclusions}
\label{sec:conclusions}
We presented an algorithmic framework suitable for reconfigurable, 
 erasure code-based atomic memory service in asynchronous,  message-passing environments. We provided experimental results of our initial prototype that can store multiple objects.
We also provided a new two-round  erasure code-based algorithm that has near optimal storage cost,  and 
%bandwidth  costs per read or write operation.
 Moreover, this algorithm is suitable specifically where during
new configuration installation  the object values passes directly from servers in older configuration to those
%in the newer configurations.
 Future work will involve adding efficient repair and reconfiguration using regenerating codes.

%%%
%%% BIBLIOGRAPHY
%%%
\bibliographystyle{acm}
%\bibliographystyle{plain}
%\bibliographystyle{abbrv}
%\bibliography{biblio,cadambe-refs,evaluation}

\begin{thebibliography}{10}
	
	\bibitem{ansible}
	Ansible.
	\newblock \url{https://www.ansible.com/overview/how-ansible-works}.
	
	\bibitem{emulab}
	Emulab network testbed.
	\newblock \url{https://www.emulab.net/}.
	
	\bibitem{intel-isal}
	Intel storage acceleration library (open source version).
	\newblock https://goo.gl/zkVl4N.
	
	\bibitem{EC-Store}
	{\sc Abebe, M., Daudjee, K., Glasbergen, B., and Tian, Y.}
	\newblock Ec-store: Bridging the gap between storage and latency in distributed
	erasure coded systems.
	\newblock In {\em 2018 IEEE 38th International Conference on Distributed
		Computing Systems (ICDCS)\/} (July 2018), pp.~255--266.
	
	\bibitem{AKMS09}
	{\sc Aguilera, M.~K., Keidar, I., Malkhi, D., and Shraer, A.}
	\newblock Dynamic atomic storage without consensus.
	\newblock In {\em Proceedings of the 28th ACM symposium on Principles of
		distributed computing (PODC '09)\/} (New York, NY, USA, 2009), ACM,
	pp.~17--25.
	
	\bibitem{aguileratutorial}
	{\sc Aguilera, M.~K., Keidary, I., Malkhi, D., Martin, J.-P., and Shraery, A.}
	\newblock Reconfiguring replicated atomic storage: A tutorial.
	\newblock {\em Bulletin of the EATCS 102\/} (2010), 84--081.
	
	\bibitem{FNP15}
	{\sc Anta, A.~F., Nicolaou, N., and Popa, A.}
	\newblock Making ``fast'' atomic operations computationally tractable.
	\newblock In {\em International Conference on Principles Of Distributed
		Systems\/} (2015), OPODIS'15.
	
	\bibitem{ABD96}
	{\sc Attiya, H., Bar-Noy, A., and Dolev, D.}
	\newblock Sharing memory robustly in message passing systems.
	\newblock {\em Journal of the ACM 42(1)\/} (1996), 124--142.
	
	\bibitem{burihabwa2016performance}
	{\sc Burihabwa, D., Felber, P., Mercier, H., and Schiavoni, V.}
	\newblock A performance evaluation of erasure coding libraries for cloud-based
	data stores.
	\newblock In {\em Distributed Applications and Interoperable Systems\/} (2016),
	Springer, pp.~160--173.
	
	\bibitem{CT06}
	{\sc Cachin, C., and Tessaro, S.}
	\newblock Optimal resilience for erasure-coded byzantine distributed storage.
	\newblock In {\em Dependable Systems and Networks, International Conference
		on\/} (Los Alamitos, CA, USA, 2006), IEEE Computer Society, pp.~115--124.
	
	\bibitem{CLMM14}
	{\sc Cadambe, V.~R., Lynch, N., M{\'e}dard, M., and Musial, P.}
	\newblock A coded shared atomic memory algorithm for message passing
	architectures.
	\newblock In {\em Network Computing and Applications (NCA), 2014 IEEE 13th
		International Symposium on\/} (Aug 2014), pp.~253--260.
	
	\bibitem{CadambeLMM17}
	{\sc Cadambe, V.~R., Lynch, N.~A., M{\'{e}}dard, M., and Musial, P.~M.}
	\newblock A coded shared atomic memory algorithm for message passing
	architectures.
	\newblock {\em Distributed Computing 30}, 1 (2017), 49--73.
	
	\bibitem{GIZA2017}
	{\sc Chen, Y. L.~C., Mu, S., and Li, J.}
	\newblock Giza: Erasure coding objects across global data centers.
	\newblock In {\em Proceedings of the 2017 USENIX Annual Technical Conference
		(USENIX ATC ’17)\/} (2017), pp.~539--551.
	
	\bibitem{CGGMS09}
	{\sc Chockler, G., Gilbert, S., Gramoli, V., Musial, P.~M., and Shvartsman,
		A.~A.}
	\newblock Reconfigurable distributed storage for dynamic networks.
	\newblock {\em Journal of Parallel and Distributed Computing 69}, 1 (2009),
	100--116.
	
	\bibitem{GD05}
	{\sc Chockler, G., and Malkhi, D.}
	\newblock Active disk paxos with infinitely many processes.
	\newblock {\em Distributed Computing 18}, 1 (2005), 73--84.
	
	\bibitem{DGL08}
	{\sc Dutta, P., Guerraoui, R., and Levy, R.~R.}
	\newblock Optimistic erasure-coded distributed storage.
	\newblock In {\em DISC '08: Proceedings of the 22nd international symposium on
		Distributed Computing\/} (Berlin, Heidelberg, 2008), Springer-Verlag,
	pp.~182--196.
	
	\bibitem{CDGL04}
	{\sc Dutta, P., Guerraoui, R., Levy, R.~R., and Chakraborty, A.}
	\newblock How fast can a distributed atomic read be?
	\newblock In {\em Proceedings of the 23rd ACM symposium on Principles of
		Distributed Computing (PODC)\/} (2004), pp.~236--245.
	
	\bibitem{FL03}
	{\sc Fan, R., and Lynch, N.}
	\newblock Efficient replication of large data objects.
	\newblock In {\em Distributed algorithms\/} (2003), F.~E. Fich, Ed., vol.~2848
	of {\em Lecture Notes in Computer Science}, pp.~75--91.
	
	\bibitem{FHN16}
	{\sc Fern{\'a}ndez~Anta, A., Hadjistasi, T., and Nicolaou, N.}
	\newblock Computationally light ``multi-speed'' atomic memory.
	\newblock In {\em International Conference on Principles Of Distributed
		Systems\/} (2016), OPODIS'16.
	
	\bibitem{GM15}
	{\sc Gafni, E., and Malkhi, D.}
	\newblock {Elastic Configuration Maintenance via a Parsimonious Speculating
		Snapshot Solution}.
	\newblock In {\em International Symposium on Distributed Computing\/} (2015),
	Springer, pp.~140--153.
	
	\bibitem{GNS08}
	{\sc Georgiou, C., Nicolaou, N.~C., and Shvartsman, A.~A.}
	\newblock On the robustness of (semi) fast quorum-based implementations of
	atomic shared memory.
	\newblock In {\em DISC '08: Proceedings of the 22nd international symposium on
		Distributed Computing\/} (Berlin, Heidelberg, 2008), Springer-Verlag,
	pp.~289--304.
	
	\bibitem{GNS09}
	{\sc Georgiou, C., Nicolaou, N.~C., and Shvartsman, A.~A.}
	\newblock Fault-tolerant semifast implementations of atomic read/write
	registers.
	\newblock {\em Journal of Parallel and Distributed Computing 69}, 1 (2009),
	62--79.
	
	\bibitem{G03}
	{\sc Gilbert, S.}
	\newblock {RAMBO II}: Rapidly reconfigurable atomic memory for dynamic
	networks.
	\newblock Master's thesis, MIT, August 2003.
	
	\bibitem{GLS03}
	{\sc Gilbert, S., Lynch, N., and Shvartsman, A.}
	\newblock {RAMBO II}: Rapidly reconfigurable atomic memory for dynamic
	networks.
	\newblock In {\em Proceedings of International Conference on Dependable Systems
		and Networks (DSN)\/} (2003), pp.~259--268.
	
	\bibitem{HW90}
	{\sc Herlihy, M.~P., and Wing, J.~M.}
	\newblock Linearizability: a correctness condition for concurrent objects.
	\newblock {\em ACM Transactions on Programming Languages and Systems 12}, 3
	(1990), 463--492.
	
	\bibitem{verapless_book}
	{\sc Huffman, W.~C., and Pless, V.}
	\newblock {\em Fundamentals of error-correcting codes}.
	\newblock Cambridge university press, 2003.
	
	\bibitem{LVM15}
	{\sc Jehl, L., Vitenberg, R., and Meling, H.}
	\newblock Smartmerge: A new approach to reconfiguration for atomic storage.
	\newblock In {\em International Symposium on Distributed Computing\/} (2015),
	Springer, pp.~154--169.
	
	\bibitem{joshi2017efficient}
	{\sc Joshi, G., Soljanin, E., and Wornell, G.}
	\newblock Efficient redundancy techniques for latency reduction in cloud
	systems.
	\newblock {\em ACM Transactions on Modeling and Performance Evaluation of
		Computing Systems (TOMPECS) 2}, 2 (2017), 12.
	
	\bibitem{SODA2016}
	{\sc Konwar, K.~M., Prakash, N., Kantor, E., Lynch, N., M{\'e}dard, M., and
		Schwarzmann, A.~A.}
	\newblock Storage-optimized data-atomic algorithms for handling erasures and
	errors in distributed storage systems.
	\newblock In {\em 2016 IEEE International Parallel and Distributed Processing
		Symposium (IPDPS)\/} (May 2016), pp.~720--729.
	
	\bibitem{radon}
	{\sc Konwar, K.~M., Prakash, N., Lynch, N., and M{\'e}dard, M.}
	\newblock Radon: Repairable atomic data object in networks.
	\newblock In {\em The International Conference on Distributed Systems
		(OPODIS)\/} (2016).
	
	\bibitem{L98}
	{\sc Lamport, L.}
	\newblock The part-time parliament.
	\newblock {\em ACM Transactions on Computer Systems 16}, 2 (1998), 133--169.
	
	\bibitem{Lynch1996}
	{\sc Lynch, N.}
	\newblock {\em Distributed Algorithms}.
	\newblock Morgan Kaufmann Publishers, 1996.
	
	\bibitem{LS02}
	{\sc Lynch, N., and Shvartsman, A.}
	\newblock {RAMBO}: A reconfigurable atomic memory service for dynamic networks.
	\newblock In {\em Proceedings of 16th International Symposium on Distributed
		Computing (DISC)\/} (2002), pp.~173--190.
	
	\bibitem{LS97}
	{\sc Lynch, N.~A., and Shvartsman, A.~A.}
	\newblock Robust emulation of shared memory using dynamic quorum-acknowledged
	broadcasts.
	\newblock In {\em Proceedings of Symposium on Fault-Tolerant Computing\/}
	(1997), pp.~272--281.
	
	\bibitem{N08bitcoin}
	{\sc Nakamoto, S.}
	\newblock Bitcoin: A peer-to-peer electronic cash system.
	
	\bibitem{ARES:Arxiv:2018}
	{\sc Nicolaou, N., Cadambe, V., Konwar, K., Prakash, N., Lynch, N., and
		M{\'{e}}dard, M.}
	\newblock Ares: Adaptive, reconfigurable, erasure coded, atomic storage.
	\newblock {\em CoRR abs/1805.03727\/} (2018).
	
	\bibitem{Raft}
	{\sc Ongaro, D., and Ousterhout, J.}
	\newblock In search of an understandable consensus algorithm.
	\newblock In {\em Proceedings of the 2014 USENIX Conference on USENIX Annual
		Technical Conference\/} (Berkeley, CA, USA, 2014), USENIX ATC'14, USENIX
	Association, pp.~305--320.
	
	\bibitem{EC-Cache}
	{\sc Rashmi, K., Chowdhury, M., Kosaian, J., Stoica, I., and Ramchandran, K.}
	\newblock Ec-cache: Load-balanced, low-latency cluster caching with online
	erasure coding.
	\newblock In {\em OSDI\/} (2016), pp.~401--417.
	
	\bibitem{SMMK2010}
	{\sc Shraer, A., Martin, J.-P., Malkhi, D., and Keidar, I.}
	\newblock {Data-centric reconfiguration with network-attached disks}.
	\newblock In {\em Proceedings of the 4th Int'l Workshop on Large Scale
		Distributed Sys. and Middleware (LADIS ’10)\/} (2010), p.~22–26.
	
	\bibitem{spiegelman:DISC:2017}
	{\sc Spiegelman, A., Keidar, I., and Malkhi, D.}
	\newblock {Dynamic Reconfiguration: Abstraction and Optimal Asynchronous
		Solution}.
	\newblock In {\em 31st International Symposium on Distributed Computing (DISC
		2017)\/} (2017), vol.~91, pp.~40:1--40:15.
	
	\bibitem{WPS}
	{\sc Wang, S., Huang, J., Qin, X., Cao, Q., and Xie, C.}
	\newblock Wps: A workload-aware placement scheme for erasure-coded in-memory
	stores.
	\newblock In {\em Networking, Architecture, and Storage (NAS), 2017
		International Conference on\/} (2017), IEEE, pp.~1--10.
	
	\bibitem{xiang2015multi}
	{\sc Xiang, Y., Lan, T., Aggarwal, V., and Chen, Y.-F.~R.}
	\newblock Multi-tenant latency optimization in erasure-coded storage with
	differentiated services.
	\newblock In {\em 2015 IEEE 35th International Conference on Distributed
		Computing Systems (ICDCS)\/} (2015), IEEE, pp.~790--791.
	
	\bibitem{xiang2016joint}
	{\sc Xiang, Y., Lan, T., Aggarwal, V., Chen, Y.-F.~R., Xiang, Y., Lan, T.,
		Aggarwal, V., and Chen, Y.-F.~R.}
	\newblock Joint latency and cost optimization for erasure-coded data center
	storage.
	\newblock {\em IEEE/ACM Transactions on Networking (TON) 24}, 4 (2016),
	2443--2457.
	
	\bibitem{yu2018sp}
	{\sc Yu, Y., Huang, R., Wang, W., Zhang, J., and Letaief, K.~B.}
	\newblock Sp-cache: load-balanced, redundancy-free cluster caching with
	selective partition.
	\newblock In {\em Proceedings of the International Conference for High
		Performance Computing, Networking, Storage, and Analysis\/} (2018), IEEE
	Press, p.~1.
	
	\bibitem{Zhang2016}
	{\sc Zhang, H., Dong, M., and Chen, H.}
	\newblock Efficient and available in-memory kv-store with hybrid erasure coding
	and replication.
	\newblock In {\em 14th {USENIX} Conference on File and Storage Technologies
		({FAST} 16)\/} (Santa Clara, CA, 2016), {USENIX} Association, pp.~167--180.
	
	\bibitem{PARS}
	{\sc Zhou, P., Huang, J., Qin, X., and Xie, C.}
	\newblock Pars: A popularity-aware redundancy scheme for in-memory stores.
	\newblock {\em IEEE Transactions on Computers\/} (2018), 1--1.
	
\end{thebibliography}

%
%\appendix
%\input{appendix.v1.tex}

\end{document}